\documentclass[10pt]{article}
\pdfoutput=1
\PassOptionsToPackage{hyphens}{url}
\usepackage[preprint]{tmlr}

\usepackage{amsmath,amsfonts,bm}

\def\eqref#1{equation~\ref{#1}}

\def\1{\bm{1}}

\DeclareMathAlphabet{\mathsfit}{\encodingdefault}{\sfdefault}{m}{sl}
\SetMathAlphabet{\mathsfit}{bold}{\encodingdefault}{\sfdefault}{bx}{n}

\usepackage{hyperref}
\usepackage{url}
\usepackage{booktabs}
\usepackage{multirow}
\usepackage{graphicx}
\usepackage{amsmath,amssymb}
\usepackage{mathtools}%
\usepackage{algorithm}
\usepackage{algpseudocode}
\makeatletter\renewcommand{\theHALG@line}{\thealgorithm.\arabic{ALG@line}}\makeatother
\usepackage{subcaption}
\usepackage{xcolor}
\usepackage{listings}
\usepackage{amsthm}
\usepackage{tikz}%
\usetikzlibrary{decorations.pathreplacing,positioning,calc}
\usepackage[capitalize,noabbrev,nameinlink]{cleveref}
\newcommand{\yes}{$\checkmark$}\newcommand{\no}{$\times$}%

\crefname{lstlisting}{Listing}{Listings}

\graphicspath{{figures/}}

\newtheorem{proposition}{Proposition}

\newtheorem{theorem}{Theorem}

\title{Computed Permutation Trellis Coding:\\
Near-Optimal Quantization from\\
Two Multiplications and a 64-Byte Table}

\author{\name Muhammed Fatih Bal{\i}n \email mfbalin@amazon.com \\
      \addr Amazon Annapurna Labs \\
      Cupertino, CA, USA}

\newcommand{\cptc}{CPTC}

\hypersetup{colorlinks=true, allcolors=blue!55!black, pdftitle={Computed Permutation Trellis Coding: Near-Optimal Quantization from Two Multiplications and a 64-Byte Table}, pdfauthor={Muhammed Fatih Bal\i n}}

\begin{document}

\maketitle

\begin{abstract}
We propose \textbf{Computed Permutation Trellis Coding
({CPTC})}, a low-bit quantization format for tensors
written once and read many times. Formats that round each value on
its own, such as INT4 and FP4, waste a large part of their
bits against the information-theoretic ideal. Trellis codes recover
this loss by encoding values jointly, but prior decoders either need
a large lookup table for their best quality or compute values with a
hash that restricts what they can decode. CPTC's decoder multiplies
its state by two coefficients and adds the top bits of the
products to index a $64$-byte table. Because that sum is not
uniformly distributed, the rarely used ends of a small table can hold
values far out in the tails. One decoder serves every bit width from
$1$ to $6$, reproduces the standard integer and floating-point
formats exactly, and comes within $0.07$ bits of the ideal at $4$
bits. It matches a random-permutation trellis code whose table is
thousands of times larger, and exceeds every QTIP codebook. In LLM weight quantization without
fine-tuning, CPTC has the lowest KL divergence to the unquantized
model at every tested bit width on Llama 3.1 8B and 70B Instruct
against QTIP's codebooks (by up to $27\%$). The decoder has been
built into the AWS Trainium~4 AI accelerator, where it runs at the
throughput of the native FP4 and FP8 formats for a
negligible fraction of the systolic array's area. Fused encoding
kernels quantize an 8B model's linear layers in
under two minutes on one GPU and, in cycle-level simulation, on one
Trainium~4 chip.
\end{abstract}

\section{Introduction}
\label{sec:intro}

When a large language model generates text for a few users at a
time, the speed limit is not arithmetic but memory: every token
requires streaming all model weights from memory into the compute
units, so halving the bits per weight roughly doubles tokens per
second. At four bits and below, though, every $0.1$ bit of wasted
precision shows up in model quality. Weights are the leading example
of a tensor that is written once and read many times, and such
tensors set the design point of this paper: the write can afford a
search, but every read decodes each value at memory bandwidth, so it
must cost no more than a few integer operations.

\begin{figure}[t]
  \centering
  \begin{subfigure}[t]{0.48\linewidth}
    \centering
    \includegraphics[width=\linewidth]{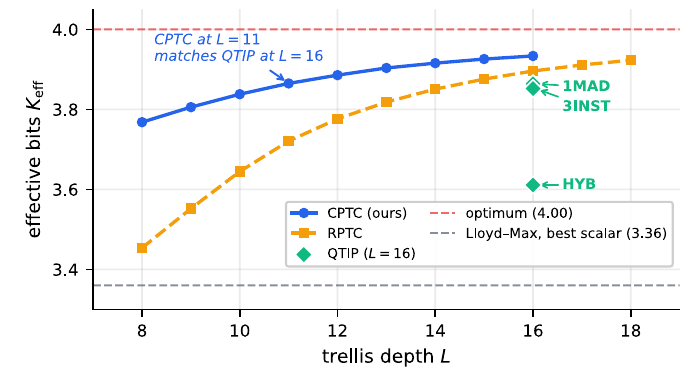}
    \caption{Effective bits ($\uparrow$) against trellis depth
      $L$ at $4$ bits: CPTC, RPTC, the QTIP
      variants at $L{=}16$, the best scalar format (Lloyd--Max) and
      the optimum $4.00$ (\cref{tab:qtip_k}).}
    \label{fig:cptc_vs_all}
  \end{subfigure}\hfill
  \begin{subfigure}[t]{0.48\linewidth}
    \centering
    \includegraphics[width=\linewidth]{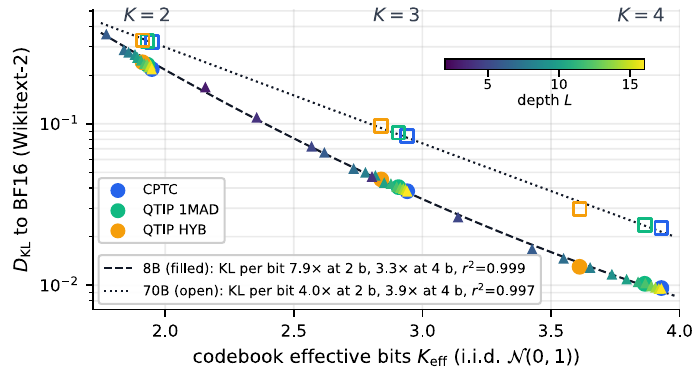}
    \caption{KL to BF16 ($\downarrow$, seed $0$) of Llama 3.1 8B (filled) and 70B
      (open) against $K_{\text{eff}}$ at $2$, $3$ and $4$ bits,
      plus the CPTC depth sweep (triangles). Across rates each bit is worth
      $4\times$ in KL at 70B and $3$--$8\times$ at 8B (curves; \cref{sec:exp_llm}).}
    \label{fig:kl_vs_keff}
  \end{subfigure}
  \caption{Effective bits against trellis depth (a) and against LLM quality (b).}
  \label{fig:keff_pair}
\end{figure}

How good can a format with $K$ bits per value be? For Gaussian data,
which LLM weights are made to resemble by standard preprocessing
(\cref{sec:background}), information theory bounds the error of any
$K$-bit quantizer: at unit variance its mean squared error $D$ is at
least $4^{-K}$ \citep{berger1971rate}. We therefore score quantizers in
\emph{effective bits}, $K_{\text{eff}} = -\log_4(D)$, so the ideal
$4$-bit quantizer scores $4.00$. Effective bits predict model quality:
the quantized LLM's KL divergence from the original falls smoothly as
$K_{\text{eff}}$ rises (\cref{fig:kl_vs_keff}). A better Gaussian quantizer is a better LLM
quantizer, so the rest of the paper compares formats on
$K_{\text{eff}}$. Formats that round each value on its own fall well
short: the best possible, the Lloyd--Max quantizer
\citep{lloyd1982lsq,max1960quantizing}, scores $3.36$, and INT4 and FP4
lower still (\cref{tab:hw_criteria}). The missing $0.6$ bits
can only be recovered by encoding values \emph{jointly}.

\emph{Trellis codes} \citep{mao2011rptc,tseng2024qtip} do that while
keeping decoding simple: each value is decoded from the last $L$ bits
of the code stream, the \emph{state}, by a fixed map ($L$ is the
trellis depth). A map built into hardware must be cheap, serve every
bit width, adapt to new distributions and deliver high quality
(\cref{sec:hw_criteria}), and prior maps each give one of these up.
Random Permutation Trellis Coding (RPTC; \citealp{mao2011rptc})
stores a table with $2^L$ entries. Linear congruential (LC) codes
\citep{eriksson2007lc} compute a table index with one multiplication
but need kilobytes of table for their best quality. QTIP
\citep{tseng2024qtip} replaces the table with a hash, which fixes the
distribution of the decoded values, or keeps a small table that limits
which values it can decode.

We show that \emph{two} multiplications shrink the table to $64$
bytes. Our decoder multiplies the state by two coefficients, adds the top
bits of the two products, and uses the sum to index the table. The sum is approximately triangular rather than uniform,
like the sum of two dice: it lands near the middle of the table often
and at its ends rarely. The rarely used end entries can therefore hold
values far out in the tails of a bell-shaped distribution, further
than a uniformly used table of the same size reaches
(\cref{sec:theory_triangular}). At $4$ bits CPTC reaches
$3.93$ effective bits, matching RPTC with a table $8{,}192\times$
larger (\cref{fig:cptc_vs_all}).

One decoder serves every bit width from $1$ to $6$ at the same cost.
Encoding searches over $2^L$ states per value. Our fused kernels run
it at tens of millions to billions of values per second on a GPU and,
in cycle-level simulation, on Trainium~4 itself (\cref{sec:encode}).
This opens the door to other tensors that are written once and read
many times: key-value (KV) caches, and the weights in
quantization-aware training, which many accelerators read for multiple
minibatches between updates.
Our main contributions are:
\begin{itemize}
\item \textbf{Decoder.} Two $16$-bit multiplications and a $64$-byte table
  come within $0.07$ bits of the ideal up to $4$ bits, at a cost that
  is the same for every bit width from $1$ to $6$
  (\cref{sec:method,tab:qtip_k}).
\item \textbf{Theory.} Every scalar format of up to $6$ bits is a
  special case of this decoder (\cref{sec:constants}). A bound on what
  any fixed table can reach, however deep the trellis, shows how much
  quality the table itself costs (\cref{thm:ceiling}). We also
  quantify what a smaller table or narrower table entries would cost
  (\cref{sec:exp_iid}).
\item \textbf{Quality.} CPTC leads every QTIP codebook at every bit
  width, and up to $4$ bits it is within $0.01$ bits of LC codes that
  store kilobytes of table (\cref{tab:qtip_k}). Loading a different $64$-byte
  table retargets it to other distributions
  (\cref{sec:exp_iid,sec:exp_matmul_synth}). On Llama 3.1 8B and 70B,
  quantized without fine-tuning in the YAQA pipeline
  \citep{tseng2025yaqa}, it has the lowest KL divergence at every bit
  width against QTIP's table-based and best hash-based codebooks
  (\cref{sec:exp_llm}).
\item \textbf{Systems.} The decoder is built into AWS Trainium~4
  \citep{aws2025trainium4}, where it runs at the throughput of the
  native FP4 and FP8 formats for a negligible fraction of the systolic
  array's area (\cref{sec:hardware}). A fused GPU encoder runs
  $2{,}400\times$ faster than QTIP's released encoder at equal quality:
  $32\times$ comes from the method, which reaches that quality with a
  trellis five bits shallower, and the remaining $75\times$ from the
  fused kernel, which is engineering. A Trainium~4 encoder, in cycle-level simulation,
  quantizes tensors on the chip itself. Either one encodes an 8B
  model's linear layers in under two minutes (\cref{sec:encode}).
\end{itemize}

\section{Background}
\label{sec:background}

\paragraph{LLM weight quantization and the standard-normal reduction.}
When few tokens are processed at once, an LLM layer's speed is
set by how fast its weights stream from memory, so weights are
compressed once and decompressed on the fly in the
weight-load path. Incoherence processing \citep{chee2023quip} with the \emph{Random
Hadamard Transform} (RHT; \citealp{tseng2024quip_sharp}) reduces this to a
clean statistical problem. With $H_n$ the orthonormal Hadamard
matrix, built by doubling for $n$ a power of two, and random sign
vectors $s_U$, $s_V$, the RHT of an $m \times n$ weight matrix $W$ is
\begin{equation*}
  \widetilde{W} = U W V^\top, \;\;
  U = H_m \operatorname{diag}(s_U), \;
  V = H_n \operatorname{diag}(s_V), \quad
  H_1 = (1), \;
  H_{2n} = \tfrac{1}{\sqrt{2}}
    \begin{pmatrix} H_n & H_n \\ H_n & -H_n \end{pmatrix}.
\end{equation*}
$U$ and $V$ are orthogonal, so the layer computes
$W x = U^\top \widetilde{W} V x$ with fast transforms on its input and
output, and every entry of $\widetilde{W}$ is a signed sum of all
entries of $W$, so after rescaling the entries closely approximate
i.i.d.\ standard-normal draws.
We adopt this preprocessing unchanged, so CPTC's quality is
directly comparable to prior work built on it, and what remains
is quantizing i.i.d.\ Gaussian samples as accurately as possible.

\paragraph{Effective bits.}
Let $X$ be the unit-variance source (the target distribution after
the preprocessing above), $\widehat{X}$ its reconstruction,
$D = \mathbb{E}\big[(X - \widehat{X})^2\big]$ the distortion
(MSE), and $R_X(D)$ the rate-distortion function of $X$
under squared error \citep{berger1971rate}: the
minimum rate, in bits per value, at which \emph{any} coder
reaches distortion $D$. We define \emph{effective bits} as
the Shannon lower bound on $R_X$ at the measured distortion, the
left end of the classical sandwich
\begin{equation}
  K_{\text{eff}}
  \;\coloneqq\; -\log_4 \frac{D}{N(X)}
  \;\le\; R_X(D) \;\le\;
  -\log_4 \frac{D}{N(X + Z)},
  \qquad
  N(X) = \frac{e^{2 h(X)}}{2\pi e},
  \label{eq:keff}
\end{equation}
where $N$ is entropy power, $h$ is differential entropy in nats,
and $Z \sim \mathcal{N}(0, D)$ is independent test-channel
noise. A quantizer spending $K$
bits per value has $K_{\text{eff}} \le R_X(D) \le K$, so the
deficit $K - K_{\text{eff}}$ never understates the wasted rate,
and one effective bit is a $4\times$ factor in distortion. For
the Gaussian, our default target, the bound is exact
($N(X) = 1$); for the other targets
\cref{app:gg} verifies that the sandwich is thin at $K \ge 3$.

\paragraph{From scalar to joint quantization.}
The scalar ceiling of \cref{sec:intro} (Lloyd--Max at $3.36$
effective bits, with INT4 and FP4 below it) binds every method that
rounds values one at a time. GPTQ \citep{frantar2023gptq}, AWQ
\citep{lin2023awq}, and SqueezeLLM \citep{kim2023squeezellm}
improve the \emph{assignment} of values to scalar codepoints and
inherit the scalar ceiling; CPTC replaces the format itself, standing
where INT4 or FP4 stand, so these methods compose with it. Joint quantization escapes the ceiling
\citep{conway1999sphere}: QuIP\# \citep{tseng2024quip_sharp}
uses the $E_8$ lattice, NestQuant \citep{savkin2025nestquant}
nests it self-similarly, LLVQ \citep{vanderouderaa2026llvq}
extends to the $24$-dimensional Leech lattice, and
learned-codebook vector quantization (VQ; AQLM, \citealp{egiazarian2024aqlm}; VPTQ,
\citealp{liu2024vptq}) trains megabyte-scale codebooks. CPTC
instead encodes values jointly as a walk through a $2^L$-state
trellis at a per-value decode cost of two multiplies and a lookup.

\paragraph{Trellis-coded quantization.}
\label{sec:tcq}
Trellis-coded quantization \citep{marcellin1990tcq} quantizes
jointly while spending exactly $K$ bits per value: each value's
reconstruction depends on its neighbors' code bits too. The
decoder keeps an $L$-bit state $q$, simply the most recent $L$
bits of the code stream, and maps it to a value through a fixed function $\phi$. In the \emph{bitshift}
subfamily relevant here,
reading the next value's $K$ bits just shifts them into the state
register, which is why decoding stays hardware-trivial. Although only $K$ bits are stored per value, consecutive values
share $L - K$ state bits, so the decoder effectively selects from
a $2^L$-entry codebook, the source of the quality beyond scalar
formats. The price is a Viterbi encoding search over all $2^L$ states
(\cref{sec:trellis}).

\paragraph{Prior bitshift trellis codes.}
\label{sec:prior_trellis}
The map $\phi$ from state to value is where prior designs differ.
\emph{Linear congruential} (LC) trellis codes
\citep{anderson2005lc,eriksson2007lc}, the earliest computed codebooks, take
branch labels from congruential maps of the state and read
values at the labels' quantiles from a stored table.
\emph{RPTC} \citep{mao2011rptc} shuffles the $2^L$ states
with a stored random permutation $P$ and assigns state $q$ the
value at quantile $(P[q]+\tfrac12)/2^L$ of the target's optimal
reproduction law; this provably
meets the marginal condition of asymptotically optimal codes and
empirically approaches the Shannon bound as $L$ grows, but the shuffle is
$2^L$ entries.
\emph{QTIP} \citep{tseng2024qtip} computes $\phi$ with a
hash: its 1MAD and 3INST variants need no table but can only
emit the approximately Gaussian law the hash produces, and its
HYB variant regains the choice of distribution through a stored
$k$-means codebook.

\paragraph{Requirements for a hardware codebook.}
\label{sec:hw_criteria}
Weight formats are decoded in a datapath fixed at tape-out and
replicated per lane beside the multiply-accumulate units, so
every lane pays for whatever the decoder holds, which sets four
requirements. It must be \emph{cheap}: a few integer
operations and register-resident state per value; a kilobyte-scale
table per lane, or one shared with a read port per lane, has no
place in the weight-load path. It must be \emph{general}
across bit widths: one circuit serving every rate, ideally
reproducing the scalar formats it replaces. It must be
\emph{flexible} in its output distribution: the target law will
change over the chip's lifetime, so retargeting must mean loading
state, not respinning silicon. Recent accelerators already expose
such state for scalar formats: AWS Trainium~3 converts MXFP4 \citep{rouhani2023mx} operands
to MXFP8 through an arbitrary programmer-defined mapping of the $16$
codes \citep{aws2025neuroncorev4}, and on NVIDIA Rubin GPUs the
tensor-core MMA can read one operand as $3$-bit indices into a lookup
table of $8$ E4M3 values per $8$-column slice
\citep{nvidia2026cuda,nvidia2026ptx}. Both tables are scalar
codebooks, so they stay below the scalar ceiling. And it must deliver \emph{high quality}, or the block
is not worth its area. Encoding is not exempt: how often a tensor is rewritten sets how
much search it can afford, so the
encoder's search cost counts too. A $T$-dimensional VQ searches $2^{TK}$
codewords unless it is a lattice with a fast closest-point algorithm;
trellis codes pay a Viterbi search over $2^L$ states, which CPTC's small-$L$ quality and fused encoders (\cref{sec:encode}) make cheap enough to open the door to online use. \Cref{tab:hw_criteria} scores
existing codebooks on these axes: each prior design gives up at
least one, and \cref{sec:method}
shows a \emph{computed} permutation meets all four.

\begin{table}[t]
  \centering
  \fontsize{7.6}{9.2}\selectfont
  \setlength{\tabcolsep}{2.5pt}%
  \caption{What each codebook costs and delivers, one column per
    requirement of \cref{sec:hw_criteria} plus the encoder's search: decoder work and storage
    per lane, whether one circuit serves every bit width, whether
    the output distribution can be changed by loading a table, how
    the encoder searches, and how much of the $4$ bits is wasted
    ($\downarrow$, bits consumed minus $K_{\text{eff}}$ on
    i.i.d.\ Gaussians; sources
    \cref{tab:qtip_k,fig:cptc_vs_all,app:iid_extended};
    n/r: not reported; NF4: \citealp{dettmers2023qlora}; NVFP4: \citealp{nvidia2025nvfp4}). $^\dagger$CPTC reaches QTIP's quality at $L{-}5$,
    with $32\times$ less search. $^\ddagger$At $4.25$ bits with a
    per-block scale index.}
  \label{tab:hw_criteria}
  \begin{tabular}{llcccc}
    \toprule
    Codebook & Decoder (ops, storage) & All $K$ & Retargetable & Encode/value & Waste at $4$ bits \\
    \midrule
    INT4 / FP4 & cast, $0$ & \no, per width & \no & round & $0.86$ / $0.85$ \\
    NF4 / Lloyd--Max & lookup, $32$\,B & \no, per width & \no{} / \yes & round & $0.67$ / $0.64$ \\
    MXFP4 / NVFP4 ($4.25$/$4.5$\,b) & cast $+$ scale, $0$ & \no, per width & \no & round & $1.09$ / $1.11$ \\
    E8P (QuIP\#) & signs $+$ $2$\,KiB LUT & \no, per rate & \no & E8 structured search & $0.56$ \\
    Nested E8 (NestQuant) & E8 closest point, $0$ & \yes & \no & E8 closest point, $O(T)$ & $0.36^{\ddagger}$ \\
    Learned VQ & MiB LUT & \no, per rate & \yes & beam search, $2^{TK}$ & n/r \\
    LC trellis & $1$ mul $+$ $8$--$128$\,KiB LUT & \yes & \yes & Viterbi, $2^L$ & $0.06$ \\
    RPTC & $128$\,KiB LUT & \yes & \yes & Viterbi, $2^L$ & $0.10$ \\
    QTIP HYB & hash $+$ $2$\,KiB LUT & \no, $K{\le}4$ & \yes & Viterbi, $2^L$ & $0.39$ \\
    QTIP 1MAD / 3INST & $1$--$3$ int ops, $0$ & \yes & \no, fixed law & Viterbi, $2^L$ & $0.14$ / $0.15$ \\
    CPTC (ours) & $2$ mul $+$ add $+$ $64$\,B LUT & \yes, $K{\in}[1,6]$ & \yes, exact scalar & Viterbi, $2^L$ ($2^{L-5}$)$^\dagger$ & $0.07$ \\
    \bottomrule
  \end{tabular}
\end{table}

\paragraph{AWS Trainium.}
\label{sec:trainium_background}
Each NeuronCore of an AWS Trainium chip pairs four compute engines
with software-managed on-chip SRAM
\citep{aws2025neuroncorev4,aws2026nkiarch}. Its main data store, the
SBUF, is a two-dimensional memory of $128$ \emph{partitions}: a
tensor's leading axis, the \emph{partition dimension}, maps onto the
partitions, and its remaining axes, the \emph{free dimension}, run
contiguously within each partition. The engines process the partitions
in parallel, one lane per partition, so work in different partitions is
independent, while moving data across partitions takes an explicit
operation. The Tensor Engine is a systolic array of $128 \times C$
processing elements, one row per partition and $C$ columns
($C = 128$ through NeuronCore-v4): it loads a \emph{stationary} operand from SBUF
into the array (the \emph{weight load}), streams a \emph{moving}
operand past it while contracting over the partition dimension, and
accumulates the results. The Vector Engine
runs operations in which each output depends on many inputs, such as
reductions; the Scalar Engine runs element-wise ones; the GpSimd
Engine's programmable vector processors run general-purpose code; and
DMA engines move data between SBUF and high-bandwidth memory (HBM).
\Cref{sec:systems} uses this vocabulary for Trainium~4.

\section{Computed Permutation Trellis Coding}
\label{sec:method}

CPTC replaces RPTC's stored random permutation with a computed
one: the $L$-bit state is multiplied by two fixed coefficients, the
top bits of the two products are added, and the sum indexes a
$64$-byte table of output values. This section gives the decoder
(\cref{sec:decode}), the encoder (\cref{sec:trellis}), how the
table is filled (\cref{sec:table}), which existing formats the
decoder reproduces (\cref{sec:constants}), why two multiplications suit bell-shaped targets (\cref{sec:theory_triangular}), and what any fixed table can reach (\cref{sec:shannon}).

\subsection{Decode primitive}
\label{sec:decode}

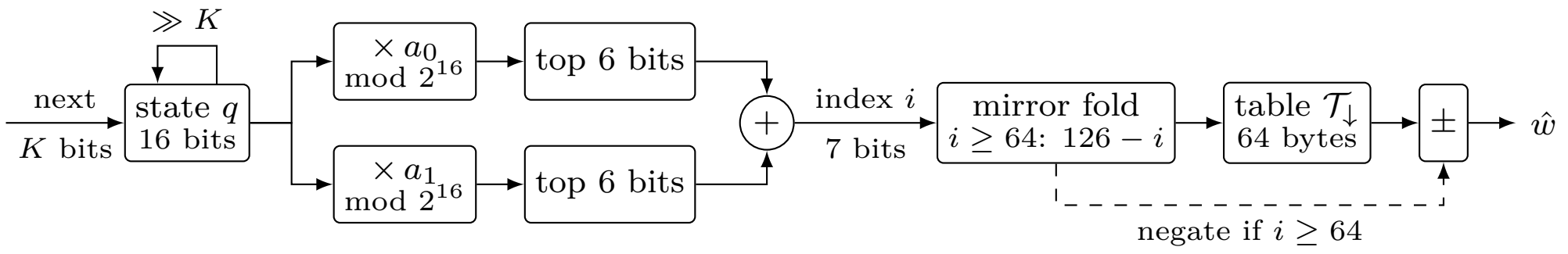
\begin{figure}[t]
  \centering
  \resizebox{\ifdim\width>\linewidth\linewidth\else\width\fi}{!}{%
  \begin{tikzpicture}[font=\footnotesize, >=latex,
      box/.style={draw, rounded corners=1.5pt, minimum height=6.5mm,
                  inner sep=2.5pt, align=center},
      sub/.style={font=\scriptsize}]
    \node[box] (q) {state $q$\\[-2pt]{\scriptsize $16$ bits}};
    \draw[<-] (q.west) -- node[above, sub] {next}
      node[below, sub] {$K$ bits} ++(-10mm,0);
    \draw[->] ([xshift=2.5mm]q.north) -- ++(0,3.2mm) -|
      node[pos=0.25, above, sub] {$\gg K$} ([xshift=-2.5mm]q.north);
    \node[box, right=7mm of q, yshift=5.2mm] (m0)
      {$\times\, a_0$\\[-2pt]{\scriptsize mod $2^{16}$}};
    \node[box, right=7mm of q, yshift=-5.2mm] (m1)
      {$\times\, a_1$\\[-2pt]{\scriptsize mod $2^{16}$}};
    \draw[->] (q.east) -- ++(3.5mm,0) |- (m0.west);
    \draw[->] (q.east) -- ++(3.5mm,0) |- (m1.west);
    \node[box, right=4mm of m0] (t0) {top $6$ bits};
    \node[box, right=4mm of m1] (t1) {top $6$ bits};
    \draw[->] (m0) -- (t0);
    \draw[->] (m1) -- (t1);
    \node[draw, circle, inner sep=1.5pt] (add)
      at ($(t0.east)!0.5!(t1.east)+(6mm,0)$) {$+$};
    \draw[->] (t0.east) -| (add.north);
    \draw[->] (t1.east) -| (add.south);
    \node[box, right=12mm of add] (fold)
      {mirror fold\\[-2pt]{\scriptsize $i \ge 64$: $126 - i$}};
    \draw[->] (add) -- node[above, sub] {index $i$}
      node[below, sub] {$7$ bits} (fold);
    \node[box, right=4mm of fold] (tab)
      {table $\mathcal{T}_{\downarrow}$\\[-2pt]{\scriptsize $64$ bytes}};
    \draw[->] (fold) -- (tab);
    \node[box, right=4mm of tab] (sgn) {$\pm$};
    \draw[->] (tab) -- (sgn);
    \node[right=4mm of sgn] (out) {$\hat{w}$};
    \draw[->] (sgn) -- (out);
    \draw[->, dashed] (fold.south) -- ++(0,-3.5mm) -|
      node[pos=0.25, below, sub] {negate if $i \ge 64$} (sgn.south);
  \end{tikzpicture}}
  \caption{The CPTC decoder for one value ($Q{=}6$,
    \cref{alg:decode}): the state shifts right to take the next $K$ stream bits,
    the top six bits of its two products are added, and the sum
    indexes the $64$-byte table through the mirror fold.}
  \label{fig:datapath}
\end{figure}

The decoder is parameterized by two coefficients $a_0, a_1$ of
$L_{\max}$ bits (we use $L_{\max} = 16$ throughout); a table
$\mathcal{T}$ of $2^{Q+1}{-}1$ entries, where the tap width $Q \le L_{\max}$
is the number of top bits kept from each product; and an optional shift
$s \in \{0, \ldots, Q{-}1\}$ (default $0$). The shift lets a decoder
built with tap width $Q$ emulate any smaller tap width $Q' = Q - s$
(\cref{prop:dominance}). \Cref{alg:decode} and
\cref{fig:datapath} show how one value is decoded from the
$L_{\max}$-bit state $q$, of which only the low $L$ bits carry
information (\cref{sec:trellis}). Every step is an integer operation.

\begin{algorithm}[t]
\caption{\cptc{} decode (per value), with multiplier width
  $L_{\max} = 16$. Lines 6--10 are an optional storage
  optimization: for mirror-symmetric codebooks only the lower
  half $\mathcal{T}_{\downarrow}$ ($2^Q$ entries) is stored and
  the upper half is served by a mirrored lookup
  (\cref{sec:table}).}
\label{alg:decode}
\label{alg:symlookup}
\begin{algorithmic}[1]
\Require $L_{\max}$-bit state $q$; coefficients
         $a_0, a_1 \in \{0,\ldots,2^{L_{\max}}{-}1\}$;
         codebook table $\mathcal{T}$, or its lower half
         $\mathcal{T}_{\downarrow}$ under symmetric storage;
         optional shift $s$ (default $s = 0$).
\State $Q \gets \log_2\!\big(|\mathcal{T}| + 1\big) - 1$
\Comment{tap width; $|\mathcal{T}| = 2^{Q+1}{-}1$,
         $s \in \{0,\ldots,Q{-}1\}$}
\State $x_0 \gets (a_0 \cdot q) \bmod 2^{L_{\max}}$, \quad
       $x_1 \gets (a_1 \cdot q) \bmod 2^{L_{\max}}$
\State $x_0 \gets x_0 \gg (L_{\max} - Q + s)$, \quad
       $x_1 \gets x_1 \gg (L_{\max} - Q + s)$
\Comment{$x_0, x_1 \in [0, 2^{Q-s}{-}1]$}
\State $i \gets x_0 + x_1$
\Comment{$i \in [0, 2^{Q-s+1}{-}2]$}
\State \textbf{return} $\mathcal{T}[i]$
\Comment{if the full table is stored; else lines 6--10}
\If{$i < 2^{Q}$}
\Comment{symmetric-storage lookup (optional)}
  \State \textbf{return} $\mathcal{T}_{\downarrow}[i]$
\Else
  \State \textbf{return} $-\mathcal{T}_{\downarrow}[2^{Q+1} - 2 - i]$
\EndIf
\end{algorithmic}
\end{algorithm}

We fix $L_{\max} = 16$ because returns diminish there: encoding cost
grows as $2^L$ while the last doubling buys only $0.008$ bits
(\cref{fig:cptc_vs_all}), and $16$-bit operands fit SIMD lanes and
hardware datapaths. When the table is mirror-symmetric
($\mathcal{T}[i] = -\mathcal{T}[2^{Q+1}{-}2{-}i]$), only its lower half
of $2^Q$ entries is stored, and lines 6--10 of \cref{alg:decode} read
it with one comparison and a conditional sign flip. We use $Q = 6$
unless stated otherwise, a $127$-entry table stored in $64$ bytes. This
is the size built into AWS Trainium~4's decode path
(\cref{sec:hardware}) and the smallest that keeps the trellis gain for
every $K \in [1, 6]$ (\cref{sec:table}).

\subsection{Trellis state and Viterbi encoder}
\label{sec:trellis}

A block of $T$ values is stored as $TK$ code bits. The decoder slides an
$L_{\max}$-bit window over them in steps of $K$ and passes each window
to \cref{alg:decode} (\cref{app:tailbiting_example} shows the bit
layout). Two properties make reading the raw window exact. First, the window needs no
masking to $L$ bits: the coefficients end in $L_{\max} - L$ zeros, so
the stream bits above the low $L$ drop out of the products modulo
$2^{L_{\max}}$. Second, the block is \emph{tail-biting}: its first
$L_{\max} - K$ bits are treated as if appended to its end, which closes
the trellis into a cycle, so every value has a full state and no bits
are spent on start-up. Values are decoded from the first to the last,
so in practice a state register replaces the window: each step shifts
it right by $K$ and takes the next $K$ stream bits at the top
(\cref{fig:datapath}).

The encoder is the standard Viterbi search for the lowest-MSE path
through the $2^L$-state trellis, in $O(T \cdot 2^L)$ time and
$O(T \cdot 2^{L-K})$ backtrace memory. Tail-biting
\citep{tseng2024qtip,ma1986tailbiting} costs a second Viterbi pass per
block. A tensor that is written once pays this cost once, and a later
update re-encodes only the blocks it touches. Our fused encoders keep
the cost modest: under two minutes for the linear weights of an 8B
model at $L{=}16$ on one GPU or, in cycle-level simulation, one
Trainium~4 chip (\cref{sec:encode}).

\subsection{Symmetric lookup table and output types}
\label{sec:table}

The index $i = x_0 + x_1$ ranges over $[0, 2^{Q+1}{-}2]$, that is
$[0, 126]$ at $Q{=}6$. For a target distribution symmetric about zero,
CPTC stores only the lower half of the table and reads the upper half
through the mirrored lookup of \cref{alg:decode}. The fold is exact
because the coefficient search keeps only pairs whose index histogram
is mirror-symmetric (\cref{app:constants}). Symmetry saves storage but
is not required: with $a_0 = 0$ the mirror branch is never taken and
the table can hold \emph{any} distribution. The table size follows from
the largest rate: to beat a scalar format at rate $K$ the table needs
at least $2^K$ stored entries, which for $K \le 6$ gives $Q = 6$.
Entries are stored one per byte as E2M5 (\cref{sec:hardware}); a
platform without E2M5 can store INT8 instead, at no cost up to
$K{=}4$ and $0.03$ bits at $K{=}6$ ($0.13$ on the Laplacian;
\cref{app:gg}).

The table starts from the target distribution and the index
distribution of the chosen pair $(a_0, a_1)$. Let $F$ be the target's
CDF, let $h[i]$ count the states whose index is $i$, and let
$G(i) = 2^{-L} \sum_{i' \le i} h[i']$ be the index CDF, with
$G(-1) = 0$. Cell $i$ owns the probability interval
$[G(i{-}1), G(i)]$, and its starting entry is the conditional mean of
the target over that interval:
\begin{equation}
  \mathcal{T}[i] = \mathbb{E}_{X \sim F}\!\big[
    X \,\big|\, F(X) \in [G(i{-}1),\, G(i)] \big].
  \label{eq:table_rule}
\end{equation}
The target enters only through the table, so retargeting changes only
the $2^Q$ stored values, and one coefficient pair serves every target.
(The reported results use one pair per rate, which at $L{=}16$ gains at
most $0.008$ bits over a single unified pair.)

\paragraph{Table generation.} \Cref{eq:table_rule} assumes each entry
represents a quantile interval, but the Viterbi encoder does not assign
values that way, least of all at small depth, where it has few paths to
choose from and the best table approaches the scalar Lloyd--Max
codebook. We therefore \emph{generate} the table with Lloyd's iteration
on the trellis \citep{stewart1982trellis}, as \citet{eriksson2007lc} do
for LC codes. Starting from \cref{eq:table_rule} rescaled by a
depth-dependent factor, we encode synthetic samples of the target with
the current table, replace each entry by the mean of the samples
assigned to it, repeat, and fit the result to the table's data type
(\cref{app:constants}). This computation can be done offline, once per
distribution, rate and trellis depth. At $L{=}K$ it recovers Lloyd--Max up to the rounding of the data type,
and every deeper trellis scores above it.

\subsection{Multiplicative coefficients and scalar formats}
\label{sec:constants}

The coefficients also set the trellis depth:
$L = 16 - \mathrm{ctz}(a_0 \mid a_1)$, where $\mathrm{ctz}$ counts
trailing zeros. For $L \le Q$ one coefficient is enough: $a_0 = 0$,
$a_1$ an odd multiple of $2^{16-L}$ (\cref{tab:best_constants}) and
$s = Q{-}L$. The choice $a_1 = 2^{16-L}$ is the scalar embedding of
\cref{prop:scalar_embedding}. For $L > Q$, two nonzero coefficients
give the near-triangular index distribution that the table exploits.

\paragraph{Best coefficients.}
We search $(a_0, a_1)$ for each depth $L$ and rate $K$ on Gaussian
samples, exhaustively at small $L$ and by sampling above
(\cref{app:constants} lists the pairs and the protocol). At $K{=}4$
quality rises with every step in $L$ and reaches $3.93$ effective bits
at $L{=}16$.

\paragraph{CPTC generalizes scalar quantization.} Every scalar
format is a special case of the decoder:
\begin{proposition}[Scalar embedding]
\label{prop:scalar_embedding}
For $K \in \{1, \ldots, Q\}$ and any $2^K$-entry codebook
$\mathcal{C} = \{c_q\}$ with $c_q$ representable in the table data
type, the CPTC decoder with $a_0 = 0$, $a_1 = 2^{16-K}$, $s = Q{-}K$,
and $\mathcal{T}[q] = c_q$ produces $\hat{w} = c_q$ for every $q$.
\end{proposition}

The proof is direct from \cref{alg:decode}: $a_0 = 0$ zeros $x_0$, and
the shift leaves $x_1 = q$. Here $L = K$ and the trellis is
degenerate. As a corollary, every INT$K$, FP$K$, Lloyd--Max or other
$K$-bit scalar format with $K \le Q$ is a CPTC instance with $L = K$.
(Irrational codepoints need a table type matching the implementation's
output precision, e.g.\ FP16.)

Adopting the decoder therefore never costs quality: $L{=}K$ recovers
the scalar format exactly. A deeper trellis was better in every
configuration we tested: every depth and rate on the Gaussian, and
every target of \cref{tab:gg_k} at $L{=}16$.

Likewise, for $L \le Q$ RPTC's $2^L$-entry codebook fits in CPTC's table verbatim (\cref{prop:scalar_embedding}
with $K \leftarrow L$). For $L > Q$ CPTC computes a structured
permutation in place of RPTC's random one, trading $2^L$ entries for
two multiplications.

An implementation also emulates smaller configurations
(\cref{app:emulation}). Say a table
data type $\mathcal{D}$ \emph{dominates} $\mathcal{D}'$ if some scale
$c > 0$ makes every value of $\mathcal{D}'$, multiplied by $c$, a value
of $\mathcal{D}$.

\begin{proposition}[Dominance]
\label{prop:dominance}
A decoder with parameters $(L_{\max}, Q, \mathcal{D})$ bit-exactly
emulates any decoder with $L'_{\max} \le L_{\max}$, $Q' \le Q$,
and $\mathcal{D}'$ dominated by $\mathcal{D}$: replace each
coefficient $a_j$ by $a_j \cdot 2^{L_{\max} - L'_{\max}}$, set the shift $s = Q - Q' + s'$ ($s'$ the emulated decoder's shift), and load the table rescaled by $c$ (outputs then agree up to the factor $c$, and exactly when $c = 1$).
\end{proposition}

\subsection{The triangular index distribution as a companding law}
\label{sec:theory_triangular}

\begin{figure}[t]
  \centering
  \includegraphics[width=0.70\linewidth]{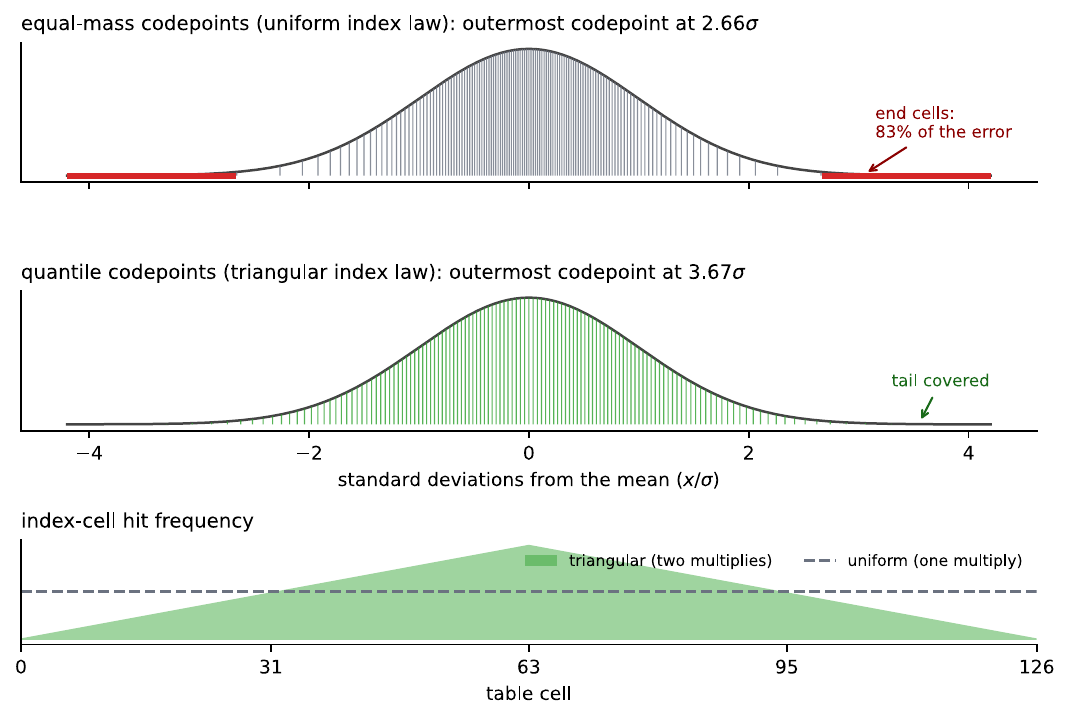}
  \caption{Where the two index laws put their codepoints ($M{=}127$,
    standard normal). \textbf{Top:} equal-mass placement from a
    uniform index law, red marking the clipped tail.
    \textbf{Middle:} the triangular law's quantile placement (the conditional-mean table of \cref{eq:table_rule} puts its outermost entry at $3.74\sigma$).
    \textbf{Bottom:} the two index laws over the $127$ table cells.}
  \label{fig:companding}
\end{figure}

With only $2^{Q+1}{-}1$ distinct codepoints, the table limits
quality at any depth (\cref{thm:ceiling}), so it matters
\emph{where the codepoints land}. The starting table of
\cref{eq:table_rule} places them according to how often each index is
hit. The index distribution thus acts as a \emph{companding law}
\citep{bennett1948spectra}: for target density $p$ and index law $g$,
the codepoint density is $\lambda(x) \propto p(x)/g(x)$. Table
generation then moves the values, but the index map still fixes how
many states share each entry.

The obvious choice is a \emph{uniform} index law, with equal
probability mass per codepoint, and it is what a single multiplication
produces. Its tails are short: with $M{=}127$ equal-mass Gaussian
codepoints the largest sits at only $2.66\sigma$, and as a scalar
codebook $83\%$ of its error comes from its two end cells. The
scalar-optimal placement, $\lambda^{\star} \propto p^{1/3}$
\citep{panter1951quantization}, instead spreads codepoints \emph{wider}
than the data, by $\sqrt{3}\times$ for a Gaussian. The second
multiplication moves the table in that direction. The \emph{sum} of the
two products' top bits is approximately \emph{triangular}, and its
rarely hit edge cells stretch the codebook to a profile about
$\sqrt{2}\times$ wider than the data, with the outermost quantile
codepoint at $3.67\sigma$ (\cref{fig:companding}). That lies between
the source width and the scalar-optimal $\sqrt{3}$, and as a scalar
codebook it stays within $7$--$17\%$ of the Lloyd--Max optimum at every
table size (\cref{app:table_law}). The table has to match the index
law: filled as if the index were uniform, it costs $0.69$ effective
bits at $L{=}16$, $K{=}4$ ($3.93 \to 3.25$, \cref{tab:gg}).

\subsection{Theoretical limits and extensions}
\label{sec:shannon}

\paragraph{Alphabet ceiling.} Every decoded value comes from a fixed
table, so no encoder and no trellis depth can beat the rate-distortion
function restricted to that table \citep{finamore1980optimal}:

\begin{theorem}[Alphabet ceiling]
\label{thm:ceiling}
Let $X_1, \ldots, X_n$ be i.i.d.\ with a density and finite variance and let a
code of rate $K$ bits per value (any block length, any encoder)
reproduce each $X_i$ by some $\hat X_i \in C$, where $C \subset
\mathbb{R}$ is a fixed finite set (more generally $C \subset \mathbb{R}^V$ for codes that emit $V$-tuples, at rate $VK$ per tuple, with distortion per value). Then
\[
  \frac1n \sum_{i=1}^{n} \mathbb{E}(X_i - \hat X_i)^2 \;\ge\;
  D_C(K) \coloneqq \min\bigl\{\mathbb{E}(X - \hat X)^2 :\ \hat X
  \in C,\ I(X;\hat X) \le K\bigr\},
\]
the alphabet-constrained rate-distortion function. In particular
\[
  K_{\text{eff}} \;\le\; K_C(K) \coloneqq -\tfrac12 \log_2 \bigl(D_C(K)/N(X)\bigr)
\]
for every trellis depth $L$ and block size $T$, and $D_C(K) \ge
N(X)\,4^{-K}$, with $N(X) \le \sigma^2$ the entropy power of
\cref{eq:keff} (equality for the Gaussian).
\end{theorem}

The ceiling is proved in \cref{app:ceiling} and can be computed with
the Blahut--Arimoto algorithm. It bounds how much of a codebook's gap
to the Shannon limit is forced by its table. CPTC's $127$-entry E2M5
table caps it at $3.998$ effective bits at $K{=}4$, so the table's
values account for at most $0.002$ bits of its $0.07$-bit gap. QTIP
HYB's $512$-pair table ($1{,}024$ pairs with its sign flip) caps HYB at
$3.687$ at $K{=}4$ and at $3.936$ for every $K \ge 5$, which is the
plateau in \cref{tab:qtip_k} (ceilings in \cref{tab:ceiling}).

Two further threads are developed in the appendices.
\emph{Achievable rate} (\cref{app:shannon}): with one multiplication
and a full table ($a_0 = 0$, $Q = L$) the decoder is RPTC with the
algebraic permutation $P_a(q) = aq \bmod 2^L$ in place of a random one.
For \emph{every} odd multiplier $a$ it provably satisfies the necessary
marginal-optimality condition of \citet{mao2011rptc}; the joint
conditions, and any such statement for the two-multiply decoder,
remain open.
\emph{In-stream metadata} (\cref{app:scale_encoding}): parity
constraints let the encoder commit auxiliary bits (block scale, zero
point, flags) into the codeword itself. Each auxiliary bit committed to
a group of $g$ values costs about $1/g$ effective bits per value.

\section{Experiments}
\label{sec:experiments}

We evaluate CPTC on (i) i.i.d.\ source quantization
(\cref{sec:exp_iid}); (ii) the bimodal
matmul-aware density of \citet{ang2026matmul}, reached by changing only
the table (\cref{sec:exp_matmul_synth}); and (iii) LLM
weight-only quantization (\cref{sec:exp_llm}).
Unless noted, CPTC uses $L{=}16$, the coefficients of
\cref{tab:best_constants}, $T{=}256$ (i.i.d.) or $T{=}128$
(matmul), tail-biting encoding, and the generated tables of
\cref{sec:table}.

\subsection{I.i.d.\ source quantization}
\label{sec:exp_iid}

We generate $2^{22}$ i.i.d.\ samples per target and seed (three seeds unless noted) and
report effective bits (\cref{eq:keff}). The spread across seeds is about
$0.0005$ bits: over ten seeds CPTC at $K{=}4$ scores
$3.9334 \pm 0.0005$ (\cref{app:iid_extended}).

\paragraph{Standard normal at $K{=}4$ (\cref{fig:cptc_vs_all}).}
CPTC at $L{=}16$ reaches $K_{\text{eff}}=3.93$. That matches RPTC run
two state bits deeper ($3.92$ at $L{=}18$, \cref{tab:scaling_L}) and
exceeds QTIP's best $4$-bit variant ($3.86$), a score CPTC already
matches at $L{=}11$. Scalar and block-scaled formats score $0.5$--$0.8$
effective bits lower (\cref{tab:hw_criteria}).

\paragraph{Codebook quality across bit widths
(\cref{tab:qtip_k}).} We reran every prior trellis codebook at
$L{=}16$ under one protocol. The QTIP rows run the authors' released
code and tables; HYB emits value pairs from a $512$-pair $k$-means
table. RPTC is the random permutation of the target's
optimal-reproduction-law quantiles \citep{mao2011rptc}. LC is the
stronger of two searched one-multiply forms per column, with its
codebook scaled and Lloyd-optimized as its authors prescribe
(\cref{app:iid_extended}). The CPTC$_{Q7}$ rows double the table (E2M5
and FP16).

CPTC leads every QTIP variant at every $K$, by a margin that grows from
$0.014$ bits at $K{=}1$ to $0.29$ at $K{=}6$; HYB's fixed alphabet caps
it past $K{=}4$ (\cref{thm:ceiling}). With
$128$--$2{,}048\times$ the storage, LC leads every column: by $0.002$--$0.004$ bits over CPTC
for $K \in [2, 4]$, $0.010$ at $K{=}1$ and $0.017$/$0.12$ at
$K{=}5$/$6$, which CPTC$_{Q7}$ in FP16 narrows to $0.004$/$0.018$.

\begin{table}[t]
  \centering
  \small
  \caption{Effective bits ($\uparrow$) of every trellis codebook
    at each bit width, all run at $L{=}16$ under one protocol
    (rows described in the text), with the table each decoder must store (CPTC in $8$-bit
    E2M5 unless marked FP16, others FP16; ``---'': nothing stored).
    Bold: within $0.003$ of the underlined column best.}
  \label{tab:qtip_k}
  \begin{tabular}{lccccccc}
    \toprule
    Codebook & Storage & $K{=}1$ & $K{=}2$ & $K{=}3$ & $K{=}4$ & $K{=}5$ & $K{=}6$ \\
    \midrule
    QTIP HYB   & $2$\,KiB & $0.932$ & $1.912$ & $2.840$ & $3.611$ & $3.936$ & $3.936$ \\
    QTIP 3INST & --- & $0.936$ & $1.931$ & $2.902$ & $3.852$ & $4.748$ & $5.469$ \\
    QTIP 1MAD  & --- & $0.933$ & $1.931$ & $2.907$ & $3.864$ & $4.765$ & $5.514$ \\
    RPTC & $128$\,KiB & $0.956$ & $1.935$ & $2.916$ & $3.896$ & $4.872$ & $5.840$ \\
    LC   & $8$--$128$\,KiB & $\underline{\mathbf{0.960}}$ & $\underline{\mathbf{1.950}}$ & $\underline{\mathbf{2.945}}$ & $\underline{\mathbf{3.938}}$ & $\underline{\mathbf{4.930}}$ & $\underline{\mathbf{5.922}}$ \\
    CPTC & $64$\,B & $0.950$ & $\mathbf{1.948}$ & $\mathbf{2.943}$ & $3.934$ & $4.913$ & $5.802$ \\
    CPTC$_{Q7}$ & $128$\,B & $0.950$ & $\mathbf{1.947}$ & $\mathbf{2.943}$ & $3.934$ & $4.919$ & $5.869$ \\
    CPTC$_{Q7\text{-FP16}}$ & $256$\,B & $0.949$ & $\mathbf{1.947}$ & $\mathbf{2.943}$ & $\mathbf{3.935}$ & $4.926$ & $5.904$ \\
    \bottomrule
  \end{tabular}
\end{table}

\paragraph{Beyond the Gaussian (\cref{tab:gg_k}).} The
unit-variance generalized Gaussian
$p_\beta(x) \propto \exp(-|x/\alpha|^{\beta})$ interpolates
from the uniform ($\beta \to \infty$) through the normal
($\beta{=}2$) to the Laplacian ($\beta{=}1$)
\citep{nadarajah2005gg,banner2019posttraining}. For each $\beta$ only
the $64$ table bytes change, and $K_{\text{eff}}$ is entropy-power
normalized (\cref{eq:keff}).

The trellis advantage over the Lloyd--Max scalar optimum of each
$(\beta, K)$ holds everywhere and \emph{grows} with tail weight, from
$0.20$ bits at $\beta{=}\infty$ to $0.97$ at $\beta{=}1$ ($K{=}4$). At
$\beta{=}1$ CPTC nears scalar quality with a full bit less rate ($2.90$
at $K{=}3$ against scalar's $2.91$ at $K{=}4$). CPTC exceeds RPTC's
quantile codebooks across the family, by $0.02$--$0.05$ bits at
$K{=}4$. A table filled as if the index were uniform trails CPTC by
more as the tails grow, from $0.42$ bits at $\beta{=}\infty$ to $1.10$
at $\beta{=}1$ (\cref{tab:gg}). CPTC meets every target with the same
coefficients. On the Gaussian ($\beta{=}2$) the gap to
the bound is $0.05$--$0.07$ bits for $K \le 4$ and grows to $0.20$ at
$K{=}6$ as the $127$-codepoint ceiling binds.

\begin{table}[t]
  \centering
  \small
  \caption{Effective bits ($\uparrow$) of CPTC on targets from
    uniform ($\beta{=}\infty$) to Laplacian ($\beta{=}1$), with the
    best scalar quantizer for each cell in parentheses; only the
    $64$ table bytes change between rows ($L{=}16$, $T{=}256$, $3$
    seeds; entropy-power normalized, so $K$ is an upper bound in every
    column; the $\beta{=}2$ row repeats \cref{tab:qtip_k}).}
  \label{tab:gg_k}
  \begin{tabular}{ccccccc}
    \toprule
    $\beta$ & $K{=}1$ & $K{=}2$ & $K{=}3$ & $K{=}4$ & $K{=}5$ & $K{=}6$ \\
    \midrule
    $\infty$ & $0.84$ ($0.75$) & $1.91$ ($1.75$) & $2.94$ ($2.75$)
          & $3.95$ ($3.75$) & $4.95$ ($4.75$) & $5.89$ ($5.75$) \\
    $2.0$ & $0.95$ ($0.73$) & $1.95$ ($1.54$) & $2.94$ ($2.43$)
          & $3.93$ ($3.36$) & $4.91$ ($4.32$) & $5.80$ ($5.30$) \\
    $1.5$ & $0.94$ ($0.63$) & $1.94$ ($1.42$) & $2.93$ ($2.29$)
          & $3.92$ ($3.21$) & $4.89$ ($4.17$) & $5.74$ ($5.15$) \\
    $1.0$ & $0.93$ ($0.40$) & $1.91$ ($1.15$) & $2.90$ ($1.99$)
          & $3.88$ ($2.91$) & $4.79$ ($3.86$) & $5.48$ ($4.84$) \\
    \bottomrule
  \end{tabular}
  
\end{table}

\paragraph{Why not a smaller decoder.} Each part of CPTC's decode cost
(the per-value map, the table size and its data type) has a measured
or provable price. One multiplication with a table of the same size
trails by $0.01$/$0.04$/$0.09$/$0.14$ bits at $K{=}3$--$6$, by more on
heavy-tailed targets, and by about twice that at $K{=}3$--$5$ with
$32$-byte tables (\cref{app:iid_extended,app:gg}). A
table with $Q = K{-}1$, or an INT($K$) table, has only $2^K{-}1$
codepoints, one short of a rate-$K$ scalar quantizer, so \emph{no
trellis depth can match scalar quality} (\cref{prop:alphabet_floor}).
This makes $Q{=}6$ the smallest table that can beat scalar at $K{=}6$.
Two doublings above that floor ($Q = K{+}1$ in INT($K{+}2$)) cost at
most $0.03$ bits against the $Q{=}6$ FP16 reference at every $K$
(\cref{tab:min_codebook}, \cref{app:dtype}).

\subsection{A bimodal target: matmul-aware tables}
\label{sec:exp_matmul_synth}

This experiment shows that CPTC's table absorbs an arbitrary symmetric
target density with no change to the decode path, coefficients or
encoder: only the $64$ table bytes differ. The target is the optimal
scalar point density $\lambda^{\star}$ of \citet{ang2026matmul} for a
matrix product whose operand pairs have correlation $\rho$. The density
turns bimodal past a phase transition at $|\rho| = 1/\sqrt{3}$. We build the
table by moving each entry of the generated Gaussian table to the same
quantile of $\lambda^{\star}$ (\cref{app:matmul_calibration}). Their
work and ours are orthogonal: they improve the \emph{codebook} for a
given rate-$K$ scalar quantizer, and we improve the \emph{coder} for a
given codebook. This is the table-design flexibility that RPTC pays
$2^L$ entries for and QTIP's hash-based variants give up. It is a
demonstration only: a deployment would estimate $\rho$ once per layer
or attention head and look up a precomputed table, which we do not
evaluate here.

We adopt the synthetic correlated-Gaussian setting of
\citet{ang2026matmul} as-is, with both operands quantized, unlike the
weight-only rest of the paper. $A$ and $B$ are drawn so each output
entry's multiplicative pairs are i.i.d.\ bivariate standard normal with
correlation $\rho$, and we report the relative Frobenius error of
$\widehat A \widehat B$ over $10$ trials for
$A \in \mathbb{R}^{m \times k}$, $B \in \mathbb{R}^{k \times n}$ at
$m{=}n{=}128$, $k{=}256$.

\begin{figure}[t]
  \centering
  \includegraphics[width=0.72\linewidth]{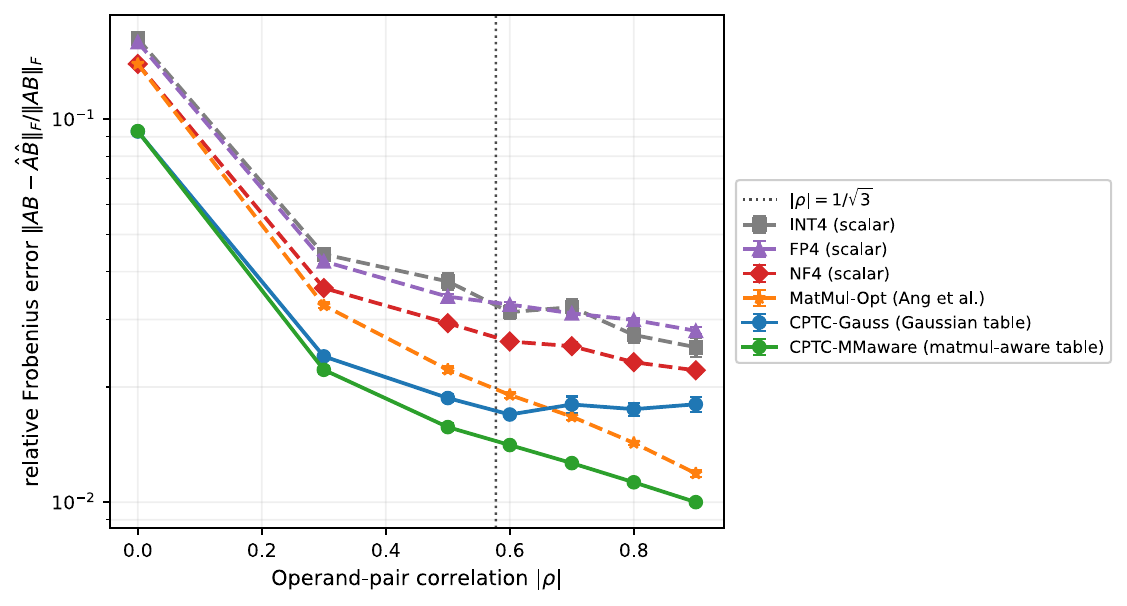}
  \caption{Matmul error against operand correlation: relative Frobenius
    error at $4$ bits per value on synthetic correlated Gaussians.
    The vertical line is the predicted phase transition
    $|\rho| = 1/\sqrt{3}$; error bars are the standard error of the mean (SEM) over $10$ trials (where not visible, smaller than the markers).}
  \label{fig:matmul_phase}
\end{figure}

\cref{fig:matmul_phase} shows that \textsc{CPTC-MMaware} (CPTC with the
$\lambda^{\star}$ table) has the lowest error at every $\rho$ tested,
tied with Gaussian-table CPTC at $\rho{=}0$. Its edge over the best
standard scalar format (NF4; \citealp{dettmers2023qlora}) widens from
$1.5\times$ at $\rho{=}0$ to $2.2\times$ at $\rho = 0.9$. There it is
also $1.8\times$ better than Gaussian-table CPTC and $1.2\times$
better than the scalar \textsc{MatMul-Opt} of \citet{ang2026matmul}.
Trellis coding alone cannot capture the matmul-tail importance that
$\lambda^{\star}$ encodes.

Baselines, scale calibration and the per-$\rho$ numbers are in
\cref{app:matmul_calibration} (\cref{tab:matmul_phase}).

\subsection{LLM weight-only quantization}
\label{sec:exp_llm}

\begin{table}[t]
  \centering
  \small
  \caption{Llama 3.1 Instruct quantized in YAQA's pipeline
    \citep{tseng2025yaqa} with only the codebook swapped, no
    fine-tuning: KL to the BF16 model ($\downarrow$), perplexities
    ($\downarrow$) and mean zero-shot accuracy ($\uparrow$; ARC-c/e,
    BoolQ, HellaSwag, PIQA; within noise, unbolded). QTIP rows are
    our runs of its HYB and 1MAD codebooks; the HYB rows match the
    published HYB numbers within $0.007$ KL. KL and
    perplexities average three RHT seeds (KL spread $0.3$--$4.1\%$); $^{\dagger}$one seed;
    Acc from seed $0$. W2: Wikitext-2. Bold: best per rate and model.}
  \label{tab:llm_main}
  \begin{tabular}{cl cccc cccc}
    \toprule
    \multirow{3}{*}{$K$} & \multirow{3}{*}{Codebook} & $D_{\mathrm{KL}}$ $\downarrow$ & \multicolumn{2}{c}{Perplexity $\downarrow$} & Acc $\uparrow$ & $D_{\mathrm{KL}}$ $\downarrow$ & \multicolumn{2}{c}{Perplexity $\downarrow$} & Acc $\uparrow$ \\
    \cmidrule(lr){3-3} \cmidrule(lr){4-5} \cmidrule(lr){6-6} \cmidrule(lr){7-7} \cmidrule(lr){8-9} \cmidrule(lr){10-10}
    & & W2 & W2 & C4 & 0-shot & W2 & W2 & C4 & 0-shot \\
    \midrule
    & & \multicolumn{4}{c}{8B-Instruct} & \multicolumn{4}{c}{70B-Instruct} \\
    \cmidrule(lr){3-6} \cmidrule(lr){7-10}
    16 & BF16 & $0$ & $6.502$ & $8.022$ & $71.4$ & $0$ & $3.524$ & $6.460$ & $77.1$ \\
    \midrule
    \multirow{4}{*}{4} & QTIP HYB    & $0.0131$ & $6.608$ & $8.117$ & $71.2$ & $0.0298^{\dagger}$ & $3.685^{\dagger}$ & $6.516^{\dagger}$ & $76.8$ \\
    & QTIP 1MAD   & $0.0100$ & $6.579$ & $8.092$ & $71.5$ & $0.0238$ & $3.647$ & $6.500$ & $76.8$ \\
    & CPTC$_{Q5}$ & $0.0096$ & $6.570$ & $8.088$ & $71.3$ & $\mathbf{0.0225}$ & $\mathbf{3.631}$ & $6.500$ & $76.9$ \\
    & CPTC        & $\mathbf{0.0095}$ & $\mathbf{6.566}$ & $\mathbf{8.086}$ & $71.3$ & $0.0227$ & $3.642$ & $\mathbf{6.499}$ & $76.9$ \\
    \midrule
    \multirow{4}{*}{3} & QTIP HYB    & $0.0450$ & $6.852$ & $8.344$ & $70.8$ & $0.0967^{\dagger}$ & $4.035^{\dagger}$ & $6.649^{\dagger}$ & $76.3$ \\
    & QTIP 1MAD   & $0.0405$ & $6.824$ & $8.314$ & $70.3$ & $0.0876$ & $3.965$ & $6.628$ & $76.4$ \\
    & CPTC$_{Q5}$ & $0.0396$ & $6.828$ & $\mathbf{8.294}$ & $70.6$ & $0.0856$ & $3.968$ & $\mathbf{6.621}$ & $76.6$ \\
    & CPTC        & $\mathbf{0.0387}$ & $\mathbf{6.797}$ & $8.297$ & $70.5$ & $\mathbf{0.0845}$ & $\mathbf{3.946}$ & $\mathbf{6.621}$ & $76.6$ \\
    \midrule
    \multirow{4}{*}{2} & QTIP HYB    & $0.2424$ & $8.402$ & $9.798$ & $66.7$ & $0.3281^{\dagger}$ & $5.259^{\dagger}$ & $7.316^{\dagger}$ & $74.3$ \\
    & QTIP 1MAD   & $0.2271$ & $8.254$ & $9.696$ & $67.3$ & $0.3221$ & $5.225$ & $7.286$ & $74.3$ \\
    & CPTC$_{Q5}$ & $0.2212$ & $\mathbf{8.228}$ & $9.653$ & $67.1$ & $\mathbf{0.3145}$ & $\mathbf{5.191}$ & $\mathbf{7.265}$ & $74.7$ \\
    & CPTC        & $\mathbf{0.2202}$ & $8.238$ & $\mathbf{9.634}$ & $67.4$ & $0.3176$ & $5.210$ & $7.269$ & $74.5$ \\
    \bottomrule
  \end{tabular}
\end{table}

We quantize Llama 3.1 Instruct models in the YAQA pipeline of
\citet{tseng2025yaqa} without fine-tuning, swapping only the codebook
(\cref{tab:llm_main}; protocol and controls in \cref{app:llm}).
CPTC$_{Q5}$ is the $63$-codepoint table of the GPU kernel
(\cref{sec:stock_decode}).

KL tracks the codebook's i.i.d.\ effective bits
(\cref{fig:kl_vs_keff}). $\log_2 D_{\mathrm{KL}}$ is a smooth function
of $K_{\text{eff}}$ (quadratic fits, $r^2 \ge 0.997$): each effective
bit is worth $4\times$ in KL at 70B, as in MSE, and $8\times$ to
$3\times$ at 8B from $2$ to $4$ bits. Within every rate the KL order of
the three codebooks is their $K_{\text{eff}}$ order, and KL falls with
depth along an 8B sweep (\cref{app:kl_vs_keff}). The i.i.d.\ gains
therefore carry to the model.

CPTC has the lowest KL divergence to the unquantized model at every
rate and scale against QTIP's HYB and 1MAD codebooks. Against HYB,
YAQA's default codebook, the margin \emph{grows} with rate ($9\%$ at
$K{=}2$, $14\%$ at $K{=}3$, $27\%$ at $K{=}4$ at 8B) as its $512$-pair
table saturates (\cref{thm:ceiling}). Against the table-free 1MAD it is
$3$--$5\%$ at 8B and $1$--$5\%$ at 70B (full-table arm), the ordering
of \cref{tab:qtip_k}. The $Q{=}5$ arm costs at most $0.001$ KL at
either scale and leads at 70B for $K \in \{2,4\}$, so the GPU kernel's
smaller table is no compromise. Zero-shot accuracy separates rates, not
formats: it is within about a point of BF16 at $K \ge 3$, and the
formats drop together at $K{=}2$. A preliminary run on Qwen3.8-27B
\citep{qwen2026qwen38}, a hybrid gated-DeltaNet model, shows the same
ordering (\cref{tab:llm_qwen}).

\section{Systems}
\label{sec:systems}

\subsection{Decoding on AWS Trainium~4}
\label{sec:hardware}

CPTC is implemented on AWS Trainium~4 \citep{aws2025trainium4}, an AI accelerator (\cref{sec:trainium_background}),
where the decode primitive sits inside the weight-load path that feeds
the systolic array. The hardware additions on top of the existing
FP4 dataflow are minimal: a state register per lane, the
two coefficient products in the shift-accumulate form of
\cref{eq:incremental_mult}, one $6$-bit adder (the sum of the two
products' top $6$ bits, a $7$-bit table index), a $64$-byte table held
in registers, and the sign flip of the mirrored lookup. Output
formatting is shared with the floating-point modes already present. Sized for throughput parity
with the native FP4 dataflow at $K \le 4$ and with FP8 at $K \le 6$,
the decoder occupies a negligible fraction of the systolic array's
area. The design as a whole, CPTC decode path included, has closed
timing and is ready for production, so these results describe a
deployable datapath rather than a synthesis estimate. The decoder RTL is
formally verified against a golden reference of
\cref{alg:decode}, and the two multiplications and the index sum fit
within one array clock cycle.

\paragraph{Integration.} The packed bitstream sits in SBUF
like any other stationary operand, and two instructions cooperate to
decode it. A table-load instruction copies the $64$-byte table from SBUF
into the decoder's table registers and sets $(a_0, a_1, T, s)$ in configuration registers from its
immediate fields, once per tensor and quadrant of the systolic array.
The weight-load instruction then takes a CPTC data type, one per bit
width $K$, as its input type and decodes with whatever the last table
load programmed, so one table load serves any number of weight loads.
Its hot path reads $32$-bit words of
packed code into a state register per lane. Each lane
decodes $8$ values per cycle via \cref{alg:decode}: at the native FP4
rate a lane is one partition, and at the FP8 rate an even/odd pair of
partitions works as one lane, $4$ values per partition (the
\emph{FP4 rate} and \emph{FP8 rate} below). The register-held table
is read by all lanes' lookup muxes in parallel, and the decoded
values enter the systolic array's multiply-accumulate pipeline in the
same cycle an FP4 or FP8 value would.

\paragraph{Shift-accumulate products.} Consecutive windows overlap, so
the window of value $t$ follows from that of value $t{+}1$:
$q_t = ((q_{t+1} \ll K) \mid c_t) \bmod 2^{16}$, with $c_t$ value
$t$'s $K$-bit code. The bits shifted out at the top contribute only
multiples of $2^{16}$ to a product, so each coefficient's product
$x_{j,t} = (a_j \cdot q_t) \bmod 2^{16}$ ($j \in \{0,1\}$) obeys
the same recurrence:
\begin{equation}
  x_{j,t} = \big( (x_{j,t+1} \ll K) + a_j \cdot c_t \big) \bmod 2^{16}.
  \label{eq:incremental_mult}
\end{equation}
A lane decodes its $8$ values per cycle as one batch and
evaluates this recurrence in parallel inside it, which replaces each
$16 \times 16$ multiplier with a $K$-bit shift
and one addition of $a_j \cdot c_t$, a $16 \times K_{\max}$ product
($K_{\max} = 6$) whose partial-product array is about $3/8$ the
size of the full one.

\begin{figure}[t]
  \centering
  \includegraphics[width=0.85\linewidth]{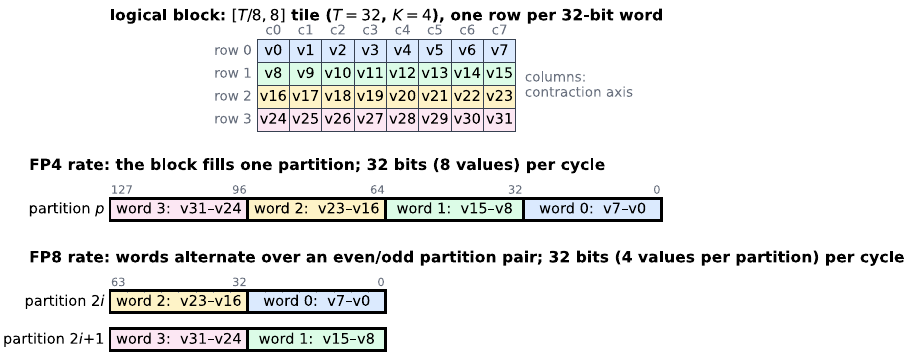}
  \caption{How one CPTC block ($K{=}4$, $T{=}32$) is laid out in SBUF. \textbf{Top:} the logical $[T/8, 8]$ tile; each row is one
    $32$-bit word of eight $4$-bit codes. \textbf{Middle:} at FP4 rate the block fills one SBUF partition.
    \textbf{Bottom:} at FP8 rate the words alternate over an even/odd
    partition pair, so the block must span an even number of words. Both
    rates read $32$ bits of packed code per cycle.}
  \label{fig:data_layout}
\end{figure}

\paragraph{Data layout (\cref{fig:data_layout}).} Each tail-biting block of $T$ values is stored
as a $[T/8, 8]$ array whose second axis lies along the systolic array's
contraction dimension (which runs over the $8$ values of a word and then across partitions), with the block's $TK$ bits packed into whole
$32$-bit words. A block never straddles a \emph{tile}, the part of the array one
load instruction fills, so each block stays inside one
decode lane (a partition at FP4 rate, an even/odd pair at
FP8 rate).
The recurrence above therefore stays inside one lane, and the lanes
run in parallel. At FP4 rate a
block occupies one SBUF partition; at FP8 rate its words
alternate across the pair, so a block must span
an even number of words. Whole-word packing therefore sets the minimum
block size: $TK$ a multiple of $32$ bits at FP4 rate and of $64$ bits at
FP8 rate, i.e.\ $T$ a multiple of $32/\gcd(32, K)$ and of
$64/\gcd(64, K)$ respectively. The minimum is $8$ to $64$ values depending on
$K$, and the $T \in \{128, 256\}$ used throughout this paper satisfy the
constraint for every $K$. The layout is
compatible with Trainium~4's microscaling support.

\paragraph{Table data type.} Every decoded value is a table entry or its
negative, so the table's data type is the decoder's output format. On
Trainium~4 every table entry must be an \emph{E2M5} value. E2M5 is an
$8$-bit float with a sign bit, $2$ exponent bits (bias $1$) and $5$
mantissa bits, with subnormals, no infinities and the all-ones magnitude reserved for NaN. Its $253$ distinct
values span $[-7.75, 7.75]$ in steps of $1/32$ for magnitudes below
$2$, $1/16$ in
$[2, 4)$ and $1/8$ in $[4, 7.75]$.

\paragraph{Throughput.} The CPTC path sustains the chip's native
per-cycle weight-load rate: parity with FP4 for $K \le 4$ and with FP8
for $K \le 6$. A memory-bound decode step therefore runs at the same
tokens per second as the native FP4 path while delivering the quality
of \cref{tab:llm_main},
and lower rates stream proportionally fewer bytes.

\subsection{Decoding on CPUs and GPUs}
\label{sec:stock_decode}

Without a native decode unit, the primitive maps to two integer
multiplies, two shifts, an add and a small indexed load on x86, Arm and
GPUs. \Cref{tab:stock_decode} times the linear layers of a Llama 3.1 8B
forward pass at batch $1$ against an INT4 kernel that shares each
pipeline. The two-multiply and CPU kernels use the $63$-codepoint
$Q{=}5$ table stored in full, which needs no mirror fold (on GPUs in
FP16, the CPTC$_{Q5}$ rows of \cref{tab:llm_main}; on CPUs in INT8).
On stock NVIDIA GPUs, a fused kernel decodes the stream straight into the
tensor-core operand registers (\cref{app:gpu_gemm}). On the
bandwidth-bound L40S it runs at $0.98\times$ the speed of the INT4
baseline ($3.6\times$ cuBLAS FP16). On the bandwidth-rich H100 it is
ALU-bound: $0.62\times$ INT4 for the $L{=}16$ two-multiply decoder, and
$0.79\times$ for a Hopper-only $L{=}10$ state-table variant at $3.84$
effective bits. This makes the case for a native decode instruction. On
CPUs, CPTC at $K{=}4$ streams the forward pass as fast as an INT4
kernel ($0.98\times$ on AMD Zen~4, $0.97\times$ on AWS Graviton5;
\cref{app:cpu_decode}).

\begin{table}[t]
  \centering
  \small
  \caption{Time per token for the linear layers of Llama 3.1 8B at
    batch $1$ (ms, $\downarrow$): CPTC at $K{=}4$ against an INT4
    kernel that shares its pipeline, with FP16 for reference. GPUs
    fuse the decode into the tensor-core GEMM; CPUs decode into the
    INT8 dot product on all cores of one node. Times cover decode and matmul only.}
  \label{tab:stock_decode}
  \begin{tabular}{llcccc}
    \toprule
    Platform & CPTC decoder & FP16 & INT4 & CPTC & CPTC speed vs.\ INT4 \\
    \midrule
    NVIDIA L40S & two multiplies, $L{=}16$ & $19.08$ & $5.12$ & $5.24$ & $0.98\times$ \\
    NVIDIA H100 & two multiplies, $L{=}16$ & $5.51$ & $2.11$ & $3.40$ & $0.62\times$ \\
    NVIDIA H100 & state table, $L{=}10$ & $5.51$ & $2.11$ & $2.67$ & $0.79\times$ \\
    AMD Zen~4, $64$ cores & AVX-512 & $51.9$ & $13.3$ & $13.6$ & $0.98\times$ \\
    AWS Graviton5, $96$ cores & NEON & $36.0$ & $8.6$ & $8.9$ & $0.97\times$ \\
    \bottomrule
  \end{tabular}
\end{table}

\subsection{Encoding on GPUs and Trainium~4}
\label{sec:encode}

\paragraph{GPU.} Our GPU kernel fuses the tail-biting Viterbi search of
\cref{sec:trellis} into one persistent Triton launch. On one L40S its
throughput runs from $60$ million values/s at $L{=}16$, $K{=}4$
($T{=}128$) to $3.3$ billion at $L{=}10$, at MSE within $0.02\%$ of a
\texttt{torch.compile} reference on identical inputs. At $L{=}16$ it
encodes the $7.0$ billion values of Llama 3.1 8B's linear layers in
under two minutes. Against QTIP's shipped encoder, equal-quality
encoding runs $2{,}400\times$ faster on the same GPU
(\cref{app:viterbi_gpu}). Of this, $32\times$ is \emph{method-level}: equal
quality at $L{=}11$ means $32\times$ less search, since work is
$\propto 2^L$. The remaining $75\times$ comes from the fused kernel,
engineering that a fused encoder for QTIP's codebooks would also gain. Encoding this fast opens the door to online uses such as
KV caches at small $L$ ($3.84$ effective bits at $L{=}10$).

\paragraph{Trainium~4.} The encoder also runs on Trainium~4 itself,
so tensors can be quantized where they are consumed. The kernel runs
the same two-pass tail-biting search as the GPU kernel, with one block per SBUF partition and the
trellis states along the free dimension. The Vector Engine runs the
forward passes, and the GpSimd Engine walks the survivors back while
other blocks run their forward passes, writing the bitstream directly
in the weight-load layout of \cref{fig:data_layout}.
In a cycle-level simulation of the pre-silicon design at $T{=}256$
and $K{=}4$, one chip encodes $69$ million values/s at $L{=}16$ and
$4.6$ billion at $L{=}10$ (a reference point, not a comparison with the GPU), so it too encodes those $7.0$ billion
values in under two minutes. For $K \in [2, 6]$
the rate stays within $0.90$--$1.04\times$ of the $K{=}4$ rate
(\cref{tab:trn4_enc_matrix}).

\section{Limitations and future work}
\label{sec:limitations}

\paragraph{Limitations.} The LLM experiments cover only Llama 3.1 8B and
70B, a model family that has since been superseded, and one preliminary
run on Qwen3.8-27B (\cref{app:llm}). The Trainium~4 encoder is measured
in cycle-level simulation, not on silicon.
Without a native decode unit, decode is ALU-bound where
bandwidth is plentiful: the H100 runs the $L{=}16$ decoder at $0.62\times$ an INT4 baseline ($0.79\times$ with an $L{=}10$ state-table variant, \cref{tab:stock_decode}),
and on Arm the NEON decoder runs $K{=}2$ and $3$ at only $1.0\times$ and
$0.91\times$ the INT4 speed although their streams are shorter
(\cref{app:cpu_decode}). The primitive is a short, fixed sequence: two $16$-bit multiplies, two shifts, an add and a $64$-byte table lookup. A single
fused decode instruction, like the dot-product and table-lookup
instructions CPUs and GPUs added for low-bit formats, would remove that
bottleneck, as the Trainium~4 datapath does natively.

\paragraph{Future work.} Fast encoding at small $L$ opens the door to
online quantization, such as KV caches and the weights in
quantization-aware training. We leave these to future work, along with
activation quantization, rotation-free quantization with retargeted
tables, and evaluation on architectures such as mixture-of-experts and multimodal
models.

\section{Conclusions}
\label{sec:conclusion}

CPTC replaces the stored trellis codebook with a computed one: two
$16$-bit multiplications and a $64$-byte symmetric table, where the
triangular index law of the two products spreads the codepoints
into the tails. One decoder serves every
$K \in [1, 6]$ and the standard scalar formats, meeting all four
requirements of \cref{sec:hw_criteria}. At $K{=}4$ it reaches
$3.93$ effective bits, matching RPTC run two state bits deeper, within $0.005$ bits of linear congruential codes that store $8$\,KiB, and exceeding QTIP. In YAQA's pipeline it has the lowest KL against
QTIP's HYB and 1MAD codebooks on Llama 3.1 8B and 70B without
fine-tuning. A timing-closed implementation in the weight-load path of the AWS
Trainium~4 AI accelerator, at FP4 ($K \le 4$) and FP8 ($K \le 6$) throughput parity for a
negligible fraction of the systolic array's area, shows the format is
realizable in hardware. On x86 and Arm CPUs the decoder runs at INT4
pace at $K{=}4$ with a $63$-entry table. The format
suits any write-once, read-many tensor; weights are the case
evaluated here, and its Viterbi encoder also runs on Trainium~4
itself in cycle-level simulation.
\label{sec:end_of_body}

\section*{Acknowledgements}

We thank Albert Tseng for many valuable discussions on trellis-coded
quantization. We thank Yida Wang and Jun Wu for encouraging us to
pursue this work. We thank Wojciech Romaszkan and Ron Diamant
for their help and support in integrating CPTC into AWS Trainium~4, and
other colleagues at Annapurna Labs for helping ship it into production.

\label{sec:end_of_main}

\bibliography{references}
\bibliographystyle{tmlr}

\appendix
\crefalias{section}{appendix}
\crefalias{subsection}{appendix}
\crefalias{subsubsection}{appendix}
\clearpage
\phantomsection\addcontentsline{toc}{section}{Appendix}
\begin{center}\LARGE\bf\sffamily Appendix\end{center}
\vspace{0.5\baselineskip}
\noindent The appendix has four parts, in the order the main text
uses them. \cref{app:part_details} gives the details of the decoder
and encoder of \cref{sec:method}: the tail-biting stream layout
(\cref{app:tailbiting_example}), the coefficient tables, search
protocol and table generation (\cref{app:constants}), the scalar-format emulation proofs
(\cref{app:emulation}), and in-stream metadata
(\cref{app:scale_encoding}). \cref{app:part_theory} is the theory
behind \cref{sec:theory_triangular,sec:shannon}: the index-law
ablation (\cref{app:table_law}), the alphabet-ceiling theorem
(\cref{app:ceiling}), and the achievable-rate reduction (\cref{app:shannon}). \cref{app:part_results} extends the
experiments of \cref{sec:experiments}: the i.i.d.\ results
(\cref{app:iid_extended,app:gg,app:dtype}), the matmul-aware methods
(\cref{app:matmul_calibration}) and the LLM details (\cref{app:llm}).
\cref{app:part_systems} covers the systems side of \cref{sec:systems}:
the GPU and CPU decode kernels (\cref{app:gpu_gemm,app:cpu_decode}) and
the fused Viterbi encoder (\cref{app:viterbi}); the accelerator is
described in \cref{sec:hardware}.

\section{Decoder and encoder details}
\label{app:part_details}

\subsection{Tail-biting construction}
\label{app:tailbiting_example}

This appendix gives the stream layout and the block decode
procedure behind \cref{sec:trellis}: \cref{fig:bitstream} shows the
layout and \cref{alg:stream} the procedure.

\paragraph{Construction.}
Tail-biting trellis codes were introduced by \citet{ma1986tailbiting}
in convolutional coding and adapted to bitshift trellis quantization
by \citet{tseng2024qtip}; we use the QTIP construction unchanged.
The block's first $L_{\max}-K$ bits are logically appended to its
end (line 1 of \cref{alg:stream}; at depth $L < L_{\max}$ only the first $L-K$ of
them affect any decoded value), so the $L$-bit windows read at
positions $0, K, \ldots, (T-1)K$ close into a cycle: the start and
end states agree by construction and no bits are spent on start-up.
The encoder runs Viterbi twice per block (once on the block rotated
by $T/2$, whose path fixes the state at the wrap seam, and once on the
block itself with its start state pinned to that state), per
\citet{tseng2024qtip}.

\begin{figure}[htbp]
  \captionsetup{hypcap=false}
  \centering
  \resizebox{\linewidth}{!}{%
  \begin{tikzpicture}[x=0.27cm, y=0.5cm, font=\scriptsize]
    \foreach \t in {0,...,7} {
      \pgfmathtruncatemacro{\xl}{28-4*\t}
      \pgfmathtruncatemacro{\parity}{mod(\t,2)}
      \ifnum\parity=0
        \draw[black!55, fill=blue!10] (\xl,0) rectangle ++(4,1);
      \else
        \draw[black!55, fill=green!13] (\xl,0) rectangle ++(4,1);
      \fi
      \node at (\xl+2,0.5) {$v_\t$};
    }
    \foreach \t in {0,1,2} {
      \pgfmathtruncatemacro{\xl}{-4*(\t+1)}
      \pgfmathtruncatemacro{\parity}{mod(\t,2)}
      \ifnum\parity=0
        \draw[black!55, dashed, fill=blue!6] (\xl,0) rectangle ++(4,1);
      \else
        \draw[black!55, dashed, fill=green!8] (\xl,0) rectangle ++(4,1);
      \fi
      \node at (\xl+2,0.5) {$v_\t'$};
    }
    \foreach \p in {0,4,8,...,32} {
      \pgfmathtruncatemacro{\x}{32-\p}
      \node[gray, anchor=south] at (\x,1.08) {\p};
    }
    \node[gray, anchor=south] at (-6,1.08)
      {$\leftarrow$ \emph{wraparound (tail-biting)}};
    \node[anchor=west, font=\scriptsize\bfseries] at (32.5,0.5)
      {bitstream};
    \foreach \t in {0,...,4} {
      \pgfmathtruncatemacro{\xl}{16-4*\t}
      \pgfmathtruncatemacro{\xr}{32-4*\t}
      \pgfmathtruncatemacro{\hi}{4*\t+15}
      \pgfmathtruncatemacro{\lo}{4*\t}
      \pgfmathsetmacro{\yt}{-0.55-\t*1.05}
      \draw[green!45!black, line width=0.8pt, rounded corners=2pt,
            fill=green!25]
        (\xl,\yt-0.8) rectangle (\xr,\yt);
      \node at ({(\xl+\xr)/2},\yt-0.4)
        {$w_\t$: bits$[\hi..\lo] \rightarrow q_\t$};
      \node[anchor=west] at (32.5,\yt-0.4) {window \t};
    }
    \foreach \t in {5,6,7} {
      \pgfmathtruncatemacro{\xr}{32-4*\t}
      \pgfmathtruncatemacro{\xw}{16-4*\t}
      \pgfmathtruncatemacro{\lo}{4*\t}
      \pgfmathtruncatemacro{\whi}{4*\t-17}
      \pgfmathsetmacro{\yt}{-0.55-\t*1.05}
      \draw[orange!75!black, line width=0.8pt, rounded corners=2pt,
            fill=orange!25]
        (0,\yt-0.8) rectangle (\xr,\yt);
      \draw[orange!75!black, line width=0.8pt, rounded corners=2pt,
            dashed, fill=orange!12]
        (\xw,\yt-0.8) rectangle (0,\yt);
      \ifnum\t=7
        \node[font=\tiny] at ({\xr/2},\yt-0.4) {$w_\t$};
        \node at ({\xw/2},\yt-0.4) {bits$[31..\lo]$, $[\whi..0]$};
      \else
        \node at ({\xr/2},\yt-0.4) {$w_\t$: bits$[31..\lo]$};
        \node[font=\tiny] at ({\xw/2},\yt-0.4) {bits$[\whi..0]$};
      \fi
      \node[anchor=west] at (32.5,\yt-0.4) {window \t};
    }
  \end{tikzpicture}}%

  \vspace{0.6em}

  \begin{minipage}{0.96\linewidth}
  \begin{algorithm}[H]
  \caption{\cptc{} block decode from the packed bitstream
    (tail-biting).}
  \label{alg:stream}
  \begin{algorithmic}[1]
  \Require code bits $b[0 \ldots TK{-}1]$ of one block ($K$ bits per
           value, $T$ values); parameters of \cref{alg:decode}.
  \State $b' \gets b \,\Vert\, b[0 \ldots L_{\max}{-}K{-}1]$
  \Comment{tail-biting: wrap the first $L_{\max}{-}K$ bits}
  \For{$t = 0, 1, \ldots, T-1$}
    \State $q \gets \sum_{k=0}^{L_{\max}-1} b'[tK{+}k] \cdot 2^{k}$
    \Comment{little-endian window; no masking to $L$ needed}
    \State $\hat{w}_t \gets \textsc{Decode}(q)$
    \Comment{\cref{alg:decode}}
  \EndFor
  \State \textbf{return} $\hat{w}_0, \ldots, \hat{w}_{T-1}$
  \end{algorithmic}
  \end{algorithm}
  \end{minipage}

  \caption{How a block is decoded from the packed bitstream.
    \textbf{Top:} bit layout for $K{=}4$, $T{=}8$, $L_{\max}{=}16$:
    value $v_t$'s code occupies bits $tK$ to $tK{+}K{-}1$ (bit $0$
    rightmost), window $t$ reads the $16$ bits starting at bit
    $tK$, and the last three windows (orange) wrap into the dashed
    tail-biting copy of the first $12$ bits. \textbf{Bottom:} the
    procedure (\cref{alg:stream}).}
  \label{fig:bitstream}
\end{figure}
\subsection{Recommended coefficients and table generation}
\label{app:constants}

\cref{tab:best_constants} gives recommended $(a_0, a_1)$ pairs,
searched over pairs in $[0, 2^L)^2$ whose OR is odd, on
$2^{19}$--$2^{20}$ Philox-generated $\mathcal{N}(0,1)$ samples (a
selection seed disjoint from every evaluation seed) at the paper's
table-size parameter $Q = 6$: per depth, a per-$K$ searched pair for
every $K \in [1, 6]$ and a single \emph{unified} pair selected on
all rates jointly (minimum mean deficit $K - K_{\text{eff}}$ over
the six rates). The $K{=}4$ column is the default pair, from a
search exhaustive through $L{=}10$ and sampled above (which the
plateau below justifies); the other columns re-score up to $2^{16}$
mirror-symmetric candidates per depth on all six rates. The pair
of each $(L, K)$ is then chosen among the best few candidates
with generated E2M5 tables under the standard protocol, and every
change is confirmed on $2^{30}$ samples.
Coefficients are in the $16$-bit form of \cref{alg:decode}
($16 - L$ trailing zeros). For $L \le 6$, $a_0 = 0$ suffices,
since all $2^L$ states map directly into the table; for $L \ge 7$
both coefficients are nonzero and, when distinct, produce the near-triangular index law
of \cref{sec:theory_triangular}. All searches at $L \ge 7$ require a perfectly
mirror-symmetric index histogram ($h[i] = h[126-i]$), which makes
the fold of \cref{sec:table} exact. At low rates and small depths
the per-$K$ winners are equal pairs ($a_0 = a_1$, which index the $64$ even
cells uniformly): there center resolution matters more than tail
protection. At $K{=}4$ quality is monotone in $L$ and saturates at
$K_{\text{eff}}{=}3.93$ ($L{=}16$, $K{=}4$); the unified pair is
within $0.017$ bits of the per-$K$ best at every rate for
$L \ge 12$.

\begin{table}[htbp]
  \centering
  \small
  \caption{The coefficient pairs to use, in hex, for each trellis depth
    $L$: the searched pair for each bit width $K$ and one unified pair
    for all rates, searched at $Q{=}6$. The $K{=}4$ column is the
    default pair; for $L \le 6$ one pair serves every $K$.}
  \label{tab:best_constants}
  \begin{tabular}{cccccccc}
    \toprule
    $L$ & $K{=}1$ & $K{=}2$ & $K{=}3$ & $K{=}4$ &
    $K{=}5$ & $K{=}6$ & Unified \\
    \midrule
     1 & \multicolumn{7}{c}{\texttt{0000,8000}} \\
     2 & \multicolumn{7}{c}{\texttt{0000,C000}} \\
     3 & \multicolumn{7}{c}{\texttt{0000,A000}} \\
     4 & \multicolumn{7}{c}{\texttt{0000,5000}} \\
     5 & \multicolumn{7}{c}{\texttt{0000,2800}} \\
     6 & \multicolumn{7}{c}{\texttt{0000,A400}} \\
    \midrule
     7 & \texttt{7A00,7A00} & \texttt{3A00,3A00} & \texttt{A200,A200} & \texttt{6200,C400} & \texttt{2E00,5C00} & \texttt{0200,0400} & \texttt{6200,C400} \\
     8 & \texttt{8700,8700} & \texttt{7A00,BD00} & \texttt{6100,6100} & \texttt{1B00,3600} & \texttt{7700,DC00} & \texttt{0400,8100} & \texttt{9700,AE00} \\
     9 & \texttt{8580,8580} & \texttt{4280,4280} & \texttt{E180,E180} & \texttt{6100,B080} & \texttt{6500,9280} & \texttt{CC80,D900} & \texttt{9900,AC80} \\
    10 & \texttt{8940,8940} & \texttt{41C0,41C0} & \texttt{8A80,9540} & \texttt{39C0,4E00} & \texttt{1A40,A400} & \texttt{1380,99C0} & \texttt{A880,C440} \\
    11 & \texttt{7920,7920} & \texttt{41E0,41E0} & \texttt{7480,AD20} & \texttt{5D00,B3A0} & \texttt{6040,A820} & \texttt{1400,C0A0} & \texttt{9BE0,A7C0} \\
    12 & \texttt{5F60,6BB0} & \texttt{77D0,C7A0} & \texttt{7DA0,E2D0} & \texttt{81D0,EE80} & \texttt{6810,A020} & \texttt{7020,9410} & \texttt{4550,82A0} \\
    13 & \texttt{61C8,6F20} & \texttt{3490,8C48} & \texttt{8320,DCC8} & \texttt{51C0,DE38} & \texttt{D820,F008} & \texttt{3008,7C10} & \texttt{43F8,83F0} \\
    14 & \texttt{6BBC,7178} & \texttt{7708,C884} & \texttt{74B0,AB2C} & \texttt{6D20,82A4} & \texttt{07F0,CFFC} & \texttt{5008,E404} & \texttt{7E48,C824} \\
    15 & \texttt{6B6E,73DC} & \texttt{792E,C75C} & \texttt{1D48,8252} & \texttt{11CA,5E50} & \texttt{9808,A002} & \texttt{5A84,9DC2} & \texttt{483E,677C} \\
    16 & \texttt{9097,97AE} & \texttt{7AE6,C8B3} & \texttt{7474,AB1D} & \texttt{6A78,85CF} & \texttt{1C04,3C01} & \texttt{0AEE,D337} & \texttt{9282,A981} \\
    \bottomrule
  \end{tabular}
  
\end{table}

\paragraph{Table generation.} The tables of \cref{sec:table}
are generated per target and $(L, K)$ at $T{=}256$ from $2^{22}$ synthetic
samples of the target per round: $30$ Lloyd rounds (a tail-biting
Viterbi encode, then each entry becomes the mean of the samples
assigned to its cell, with mirrored cells pooled for $L > 6$) and $20$
damped ones, keeping the round that scores best on a held-out set.
Generation starts from \cref{eq:table_rule} rescaled by
$\sqrt{(1 + 2/n)(1 - 4^{-K})}$ with $n = L - K + 1$, a heuristic starting
width from the point-density
law of fixed-rate high-resolution quantization in $n$ dimensions
\citep{gersho1979asymptotically} with a rate correction (for the
generalized Gaussians of \cref{app:gg}, $(1 + 2/n)^{1/\beta}$); for
$L \le 6$, where every state has its own entry, it is run from the
\cref{eq:table_rule} table of every odd multiplier $a_1$, each of
which orders the states differently, prescreened with
$12$ rounds on $2^{20}$ samples with the best three kept. The
result is fitted to its data type by $16$ further rounds with every
entry snapped to the grid. A final pass then moves single entries to
a neighboring grid value while the measured error falls. Reported
numbers use seeds disjoint from generation.

\paragraph{Bad pairs and the plateau.} Exhaustive Viterbi evaluation
at $K{=}4$, with \cref{eq:table_rule} tables, of every unordered pair with odd OR
(histogram symmetry not imposed) at
$L{=}8$ ($24{,}640$ pairs) and $L{=}10$ ($393{,}472$), plus a
$40{,}000$-pair sample at $L{=}12$, shows that failure is governed by
the sum $\Sigma = a_0 + a_1 \bmod 2^L$, with $|\Sigma| = \min(\Sigma, 2^L - \Sigma)$: no pair with
$|\Sigma| < 2^{L-5}$ or $2$-adic valuation $\nu_2(\Sigma) > 2$
(${\approx}14\%$ of the space) is within $0.05$ bits of the optimum
at $L{=}8$ or $10$. Outside this set the landscape is a broadening
plateau: the fraction of pairs within $0.05$ bits of the optimum
grows from $15\%$ ($L{=}8$) to $30\%$ ($L{=}10$) to $63\%$
($L{=}12$), and the max-to-median gap shrinks $0.11 \to 0.07 \to
0.04$ bits, so the search only has to avoid the bad set and pick
from a large near-optimal one.

\paragraph{All-$K$ selection.} At $L{=}16$ the unified pair scores
$(0.948, 1.945, 2.938, 3.931, 4.906, 5.795)$ for $K = 1, \ldots, 6$
(generated E2M5 tables), within $0.008$ bits of the per-$K$ pair at
every rate, so joint selection is essentially free at the design
point. The stakes are larger at small depths: at $L{=}10$ the
unified pair gives up $0.02$ bits to the default pair at $K{=}4$ and gains $0.08$--$0.09$ over it at
$K \in \{3, 5, 6\}$, the relevant trade for online small-$L$ uses
(\cref{sec:encode}). The paper's quality tables load the per-$K$
column for each rate; the unified column serves deployments that
fix one pair for all rates.

\paragraph{Coefficients for $Q = 7$.} Where the $127$-codepoint
alphabet ceiling binds ($K \ge 5$), one table doubling recovers about a third
of the $K{=}6$ deficit in either data type (about half if the entries are also widened to FP16) while keeping the symmetric fold. Re-searching pairs
whose $255$-cell $Q{=}7$ histogram is full and mirror-symmetric
yields, with FP16 tables, $4.926$ at $K{=}5$ (pair
$(\texttt{0xBEC5}, \texttt{0xC514})$) and $5.904$ at $K{=}6$ (pair
$(\texttt{0x06CE}, \texttt{0x13E7})$), with $K \le 4$ unchanged; in
E2M5 at the scale of \cref{app:dtype} the same
pairs score $4.919$/$5.869$, the CPTC$_{Q7}$ row of
\cref{tab:qtip_k}. A single unified $Q{=}7$ pair
$(\texttt{0x518D}, \texttt{0x781A})$ is within $0.005$ bits of the
per-$K$ winners at every rate. By \cref{prop:dominance}, $Q{=}7$
hardware serves every $Q \le 7$ configuration.

\subsection{Scalar format emulation}
\label{app:emulation}

This appendix gives the constructions behind the scalar-format
reductions of \cref{sec:constants}, and proves the dominance rule
(\cref{prop:dominance}).

\paragraph{Integer formats via CPTC.} Setting
$a_0 = \texttt{0x0000}$, $a_1 = 2^{16-K}$, $s = 6 - K$, and
$\mathcal{T}[q] = (q - 2^{K-1})/8$ for $q \in [0, 2^K{-}1]$ (a signed
$K$-bit integer ramp at scale $\tfrac18$, every entry an E2M5 value for
$K \le 6$) reduces \cref{alg:decode} to a direct integer lookup.
Since $L = 16 - \mathrm{ctz}(a_1) = K$, the trellis is
degenerate. All $2^K$ indices fall in the lower half of the table; the
symmetric upper half is unused.

\paragraph{FP4 via CPTC at $K{=}4$.} Setting
$K = 4$, $a_0 = \texttt{0x0000}$, $a_1 = \texttt{0x1000}$, $s = 2$,
and populating the first $16$ entries of $\mathcal{T}$ with the
FP4 codepoints, each of which is an E2M5 value, reduces the
decoder to an FP4 lookup.

\paragraph{Dominance-rule mechanics (proof of
\cref{prop:dominance}).} The bit-exact emulation
works direction by direction. \emph{Table direction:} the shift $s$ lets a $Q$-sized
implementation emulate one built with a smaller table parameter
$Q' = Q - s$: for $s \ge 1$ every index in \cref{alg:decode}
falls below $2^Q$, so the mirror branch of \cref{alg:decode}
(lines 6--10) is inactive and the
emulated codebook, signs stored explicitly, occupies the first
$2^{Q-s+1}{-}1$ entries of $\mathcal{T}$ (the FP4 embedding above
uses this). \emph{Width direction:} replacing $a_j$ (mod
$2^{L'_{\max}}$) by $a_j \cdot 2^{L_{\max} - L'_{\max}}$ shifts
every product's bits up in lockstep, and the top-bits extraction
of \cref{alg:decode} line 3 removes the difference; this is
checked against the golden reference decoder (\cref{app:viterbi}). \emph{Data-type direction:} the
dominated table loads after the common rescaling $c$ that a
per-table scale absorbs, e.g.\ FP4's
codepoints are E2M5 values (\cref{sec:hardware}).
\subsection{Encoding auxiliary information into the CPTC bitstream}
\label{app:scale_encoding}

This appendix develops the in-stream metadata thread of
\cref{sec:shannon}. CPTC's encoder is free to choose, among all bit sequences that decode
to a low-MSE reproduction of the input, one whose bits also satisfy a
chosen linear constraint. This lets the encoder commit auxiliary
metadata (a per-block scale, a zero point, an exponent, a flag)
\emph{into the CPTC codeword itself}, with no separate storage and
no change in the encoder's complexity.

\paragraph{Construction.} Commit auxiliary bit $b_j$ at the boundary of group $j$ of $g$ values by constraining the parity (XOR) of that group's $gK$ emitted code bits to equal $b_j$. The constraint is
enforced inside Viterbi without enlarging the state space: track the
cumulative parity as one extra bit alongside each state's accumulated
cost, and at every group boundary set the cost of any state whose
recorded parity disagrees with the target bit to $+\infty$.
Subsequent steps propagate only consistent paths, so the encoder
stays $O(T \cdot 2^L)$, the same complexity as auxiliary-free CPTC.
This is a heuristic rather than the exact constrained optimum,
which would double the state space with the parity bit: when two
paths of different parity merge into one state, only the
survivor's parity is kept. The measured cost below includes
whatever optimality this forfeits.

\paragraph{Rate cost.} Each auxiliary bit fixes one linear
constraint on the $gK$ code bits committed to it, removing one
degree of freedom from the encoder's search space: $1/g$
effective bits per value per auxiliary bit. The canonical use is the
per-block exponent of an MX-style format \citep{rouhani2023mx}: at
$L{=}16$, $K{=}4$, an $8$-bit scale per $G{=}64$ values ($g{=}8$) costs
$1/8 = 0.125$
effective bits per value in theory and $0.132$ measured,
leaving CPTC at $K{=}4$ with an in-stream scale at $3.80$ effective bits,
still well above the $3.36$ of the
Lloyd--Max scalar.
The same construction carries per-block zero points (asymmetric
quantization), per-block format flags, or any other
small-bit-budget metadata.
\section{Theory}
\label{app:part_theory}

\subsection{Index-law ablation: uniform vs.\ triangular composition}
\label{app:table_law}

This appendix quantifies the statement of \cref{sec:theory_triangular}
that the table has to match the index law: filled as if the index
were uniform, a small table loses its trellis gain. \cref{fig:companding} shows the two placements and
the two laws side by side.

\paragraph{Relation to 1MAD and LC.} QTIP's 1MAD also sums
equidistributed quantities (hashed bytes) to obtain a bell-shaped
law, but it uses the sum \emph{as the value}: the decoded number is
the sum itself, so the reproduction distribution is whatever the
central-limit sum gives and cannot be retargeted. CPTC uses the sum
as an \emph{index} into a table composed with that sum's law
(\cref{eq:table_rule}), which is a companding rule: the law
determines where codepoints are dense and the table determines what
they are. That separation is what lets a $64$-byte table follow a
Gaussian, a generalized Gaussian or a bimodal matmul-aware target
(\cref{sec:exp_iid,sec:exp_matmul_synth}). LC has the same
separation of labels from reproducers, but one congruential map
gives a near-uniform index law (\cref{prop:reduction}, \cref{app:shannon}).

\paragraph{Codebook level.} \cref{tab:law_scalar} reports exact
scalar nearest-codepoint MSE on $\mathcal{N}(0,1)$ for $M$-point
codebooks $v_i = F^{-1}\bigl((G(i{-}1)+G(i))/2\bigr)$, the midpoint
quantile of each cell of the index law's CDF $G$, under three index laws: uniform
(equal probability mass), symmetric triangular, and the Lloyd--Max
optimum \citep{lloyd1982lsq,max1960quantizing} as the reference.
Triangular tracks Lloyd--Max within $7$--$17\%$ at every size, while
uniform degrades from $2.2\times$ (at $M{=}15$) to $29\times$ (at
$M{=}511$) worse than triangular. The mechanism is overload: the
uniform law's largest codepoint sits at the $(M-0.5)/M$
quantile ($2.66$ at $M{=}127$), and the two unbounded tail
cells contribute $83\%$ of its MSE; the triangular law reaches
$3.67$ (tail share $16\%$) and Lloyd--Max $4.18$ (tail share $2\%$).
Because overload decays only as about $1/(M \ln M)$ while granular
error decays as $M^{-2}$ \citep{panter1951quantization,
bennett1948spectra}, the uniform law also loses the \emph{rate}:
doubling $M$ buys only ${\sim}2.3\times$ instead of ${\sim}4\times$.
Matching the $127$-entry triangular codebook with a uniform-composed
one takes $M > 511$: the composition is worth more than $4\times$
in table size.

\begin{table}[htbp]
  \centering
  \small
  \caption{Exact MSE ($\downarrow$) of $M$-point Gaussian codebooks built
    from a uniform vs.\ a triangular index law, against the
    Lloyd--Max optimum.}
  \label{tab:law_scalar}
  \begin{tabular}{rccccc}
    \toprule
    $M$ & Uniform & Triangular & Lloyd--Max &
    $\tfrac{\text{Unif}}{\text{Tri}}$ &
    $\tfrac{\text{Tri}}{\text{LM}}$ \\
    \midrule
     15 & $2.52\text{e-}2$ & $1.14\text{e-}2$ & $1.07\text{e-}2$ & $2.2$ & $1.07$ \\
     31 & $1.00\text{e-}2$ & $2.90\text{e-}3$ & $2.66\text{e-}3$ & $3.5$ & $1.09$ \\
     63 & $4.17\text{e-}3$ & $7.37\text{e-}4$ & $6.65\text{e-}4$ & $5.7$ & $1.11$ \\
    127 & $1.79\text{e-}3$ & $1.88\text{e-}4$ & $1.66\text{e-}4$ & $9.6$ & $1.13$ \\
    255 & $7.85\text{e-}4$ & $4.77\text{e-}5$ & $4.15\text{e-}5$ & $16.5$ & $1.15$ \\
    511 & $3.50\text{e-}4$ & $1.21\text{e-}5$ & $1.04\text{e-}5$ & $28.9$ & $1.17$ \\
    \bottomrule
  \end{tabular}
  
\end{table}

\paragraph{Beyond the Gaussian.} The same comparison at $M{=}127$
(all sources unit variance): uniform-over-triangular MSE ratio is
$9.6\times$ for $\mathcal{N}(0,1)$, $15.6\times$ for Laplacian, and
$7.4\times$ for Student-$t_5$. The ratio exceeds $7\times$ for all three targets. The composition benefit is a
property of mean-concentrated targets generally, not of the Gaussian.

\paragraph{End-to-end trellis ablation.} \cref{tab:law_trellis}
repeats the comparison inside the full CPTC pipeline ($L{=}16$,
$K{=}4$, $T{=}128$, the paper's $(a_0, a_1)$, Viterbi encoding of
$2^{22}$ i.i.d.\ $\mathcal{N}(0,1)$ samples), varying only how the
table is populated. Two conclusions. First, filling the table as if
the index were uniform costs
$0.69$ effective bits ($3.92 \to 3.23$) at $T{=}128$ with quantile-midpoint tables (the same at $T{=}256$ with conditional-mean entries, \cref{tab:gg}) and lands \emph{below the Lloyd--Max scalar}
($3.36$).
Second, replacing the triangular composition with the scalar
Lloyd--Max table of the same size changes nothing ($3.917$ vs.\
$3.919$): the triangular composition already places each entry
inside its cell, and at $L{=}16$ the partition itself is
also near-optimal.
The composition assumes uniformly random states, while the encoder visits
them non-uniformly; the two index distributions differ by only
$0.002$ in total variation at $L{=}16$, $K{=}4$. This compares visit
frequencies only; which samples an entry receives is a separate
matter, addressed by table generation (\cref{sec:table}).

\begin{table}[htbp]
  \centering
  \small
  \caption{Effective bits ($\uparrow$) of the full CPTC pipeline when
    only the law that fills the table changes ($L{=}16$, $K{=}4$,
    $T{=}128$, $2^{22}$ samples; quantile-midpoint tables, before
    table generation).}
  \label{tab:law_trellis}
  \begin{tabular}{clc}
    \toprule
    Effective entries & Table law & $K_{\text{eff}}$ \\
    \midrule
    $127$ ($s{=}0$) & Uniform composition & $3.233$ \\
    $127$ ($s{=}0$) & Scalar Lloyd--Max table & $3.917$ \\
    $127$ ($s{=}0$) & Triangular composition & $\mathbf{3.919}$ \\
    \midrule
    $63$ ($s{=}1$) & Uniform composition & $3.197$ \\
    $63$ ($s{=}1$) & Triangular composition & $\mathbf{3.903}$ \\
    \midrule
    $31$ ($s{=}2$) & Uniform composition & $3.048$ \\
    $31$ ($s{=}2$) & Triangular composition & $\mathbf{3.821}$ \\
    \bottomrule
  \end{tabular}
  
\end{table}

\subsection{Alphabet ceilings: what the table can and cannot cost}
\label{app:ceiling}

CPTC's decoder emits values from a fixed table, $127$ entries at
$Q{=}6$. \cref{prop:alphabet_floor} bounds what such a table can
achieve when it has at most $2^K$ values. \cref{thm:ceiling} in \cref{sec:shannon}
gives the general form of that bound, a restatement of the classical
alphabet-constrained rate-distortion function
\citep{finamore1980optimal,berger1971rate}, which
\citet{eriksson2007lc} also use as the reference for LC codes. This appendix proves it, computes it for the CPTC and HYB alphabets of
\cref{tab:qtip_k}, and uses it to bound how much of
CPTC's gap to the Shannon limit the set of table values forces.

\begin{proof}[Proof of \cref{thm:ceiling}]
Let $B$ be the $nK$-bit codeword, so $I(X^n;\hat X^n) \le
I(X^n; B) \le H(B) \le nK$. Since the $X_i$ are independent and
$\hat X^n$ is a function of $B$, $I(X^n;\hat X^n) \ge \sum_i
I(X_i;\hat X_i)$ (independence and the chain rule; \citealp{berger1971rate}). Each $\hat
X_i$ lies in $C$, so $I(X_i;\hat X_i) \ge R_C(D_i)$ with $D_i =
\mathbb{E}(X_i - \hat X_i)^2$ and $R_C$ the rate-distortion
function with reproduction alphabet $C$, which is convex and
nonincreasing because the alphabet constraint is per letter and
preserved under time sharing. Hence $nK \ge \sum_i R_C(D_i) \ge
n R_C(\bar D)$, i.e.\ $\bar D \ge D_C(K)$. Dropping the alphabet
constraint gives the Shannon bound.
\end{proof}

In the classical random-coding setting the ceiling is approached:
for a memoryless source and a bounded distortion measure there exist
time-varying trellis codes whose distortion approaches the
rate-distortion function exponentially in the constraint
length \citep{viterbi1974trellis}. RPTC and CPTC are
time-invariant, and squared error on a Gaussian is unbounded, so for
them $K_C(K)$ is a bound, not a proven asymptote. It is also computable: $D_C(K)$ is a
convex program over the output law on $C$, and the Blahut--Arimoto
iteration \citep{blahut1972computation} converges to it. We
discretize $\mathcal{N}(0,1)$ on $17{,}001$ points of $[-8.5,
8.5]$ (step $0.001$, far below every codepoint spacing) and
bisect the log-slope to a bracket of $10^{-4}$. As
a check, the $16$-point Lloyd--Max alphabet at $K{=}4$ gives
$K_C = 3.3589$, the scalar Lloyd--Max value to four decimals, as
it must (with $|C| = 2^K$ the optimal assignment is
deterministic). For HYB the alphabet is a set of value pairs, so
the same computation runs in two dimensions at rate $2K$ per
pair on a $401 \times 401$ grid over $[-6, 6]^2$, with the pair
table normalized to unit variance as in \cref{tab:qtip_k}.

\begin{table}[htbp]
  \centering
  \small
  \caption{The most effective bits ($\uparrow$) any code of rate $K$ can
    reach with a given reproduction alphabet, at any trellis depth
    (\cref{thm:ceiling}, standard normal). Measured $L{=}16$ rows
    repeat \cref{tab:qtip_k}; the CPTC row uses the generated $L{=}16$ table of each $K$. HYB's alphabet is $512$ $k$-means
    pairs with a sign flip on the first component ($1{,}024$ pairs at
    rate $2K$ per pair).}
  \label{tab:ceiling}
  \begin{tabular}{lccccccc}
    \toprule
    Alphabet & $|C|$ & $K{=}1$ & $K{=}2$ & $K{=}3$ & $K{=}4$ & $K{=}5$ & $K{=}6$ \\
    \midrule
    Shannon limit & $\infty$ & $1.000$ & $2.000$ & $3.000$ & $4.000$ & $5.000$ & $6.000$ \\
    Lloyd--Max optimal $127$-point table & $127$ & $1.000$ & $2.000$ & $3.000$ & $3.999$ & $4.993$ & $5.930$ \\
    CPTC generated table, E2M5 (deployed) & $\le 127$ & $1.000$ & $2.000$ & $3.000$ & $3.998$ & $4.987$ & $5.871$ \\
    \quad CPTC measured, $L{=}16$ & & $0.950$ & $1.948$ & $2.943$ & $3.934$ & $4.913$ & $5.802$ \\
    Equal-mass table, midpoints & $127$ & $1.000$ & $1.995$ & $2.955$ & $3.786$ & $4.316$ & $4.521$ \\
    Equal-mass table, midpoints & $64$ & $1.000$ & $1.987$ & $2.898$ & $3.577$ & $3.894$ & $3.966$ \\
    QTIP HYB pair table & $1{,}024$ pairs & $1.000$ & $1.994$ & $2.944$ & $3.687$ & $3.936$ & $3.936$ \\
    \quad HYB measured, $L{=}16$ & & $0.932$ & $1.912$ & $2.840$ & $3.611$ & $3.936$ & $3.936$ \\
    \bottomrule
  \end{tabular}
\end{table}

\paragraph{What the table costs.} Every measured value in
\cref{tab:ceiling} lies at or below its ceiling (HYB attains it for $K \ge 5$), as the theorem
requires, and the ceilings split CPTC's gap to Shannon into a
part its table values force ($K - K_C$) and a remainder ($K_C - $ measured). At
$K \le 4$ at most $0.002$ bits of the $0.05$--$0.07$-bit gap of
\cref{tab:qtip_k} is forced by the set of table values; the
remainder depends on depth and on the map from states to entries. At
$K{=}5$ the table values force $0.013$ of the $0.087$-bit gap, at
$K{=}6$ $0.129$ of $0.198$, the rigorous form of the
alphabet ceiling that \cref{sec:exp_iid} and \cref{app:constants}
observe. The Lloyd--Max $127$-point row is a reference alphabet:
the deployed table is within $0.001$ bits of it through $K{=}4$
and $0.06$ below at $K{=}6$. For HYB the ceiling saturates at $3.936$ from
$K{=}5$ on (the rate constraint stops binding once it exceeds
the alphabet's nearest-neighbor entropy of $9.92$ bits per
pair), and the measured plateau is $3.936$: HYB's fixed
$512$-pair table is provably what caps it past $K{=}4$, and
no trellis depth can lift it.

\paragraph{One multiplication, scale, and depth.} The equal-mass
rows quantify the tail argument of \cref{sec:theory_triangular}.
With the equal-mass composition at its natural scale the extreme
codepoint sits at the quantile $1 - 1/(2M)$, and the overload error
alone bounds $K_{\text{eff}} \le \tfrac12 \log_2 (M \ln M) + o(1)$
regardless of $K$ or $L$ (the exact overload integral gives
$4.16$ at $M{=}64$ and $4.74$ at $M{=}127$; the closed form alone
$4.03$ and $4.63$),
while the two-multiply law places the extreme cell at mass $2^{-2Q}$, an
overload of $2.7 \times 10^{-5}$ at $Q{=}6$, below the Shannon
distortion for every $K \le 7$. This bound is a property of the
composition, not of the multiplication count: stretching the
$127$-point uniform-law table by a factor $\kappa$ raises its ceiling at
$K{=}4$ from $3.82$ at $\kappa{=}1$ (conditional-mean entries; $3.79$
with quantile midpoints, \cref{tab:ceiling}) to $3.92$ ($\kappa{=}1.1$)
and $3.98$ ($\kappa{=}1.3$).
What remains for a one-multiply table is a finite-depth cost
(\cref{tab:scale_L16}): with a tuned stretch it recovers from
$3.78$ to $3.88$ at $K{=}4$ and from $4.37$ to $4.80$ at $K{=}5$,
while the two-multiply table is nearly scale-insensitive (${+}0.005$ and
${+}0.016$ at its best stretch, close to the rescaling table
generation starts from) and still leads by $0.06$ and $0.12$ bits
at matched table size ($128$ vs.\ $127$ cells; the generated $64$-byte tables are $0.04$ and $0.09$ apart, \cref{app:iid_extended}), and by $0.11$ and
$0.25$ against a $64$-cell one-multiply table. One multiplication can protect the tails, but only with a
stretch worth $0.1$--$0.4$ bits at $128$ cells, and then trails at
matched table size; for the triangular law the stretch is a
$0.01$-bit refinement.

\begin{table}[htbp]
  \centering
  \small
  \caption{Effective bits ($\uparrow$) at $L{=}16$ when the table is
    stretched by a factor $\kappa$: one-multiply tables with conditional-mean entries (best of $16$
    random odd multipliers, chosen at $\kappa{=}1$ on held-out data)
    against the two-multiply table of \cref{eq:table_rule} in FP16,
    before table generation ($T{=}256$, $3$ seeds).}
  \label{tab:scale_L16}
  \begin{tabular}{lccccccc}
    \toprule
    Table & $K$ & $\kappa{=}1.0$ & $1.05$ & $1.1$ & $1.2$ & $1.3$ & $1.4$ \\
    \midrule
    One multiply, $64$ cells & $4$ & $3.612$ & $3.697$ & $3.755$ & $3.817$ & $\mathbf{3.825}$ & $3.802$ \\
    One multiply, $128$ cells & $4$ & $3.777$ & $3.832$ & $3.862$ & $\mathbf{3.878}$ & $3.858$ & $3.820$ \\
    Two multiplies (CPTC), $127$ cells & $4$ & $3.930$ & $\mathbf{3.935}$ & $3.932$ & $3.908$ & $3.871$ & $3.826$ \\
    \midrule
    One multiply, $64$ cells & $5$ & $3.986$ & $4.154$ & $4.292$ & $4.502$ & $4.622$ & $\mathbf{4.672}$ \\
    One multiply, $128$ cells & $5$ & $4.374$ & $4.526$ & $4.629$ & $4.760$ & $\mathbf{4.796}$ & $4.790$ \\
    Two multiplies (CPTC), $127$ cells & $5$ & $4.902$ & $4.916$ & $\mathbf{4.918}$ & $4.901$ & $4.866$ & $4.822$ \\
    \bottomrule
  \end{tabular}
\end{table}

\subsection{Achievable rate and the limit \texorpdfstring{$L \to \infty$}{L to infinity}}
\label{app:shannon}

The classical achievability results are \citet{viterbi1974trellis}
(time-varying trellises with random branch labels achieve the
rate-distortion function of an i.i.d.\ source under a bounded
distortion measure) and
\citet{gray1977trellis} (a non-constructive existence proof for
time-invariant trellis codes). \citet{mao2011rptc} complement these
with \emph{necessary} conditions on every asymptotically optimal
stationary code (the marginal reproduction distribution converges
to a Shannon-optimal one; finite-order joints factor; the
reproduction whitens; the code bits approach fair coins) and prove
that RPTC satisfies the marginal condition. Their proof uses one
property of the random permutation, bijectivity, so it transfers
deterministically to the one-multiply, full-table case of CPTC,
for every odd multiplier:

\begin{proposition}[Reduction and marginal optimality]
\label{prop:reduction}
With $a_0 = 0$, $a_1 = a \cdot 2^{16-L}$ for odd $a$, $Q = L$, $s = 0$, and $\mathcal{T}$
holding the target's $2^L$ quantile values (RPTC's rule;
\cref{eq:table_rule} gives the cell means instead, with the same
limit), CPTC is RPTC with the specific
permutation $P_a(q) = a q \bmod 2^L$ in place of a random one:
$P_a$ is a bijection for every odd $a$, and the decoder computes
$\phi(q) = \mathcal{T}[P_a(q)]$. Consequently, for any target
with CDF $F$ and finite variance, the decoded-state marginal
under uniform states is the exact $2^L$-point quantile
discretization of $F$, independent of $a$, and converges weakly
(and in Wasserstein-$2$) to the target as $L \to \infty$, so
under uniformly distributed states this case satisfies the necessary marginal condition of
\citet{mao2011rptc} for every odd multiplier.
\end{proposition}

The condition is necessary, not sufficient, and does not
discriminate among coefficients: the identity permutation
($a = 1$) satisfies it and yields a poor code. It concerns the \emph{optimal
reproduction} distribution, $\mathcal{N}(0, \sigma^2 - D)$ for
Gaussian MSE, not the source; since \cref{prop:reduction} holds
for any loaded $F$, the table can be composed to the test channel,
which at our rates is numerically indistinguishable from the source
($\sigma^2 - D \approx 0.996$ at $K{=}4$). The same reduction places one-multiply CPTC
inside the linear congruential trellis family of
\citet{eriksson2007lc}: $P_a$ carries the bitshift dynamics to an affine recursion,
$P_a((q \ll K) \mid u) = (P_a(q) \ll K) + a u \bmod 2^L$, so the
permuted state follows their generating LC recursion and the
destination-state labels realize their affine congruential label
rule. In neither family is a decode-time recursion implied; for
CPTC the incremental update (\cref{eq:incremental_mult}) is an
optional optimization.

What neither family has is an achievability theorem. The joint conditions are open even for RPTC: \citet{mao2011rptc}
anticipate ``a source coding theorem using random coding ideas'',
and none has appeared; for CPTC with two coefficients and a small
table both the marginal and the joint conditions are open. All
CPTC results are at finite $L \le 16$, where the per-$L$
comparison with RPTC (\cref{tab:scaling_L}) is direct evidence and no such
theorem is needed.

\section{Extended results}
\label{app:part_results}

\subsection{Extended i.i.d.\ source-quantization results}
\label{app:iid_extended}

This appendix collects the i.i.d.\ results that \cref{sec:exp_iid}
summarizes: the trellis-depth sweep behind \cref{fig:cptc_vs_all}
(\cref{tab:scaling_L}), the block-size sweep, the block-scaled scalar
formats, the RPTC
search-budget control, the LC baseline construction with the
headline-margin precision, and the lattice VQ
baselines of \cref{tab:hw_criteria}. Each baseline follows its own
paper: RPTC takes the quantiles of $\mathcal{N}(0, 1 - 4^{-K})$ under a
random permutation, in $16$ bits and with no Lloyd step; QTIP runs the
authors' code and tables, with codebook and data matched in scale as
in its pipeline; LC is described below. The CPTC and LC tables are
Lloyd-iterated.

\begin{table}[htbp]
  \centering
  \small
  \caption{Effective bits ($\uparrow$) at each trellis depth $L$ for
    $K{=}4$ ($T{=}256$, $2^{22}$ samples): RPTC, the $64$-byte LC code
    and CPTC, with QTIP's 1MAD at $L{=}16$ (\cref{tab:qtip_k}). Scalar
    formats stop at $3.36$ (Lloyd--Max; symmetric INT4 $3.14$, FP4 $3.15$)
    and the limit is $4.00$.}
  \label{tab:scaling_L}
  \begin{tabular}{lccccccccccc}
    \toprule
    $L$ & $8$ & $9$ & $10$ & $11$ & $12$ & $13$ & $14$ & $15$ &
    $16$ & $17$ & $18$ \\
    \midrule
    QTIP 1MAD &
      --- & --- & --- & --- & --- & --- & --- & --- &
      $3.86$ & --- & --- \\
    RPTC &
      $3.45$ & $3.55$ & $3.64$ & $3.72$ & $3.78$ &
      $3.82$ & $3.85$ & $3.88$ & $3.90$ & $3.91$ & $3.92$ \\
    LC, $64$\,B &
      $3.36$ & $3.70$ & $3.81$ & $3.86$ & $3.87$ &
      $3.88$ & $3.89$ & $3.89$ & $3.90$ & --- & --- \\
    CPTC &
      $3.77$ & $3.81$ & $3.84$ & $3.86$ & $3.89$ &
      $3.90$ & $3.92$ & $3.93$ & $\mathbf{3.93}$ & --- & --- \\
    \bottomrule
  \end{tabular}
\end{table}

\paragraph{Block size $T$.} Larger $T$ gives the Viterbi encoder more
freedom to find a good cyclic walk; smaller $T$ adds tail-biting
boundaries. Empirically (\cref{tab:block_size}) quality saturates
near $T = 128$ on the standard normal at $K{=}4$, $L{=}16$.

\begin{table}[htbp]
  \centering
  \small
  \caption{How effective bits ($\uparrow$) depend on the block length
    $T$ ($K{=}4$, $L{=}16$, standard protocol).}
  \label{tab:block_size}
  \begin{tabular}{cccccc}
    \toprule
    Block size $T$ & $16$ & $32$ & $64$ & $128$ & $256$ \\
    \midrule
    $K_{\text{eff}}$ & $3.24$ & $3.71$ & $3.87$ & $3.92$ & $3.93$ \\
    \bottomrule
  \end{tabular}
\end{table}

\paragraph{Block-scaled scalar formats.} The $4$-bit formats that
hardware executes natively attach one $8$-bit scale to each group
of $G$ values: MXFP4 \citep{rouhani2023mx} an E8M0 (power-of-two)
scale with $G{=}32$, $4.25$ bits per value, and NVFP4 \citep{nvidia2025nvfp4} an E4M3
scale with $G{=}16$, $4.5$ bits. \cref{tab:microscaling} scores
both on the standard normal under the protocol of
\cref{sec:exp_iid} ($2^{22}$ samples, round-to-nearest, the
accelerator's bit-exact microscaling library, each format's own
per-group max-abs scale rule: the shared E8M0 exponent of the
group maximum for MX, an E4M3 scale for NVFP4), together with the
INT4 element type and the $6$-bit variants. A power-of-two scale
buys nothing on the whitened i.i.d.\ source (MXFP4 is within
$0.01$ bits of per-tensor FP4); an E4M3 scale acts as a mild compander, worth $0.1$--$0.4$ bits, but costs $0.25$--$0.5$ bits of rate.
CPTC carries a block scale without leaving its $K$ bits per value:
the in-stream $8$-bit scale per $64$ of \cref{app:scale_encoding}
keeps $3.80$ effective bits at exactly $4$ bits, $0.4$ above NVFP4
at $4.5$. Block scaling earns its keep on unrotated weights with
outliers, which the RHT removes, so these are statements about the
i.i.d.\ regime only.

\begin{table}[htbp]
  \centering
  \small
  \caption{Block-scaled scalar formats on a standard normal: element
    type, scale type and group size $G$, total bits per value
    including the scale, and effective bits ($\uparrow$; standard
    protocol, round-to-nearest, bit-exact microscaling library).
    Reference rows: per-tensor scalar formats, and CPTC from \cref{tab:qtip_k} and \cref{app:scale_encoding}.}
  \label{tab:microscaling}
  \begin{tabular}{llcccl}
    \toprule
    Element & Scale & $G$ & bits/value & $K_{\text{eff}}$ & Format \\
    \midrule
    FP4 E2M1 & E8M0 & $32$ & $4.25$ & $3.16$ & MXFP4 \\
    FP4 E2M1 & E8M0 & $16$ & $4.50$ & $3.18$ & \\
    FP4 E2M1 & E4M3 & $32$ & $4.25$ & $3.30$ & \\
    FP4 E2M1 & E4M3 & $16$ & $4.50$ & $3.39$ & NVFP4 \\
    INT4     & E8M0 & $32$ & $4.25$ & $2.86$ & \\
    INT4     & E4M3 & $16$ & $4.50$ & $3.53$ & \\
    FP6 E2M3 & E8M0 & $32$ & $6.25$ & $5.15$ & MXFP6 \\
    FP6 E2M3 & E4M3 & $16$ & $6.50$ & $5.33$ & \\
    \midrule
    FP4 E2M1 & per-tensor & --- & $4.00$ & $3.15$ & \\
    NF4 & per-tensor & --- & $4.00$ & $3.33$ & \\
    Lloyd--Max & per-tensor & --- & $4.00$ & $3.36$ & scalar optimum \\
    CPTC $K{=}4$ & in-stream $8$-bit & $64$ & $4.00$ & $3.80$ & \cref{app:scale_encoding} \\
    CPTC $K{=}4$ & per-tensor & --- & $4.00$ & $3.93$ & \\
    CPTC $K{=}6$ & per-tensor & --- & $6.00$ & $5.80$ & \\
    \bottomrule
  \end{tabular}
\end{table}

\paragraph{Is CPTC's edge just search budget?} The trellis-depth
sweep of \cref{tab:scaling_L} (plotted as \cref{fig:cptc_vs_all})
shows CPTC ahead of our RPTC reimplementation
at every $L$. CPTC's coefficients are the best of a search (exhaustive through $L{=}10$, sampled above), while each RPTC number is a random draw, so we gave RPTC a
search too: the best of $100$ random permutations on a held-out
$2^{20}$-sample selection set, evaluated on the standard
$2^{22}$-sample set. Selection gains far less than the gap to CPTC: best-of-$100$ RPTC reaches $3.49$ at $L{=}8$ (vs.\
$3.45$ single-draw, $3.77$ CPTC), $3.78$ at $L{=}12$ (vs.\ $3.78$,
$3.89$), and $3.90$ at $L{=}16$ (vs.\ $3.90$, $3.93$).
Random-permutation distortion concentrates tightly (worst-of-$100$
to best-of-$100$ spans $0.08$ bits at $L{=}8$ and under $0.02$ from $L{=}12$), so selecting
among random permutations does not close the gap to CPTC. Extending the reimplementation to $L \in \{17, 18\}$
gives the RPTC $L{=}17$/$18$ entries of \cref{tab:scaling_L}
($3.91$ and $3.92$), so CPTC at $L{=}16$ ($3.93$) matches RPTC two
state bits deeper. CPTC's edge at equal depth has no
theorem yet.

\paragraph{The LC baseline and headline-margin precision.} The LC
row of \cref{tab:qtip_k} reports, per rate, the stronger of two
searched one-multiply constructions, both run under the
standard protocol with our tail-biting encoder. \emph{(i) Axiom
SR/LC}, the construction of \citet{eriksson2007lc}: $M_S = 2^{16-K}$
shift-register states, per-branch labels $l_i^{(u)} = (g \cdot i +
a^{(u)}) \bmod M_L$ with $M_L = M_S$, one shared odd generator and
maximally separated offsets $a^{(u)} = \alpha + u M_L/2^K$ (their
Theorems 1--2, extended to $K \ge 3$), best of $150$ random
$(g, \alpha)$ on a held-out seed. Its reproduction values follow their paper:
the source quantiles at the labels times a scale factor chosen on
held-out data (their Eq.~17), then optimized by the generalized Lloyd
algorithm on the trellis (their Section~IV-A), the procedure
\cref{sec:table} uses for CPTC. It supplies the LC row at $K \le 4$
($0.960$/$1.950$/$2.945$/$3.938$). \emph{(ii) Full-codebook computed
permutation} (\cref{prop:reduction}: $a_0 = 0$, $P_a(q) = aq \bmod
2^{16}$, $2^{16}$-entry codebook built the same way, best of $200$ odd
multipliers per rate), a strengthening beyond their paper; it supplies
$K{=}5$ ($4.930$) and $K{=}6$ ($5.922$, against the axiom form's
$5.892$). On the winning forms the scale factor is worth up to
$0.014$ bits and the Lloyd step up to $0.005$ ($0.03$\,dB, in line
with the ${<}0.04$\,dB their paper reports for Gaussians). At CPTC's storage and data type
(the labels reduced to $128$ cells by their top bits and a
mirror-symmetric E2M5 table, $64$ bytes) the axiom construction,
searched at that size (best of $600$ random $(g, \alpha)$), with
the same two steps reaches $0.957$/$1.948$/$2.932$/$3.898$/$4.820$/$5.661$
at $K = 1, \ldots, 6$: ahead of CPTC at $K{=}1$, level at $K{=}2$ and
behind by $0.011$/$0.036$/$0.093$/$0.141$ at $K{=}3$--$6$. With both
tables halved ($64$ labels against CPTC's $63$-entry $Q{=}5$ table,
$32$ bytes each) it reaches $0.957$/$1.944$/$2.914$/$3.848$/$4.705$ at
$K = 1, \ldots, 5$ against CPTC's
$0.948$/$1.948$/$2.942$/$3.928$/$4.868$: the gap at $K{=}3$--$5$
roughly doubles, to $0.027$/$0.080$/$0.163$ (at $K{=}6$ neither $63$
nor $64$ values can beat the $64$-level scalar quantizer,
\cref{prop:alphabet_floor}). Across
depth at $K{=}4$ it is the LC row of \cref{tab:scaling_L}; below $L{=}11$ it has fewer than $128$
labels, one per state, and its $L{=}16$ cell is the $K{=}4$
value above. Their Definition~3 read literally, with labels
taken modulo $128$, depends on only $K{+}7$ state bits and scores
lower ($3.861$ at $K{=}4$).

For the headline
margins we also report $10$-seed precision: CPTC $L{=}16$, $K{=}4$
scores $3.9334 \pm 0.0005$ and RPTC $L{=}18$, $K{=}4$ scores
$3.9228 \pm 0.0004$. A random permutation beats CPTC's $Q{=}6$ table in two cells of
\cref{tab:qtip_k}: $K{=}1$ ($0.956$ against CPTC's
$0.9497 \pm 0.0004$), where its reproduction law
$\mathcal{N}(0, 1 - 4^{-K})$ is narrowest, and $K{=}6$
($5.840$ vs.\ $5.802$).

\paragraph{Lattice VQ baselines (\cref{tab:hw_criteria}).}
\label{app:lattice_baselines}
The E8P and nested-E8 entries are our runs of the authors'
released code on the protocol of \cref{tab:qtip_k} ($2^{22}$
samples, three seeds). QuIP\#'s E8P is its $2$-bit codebook
($256$ absolute-value entries with sign flips and a $\pm\tfrac14$
shift, $2^{16}$ points in $8$ dimensions); its $4$-bit
codebook quantizes the residual with a second, rescaled E8P
stage, and its $3$-bit codebook with a $1$-bit E8 stage. We sweep the input and residual scales around the released
values: $K{=}2$ gives $1.727$ at the released scale ($0.0913$
MSE, against the $0.089$ \citet{tseng2024qtip} report for the
same codebook), $K{=}3$ gives $2.541$, and $K{=}4$ gives $3.444$ ($3.351$ at
the released scales). NestQuant's nested Gosset code maps each
$8$-vector to the E8 closest point, reduces modulo $m$ ($m{=}16$
at $4$ bits) and decodes with a second closest-point search; its
encoder picks one of $k$ scales per block, a side channel
of $\log_2 k/8$ bits per value, under the shared pseudorandom
dither of the released code. With a single scale (exactly $4$ bits, scale swept,
overload rate $10^{-4}$) it reaches $3.03$ effective bits, below
the Lloyd--Max scalar ceiling, because one scale's Voronoi shaping
region must enclose the Gaussian tails; with $k{=}2$ and $k{=}4$
scales chosen by exhaustive subset search it reaches $3.60$ at
$4.125$ bits and $3.89$ at $4.25$ bits, deficits of $0.53$ and
$0.36$ bits against its own rate. At $2$ bits ($m{=}4$) the same
runs give $1.09$, $1.54$ ($2.125$ bits) and $1.77$ ($2.25$ bits),
against $1.727$ for E8P and $1.948$ for CPTC at exactly $2$ bits.
The Leech-lattice codebook of \citet{vanderouderaa2026llvq} has
no released implementation; its paper reports $0.078$ MSE at
$2$ bits ($1.84$ effective bits) and no $4$-bit figure.

\subsection{Generalized-Gaussian targets: normalization and extended results}
\label{app:gg}

This appendix supports the beyond-the-Gaussian experiment of
\cref{sec:exp_iid} (\cref{tab:gg_k}): the entropy-power
normalization and its verified tightness, the $K{=}4$
method comparison of \cref{tab:gg}, the $64$-byte LC comparison
(\cref{tab:gg_lc}) and INT8 tables (\cref{tab:gg_int8}). For each
$\beta$ the generation of \cref{sec:table} starts from
\cref{eq:table_rule} (empirical index intervals, per-cell
conditional mean under $p_\beta$ in closed form through incomplete
gamma functions) and its result is fitted to E2M5 at a per-$\beta$
scale chosen on one held-out seed, which keeps it
within $0.01$ bits of the FP16 one at $K{=}4$ for every $\beta$, with the
same $(a_0, a_1)$ and encoder throughout.

\paragraph{Normalization and its tightness.} For the unit-variance
generalized Gaussian the differential entropy has the closed form
$h(X) = \beta^{-1} + \ln\!\big(2 \alpha \Gamma(1/\beta) / \beta\big)$
with $\alpha = \bigl(\Gamma(1/\beta)/\Gamma(3/\beta)\bigr)^{1/2}$
\citep{nadarajah2005gg}, which is what \cref{eq:keff}'s $N(X)$ uses;
as $\beta \to \infty$ it gives the uniform's $N(X) = 12/(2\pi e)
\approx 0.703$, the classical quarter-bit advantage of the uniform
source over the Gaussian at equal variance. For $\beta \ne 2$ the
Shannon ceiling is not exactly achievable at finite rate, but
nearly so at $K \ge 3$: bisecting \cref{eq:keff}'s upper end for the
distortion it certifies achievable (with $N(X + Z_D)$ by FFT
convolution) places the true limit at $K{=}4$ above $3.995$
effective bits across $\beta \in [1, 2]$ (above $2.98$ at $K{=}3$)
and above $3.95$ for the uniform, whose discontinuous density slows
the bound's convergence. For $\beta \in [1, 2]$ the gaps in \cref{tab:gg} are therefore
genuine coding and table loss, not slack in the ceiling.

\begin{table}[t]
  \centering
  \small
  \caption{Effective bits ($\uparrow$) at $K{=}4$, $L{=}16$ on
    generalized-Gaussian targets, entropy-power normalized so
    $4.00$ is an upper bound in every row ($T{=}256$, $3$ seeds, std below
    $0.01$). ``Uniform fill'' is the CPTC table of \cref{eq:table_rule}
    filled as if the index were uniform, with no table generation;
    ``LC, $64$\,B'' is the one-multiply code of \cref{tab:gg_lc}; RPTC
    uses the quantiles of each target's optimal reproduction law
    \citep{mao2011rptc}, with no Lloyd step.}
  \label{tab:gg}
    \begin{tabular}{ccccccc}
    \toprule
    $\beta$ & Lloyd--Max &
    Uniform fill & RPTC & LC, $64$\,B & CPTC (E2M5) & CPTC (FP16) \\
    \midrule
    $\infty$ (uniform) & $3.75$ & $3.53$ & $3.93$ & $3.95$ & $3.95$ & $3.95$ \\
    $2.0$ (normal)    & $3.36$ & $3.25$ & $3.90$ & $3.90$ & $3.93$ & $3.93$ \\
    $1.5$             & $3.21$ & $3.11$ & $3.87$ & $3.86$ & $3.92$ & $3.92$ \\
    $1.0$ (Laplacian) & $2.91$ & $2.78$ & $3.83$ & $3.77$ & $3.88$ & $3.88$ \\
    \bottomrule
  \end{tabular}
  
\end{table}

\paragraph{One multiplication at equal storage.} \cref{tab:gg_lc}
repeats the $64$-byte LC comparison of \cref{app:iid_extended} on
these targets. For each target and rate the LC code is built as its
paper prescribes: a generator and offset searched on the target
(best of $600$), the source quantiles at the labels times a scale
factor chosen on held-out data, the Lloyd step on the trellis, and a
mirror-symmetric E2M5 table at the best of four storage scales. CPTC
keeps one coefficient pair per rate for every target. On the uniform
target the one-multiply code is level or ahead at every rate; on the
other three CPTC's lead grows with tail weight at every rate from $2$
to $5$ and, except on the Laplacian at $K{=}6$, with the rate. The
$\beta{=}2$ row is a run independent of \cref{app:iid_extended} and
agrees with it within $0.005$ bits; CPTC's values are those of the run
behind \cref{tab:gg_k}.

\begin{table}[htbp]
  \centering
  \small
  \caption{Effective bits ($\uparrow$) of the one-multiply LC code with
    a $64$-byte E2M5 table on generalized-Gaussian targets, with
    CPTC's lead over it in parentheses (negative: LC ahead).
    $L{=}16$, $T{=}256$, $3$ seeds.}
  \label{tab:gg_lc}
  \begin{tabular}{ccccccc}
    \toprule
    $\beta$ & $K{=}1$ & $K{=}2$ & $K{=}3$ & $K{=}4$ & $K{=}5$ & $K{=}6$ \\
    \midrule
    $\infty$ & $0.845$ ($-0.007$) & $1.910$ ($-0.001$) & $2.939$ ($-0.002$)
          & $3.953$ ($-0.004$) & $4.956$ ($-0.006$) & $5.934$ ($-0.043$) \\
    $2.0$ & $0.956$ ($-0.009$) & $1.947$ ($0.000$) & $2.931$ ($0.011$)
          & $3.896$ ($0.037$) & $4.816$ ($0.096$) & $5.660$ ($0.140$) \\
    $1.5$ & $0.945$ ($-0.004$) & $1.932$ ($0.005$) & $2.908$ ($0.022$)
          & $3.858$ ($0.060$) & $4.749$ ($0.137$) & $5.548$ ($0.194$) \\
    $1.0$ & $0.916$ ($0.010$) & $1.894$ ($0.019$) & $2.854$ ($0.049$)
          & $3.767$ ($0.112$) & $4.590$ ($0.203$) & $5.313$ ($0.164$) \\
    \bottomrule
  \end{tabular}
\end{table}

\paragraph{INT8 tables.} \cref{tab:gg_int8} repeats the sweep of
\cref{tab:gg_k} with INT8 tables, fitted like the E2M5 ones at a
per-$\beta$ scale chosen on one held-out seed. INT8 matches E2M5 within
$0.003$ bits up to $K{=}4$ on every target. At $K{=}6$ it costs
$0.03$ bits on the uniform and the Gaussian and $0.13$ on the
Laplacian, where E2M5's finer steps near zero help.

\begin{table}[htbp]
  \centering
  \small
  \caption{Effective bits ($\uparrow$) of CPTC with an INT8 table on
    generalized-Gaussian targets, with its deficit against the E2M5
    table in parentheses. $L{=}16$, $T{=}256$, $3$ seeds.}
  \label{tab:gg_int8}
  \begin{tabular}{ccccccc}
    \toprule
    $\beta$ & $K{=}1$ & $K{=}2$ & $K{=}3$ & $K{=}4$ & $K{=}5$ & $K{=}6$ \\
    \midrule
    $\infty$ & $0.838$ ($0.000$) & $1.909$ ($0.000$) & $2.937$ ($0.000$)
          & $3.948$ ($0.001$) & $4.949$ ($0.001$) & $5.857$ ($0.034$) \\
    $2.0$ & $0.947$ ($0.000$) & $1.947$ ($0.000$) & $2.942$ ($0.000$)
          & $3.933$ ($0.000$) & $4.910$ ($0.001$) & $5.767$ ($0.032$) \\
    $1.5$ & $0.941$ ($0.000$) & $1.937$ ($0.000$) & $2.930$ ($0.000$)
          & $3.918$ ($0.000$) & $4.882$ ($0.004$) & $5.736$ ($0.006$) \\
    $1.0$ & $0.926$ ($-0.001$) & $1.913$ ($0.000$) & $2.903$ ($0.000$)
          & $3.876$ ($0.003$) & $4.782$ ($0.011$) & $5.346$ ($0.131$) \\
    \bottomrule
  \end{tabular}
\end{table}

\subsection{Table size and data type}
\label{app:dtype}

This appendix supports the table-data-type knob of \cref{sec:exp_iid}
(\cref{tab:min_codebook}) and proves the alphabet floor used there.
The $8$-bit table entries are themselves quantized codepoints, so
the table data type sets a second floor on quality: rounding the
table entries to the nearest representable value perturbs each by up to half an output unit in
the last place. \cref{tab:dtype} sweeps table data types at
$L{=}16$ across $K$ under the identical encoder and protocol
($T{=}256$, $2^{22}$ samples, $3$ seeds; seed std $< 10^{-3}$
bits): signed integers INT3--INT8, the mini-floats FP4 (E2M1, $15$
values), FP6 (E2M3, $63$ values) and E4M3 ($253$ values), the
deployed type E2M5 ($253$ values), and the codepoints kept in BF16
or FP16 as references. Each type gets a single storage scale for
all $K$, chosen
deterministically from the maximum entry of the $K{=}4$ starting
table of \cref{eq:table_rule}: the largest
$3$-mantissa-bit float such that after rounding no codepoint of that
table saturates. The generated table of each rate is then fitted to
the type at that scale (\cref{app:constants}). For E2M5 the rule gives a
scale of $2$: the outermost entry of the starting table sits
at $3.74\sigma$ (the conditional mean of the triangle's edge cell;
its quantile midpoint in \cref{sec:theory_triangular} is
$3.67\sigma$), and $3.74 \times 2$ stays inside the type's span of
$7.75$, which the next scale would exceed. For INT6 it
gives $8$, and its table scores $3.926$ at $K{=}4$ against E2M5's
$3.934$.

Three observations. First, \emph{the rule of thumb is
$b \approx K + 2$ integer bits}: INT4 saturates beyond $K{=}2$,
INT5 beyond $K{=}3$, INT6 beyond $K{=}4$. Second, \emph{level
placement sets the plateau height}: with the same $15$ levels FP4
plateaus at $3.12$ against INT4's $2.79$, FP6 is ahead of INT6 by
$0.05$ at $K{=}5$ and $0.34$ at $K{=}6$, and E2M5, whose $253$ values
keep all $127$ codepoints distinct for $K \ge 3$, stays within $0.07$
bits of FP16 everywhere, while E4M3's coarse mantissa caps it at
$5.19$ at $K{=}6$. Third, wherever a type
saturates, the CPTC rows flatten at the type's own scalar quality
(the ``at $c$'' column): the coder degenerates to
the format it stores. The design
guidance: E2M5 is the right default for the full $K \in [1, 6]$
range; a platform without it can default to INT8, which matches E2M5
up to $K{=}5$ and costs $0.03$ bits at $K{=}6$ (other targets in
\cref{tab:gg_int8}); INT6 or FP6 are valid
cost reductions for $K \le 4$, where they give up at most $0.01$ bits. (This table's E2M5 row uses the
deterministic no-clipping scale; \cref{tab:gg_k} uses a
per-$\beta$ calibrated scale for $\beta \ne 2$, and its
$\beta{=}2$ row repeats \cref{tab:qtip_k}.)

\begin{table}[htbp]
  \centering
  \small
  \caption{Effective bits ($\uparrow$) by table data type at $L{=}16$
    ($T{=}256$, $2^{22}$ samples, $3$ seeds). Each unit-variance
    codepoint $v$ is stored as $\mathrm{round}(c\,v)$ in the type's
    grid, with the fixed scale $c$ of the Scale column, and
    decoded as the stored value$/c$. FP4 is E2M1 ($15$ values),
    FP6 is E2M3 ($63$); E4M3 and E2M5 have $253$ values each
    (all-ones reserved for NaN). The last two columns use the
    type's grid as a plain scalar quantizer, at the same scale and
    at the MSE-optimal scale with clipping; the $16$-bit rows are
    native round-to-nearest.}
  \label{tab:dtype}
  \begin{tabular}{lccccccccc}
    \toprule
    & & \multicolumn{6}{c}{CPTC ($L{=}16$)} &
        \multicolumn{2}{c}{Scalar (type alone)} \\
    \cmidrule(lr){3-8} \cmidrule(lr){9-10}
    Table type & Scale & $K{=}1$ & $K{=}2$ & $K{=}3$ & $K{=}4$ &
    $K{=}5$ & $K{=}6$ & at $c$ & opt.\ $c$ \\
    \midrule
    INT3 & $0.875$ & $0.95$ & $1.60$ & $1.60$ & $1.60$ & $1.60$ & $1.60$ & $1.60$ & $2.21$ \\
    INT4 & $2$     & $0.96$ & $1.94$ & $2.77$ & $2.79$ & $2.79$ & $2.79$ & $2.79$ & $3.14$ \\
    FP4  & $1.75$  & $0.95$ & $1.94$ & $2.84$ & $3.12$ & $3.12$ & $3.12$ & $3.12$ & $3.15$ \\
    INT5 & $4$     & $0.95$ & $1.95$ & $2.94$ & $3.76$ & $3.79$ & $3.79$ & $3.79$ & $4.04$ \\
    INT6 & $8$     & $0.95$ & $1.95$ & $2.94$ & $3.93$ & $4.75$ & $4.78$ & $4.79$ & $4.93$ \\
    FP6  & $2$     & $0.95$ & $1.95$ & $2.94$ & $3.92$ & $4.80$ & $5.12$ & $5.12$ & $5.12$ \\
    E4M3 & $120$   & $0.95$ & $1.95$ & $2.94$ & $3.93$ & $4.81$ & $5.19$ & $5.22$ & $5.24$ \\
    INT7 & $16$    & $0.95$ & $1.95$ & $2.94$ & $3.93$ & $4.90$ & $5.69$ & $5.77$ & $5.83$ \\
    INT8 & $32$    & $0.95$ & $1.95$ & $2.94$ & $3.93$ & $4.91$ & $5.77$ & $6.73$ & $6.73$ \\
    E2M5 & $2$     & $0.95$ & $1.95$ & $2.94$ & $3.93$ & $4.91$ & $5.80$ & $6.99$ & $7.08$ \\
    BF16 & $1$ & $0.95$ & $1.95$ & $2.94$ & $3.93$ & $4.92$ & $5.86$ & $9.23$ & $9.23$ \\
    FP16 & $1$ & $0.95$ & $1.95$ & $2.94$ & $3.93$ & $4.92$ & $5.87$ & $12.23$ & $12.23$ \\
    \bottomrule
  \end{tabular}
  
\end{table}

\paragraph{A minimal codebook.} The sweep above holds the
table-size parameter at $Q{=}6$ and varies only the storage type;
the two knobs can be driven down together. Setting $Q = K{+}1$
gives a codebook of $2^{K+2}{-}1$ codepoints, and INT($K{+}2$)
storage offers exactly $2^{K+2}{-}1$ symmetric levels, one per
codepoint, so both the entry count and the entry width are at
their matched minimum. Two bits further down on either knob is a
provable floor, formalizing the table-sizing argument of
\cref{sec:table}:
\begin{proposition}[Alphabet floor]
\label{prop:alphabet_floor}
If the decoded-state codebook takes at most $M$ distinct values,
then for every trellis depth $L$, block size $T$, and encoder,
the per-sample MSE is at least that of the $M$-point Lloyd--Max
scalar quantizer.
\end{proposition}
The proof is two lines: every decoded value lies in the codebook,
so each sample's squared error is at least its squared distance to
the nearest codepoint, and the $M$-point Lloyd--Max codebook
minimizes the expectation of exactly that quantity. At
$Q = K{-}1$ the codebook has $2^{K}{-}1$ distinct
codepoints, and a symmetric INT($K$) table has $2^{K}{-}1$
representable levels whatever $Q$ is, in either case one value
short of the $2^K$ a rate-$K$ scalar quantizer uses, so by the
proposition the decoder is strictly worse than the scalar optimum
no matter how deep the trellis. Any trellis gain therefore
requires $Q \ge K$ and strictly more than $K$ storage bits, and
the matched configuration sits two doublings above each floor;
\cref{tab:min_codebook} shows those two bits buy back essentially
everything. By \cref{prop:dominance}, every row
with $K \le 5$ runs bit-exactly on the fixed $(L_{\max}{=}16,
Q{=}6)$ datapath via the shift $s = 5 - K$; the $K{=}6$ row has
$Q{=}7$ and needs the wider table of CPTC$_{Q7}$.

\begin{table}[htbp]
  \centering
  \small
  \caption{The smallest matched codebook, $Q = K{+}1$ stored in
    INT($K{+}2$), and its effective bits ($\uparrow$; protocol of
    \cref{tab:qtip_k}): ``Scale'' is the no-clipping rule of \cref{app:dtype} applied to each row's own starting table, ``Table bits'' the stored half ($2^{K+1}$
    entries of $K{+}2$ bits), and the last two columns the
    same-size and the $Q{=}6$ codebooks in FP16.}
  \label{tab:min_codebook}
  \begin{tabular}{ccccccccc}
    \toprule
    & & & & & Table & \multicolumn{3}{c}{$K_{\text{eff}}$} \\
    \cmidrule(lr){7-9}
    $K$ & $Q$ & Codepoints & Type & Scale & bits &
    INT($K{+}2$) & same $Q$, FP16 & $Q{=}6$, FP16 \\
    \midrule
    1 & 2 & $7$   & INT3 & $1.75$ & $12$  & $0.95$ & $0.95$ & $0.95$ \\
    2 & 3 & $15$  & INT4 & $2.75$ & $32$  & $1.93$ & $1.94$ & $1.95$ \\
    3 & 4 & $31$  & INT5 & $5$    & $80$  & $2.92$ & $2.94$ & $2.94$ \\
    4 & 5 & $63$  & INT6 & $9$    & $192$ & $3.91$ & $3.93$ & $3.93$ \\
    5 & 6 & $127$ & INT7 & $16$   & $448$ & $4.90$ & $4.92$ & $4.92$ \\
    6 & 7 & $255$ & INT8 & $30$   & $1{,}024$ & $5.87$ & $5.90$ & $5.87$ \\
    \bottomrule
  \end{tabular}
  
\end{table}

\subsection{Matmul-aware experiment: methods and per-\texorpdfstring{$\rho$}{rho} results}
\label{app:matmul_calibration}

This appendix gives the table design, baselines, scale calibration and per-$\rho$
numbers behind \cref{sec:exp_matmul_synth}.

\paragraph{Table design.} The table contents encode the
\emph{reproduction distribution}, not the decode path.
\citet{ang2026matmul} show that for a multiplicative pair with
operand correlation $\rho$, the asymptotically optimal scalar
quantizer has point density
\begin{equation}
  \lambda^{\star}(u) \;\propto\;
    \exp(-u^{2}/6) \cdot
    \big( (1{-}\rho^{2}) + \rho^{2} u^{2} \big)^{1/3},
  \quad u = x/\sigma,
  \label{eq:ang_density}
\end{equation}
exhibiting a unimodal-to-bimodal phase transition at
$|\rho| = 1/\sqrt{3}$. We move each entry of the generated Gaussian
table $\mathcal{T}$ of \cref{sec:table} to the same quantile of the
normalized $\lambda^{\star}$,
\begin{equation}
  \mathcal{T}^{\star}[i] = (F^{\star})^{-1}\big(\Phi(\mathcal{T}[i] / \sigma_{\mathcal{T}})\big),
  \label{eq:matmul_table}
\end{equation}
where $F^{\star}$ is the CDF of the normalized $\lambda^{\star}$,
$\Phi$ the standard normal CDF and $\sigma_{\mathcal{T}}$ the standard
deviation of the decoded values.

\paragraph{Methods.} Scalar baselines INT4, FP4,
and NF4 use $\max\lvert c\rvert{=}1$ codebooks with a $20$-point
grid-searched per-matrix scale. \textsc{MatMul-Opt}
\citep{ang2026matmul} uses a $16$-level scalar codebook at
$\lambda^{\star}$-quantile midpoints. \textsc{CPTC-Gauss} is CPTC
with the Gaussian-target table; \textsc{CPTC-MMaware} uses
$\lambda^{\star}$ at the data-generating $\rho$
(\cref{eq:matmul_table}). For matmul-aware methods the codebook is
in u-space (variance $\approx 3$ at $\rho{=}0$, $\approx 4.6$ at
$\rho{=}0.9$); the shared scale calibration is described
below.

\paragraph{Per-$\rho$ results.} \cref{tab:matmul_phase} gives the
numbers plotted in \cref{fig:matmul_phase}. At $\rho = 0$, where
$\lambda^{\star}$ degenerates to a Gaussian wider than the source,
\textsc{CPTC-MMaware} ties \textsc{CPTC-Gauss} after shared-scale
calibration; the matmul-aware codebook pays off from $\rho = 0.3$
onward, and \textsc{CPTC-Gauss} plateaus around $0.017$--$0.019$ for
$|\rho| \ge 0.5$.

\begin{table}[htbp]
  \centering
  \small
  \caption{Relative Frobenius error ($\downarrow$) of the quantized
    product against the exact one at $K{=}4$ ($m{=}n{=}128$,
    $k{=}256$, mean of $10$ trials, SEM $\le 0.0021$). Bold: best
    per column.}
  \label{tab:matmul_phase}
  \begin{tabular}{lccccccc}
    \toprule
    & \multicolumn{7}{c}{Operand-pair correlation $|\rho|$} \\
    \cmidrule(lr){2-8}
    Method & $0.0$ & $0.3$ & $0.5$ & $0.6$ & $0.7$ & $0.8$ & $0.9$ \\
    \midrule
    INT4              & $0.162$ & $0.044$ & $0.038$ & $0.031$ & $0.032$ & $0.027$ & $0.025$ \\
    FP4               & $0.159$ & $0.043$ & $0.034$ & $0.033$ & $0.031$ & $0.030$ & $0.028$ \\
    NF4               & $0.139$ & $0.036$ & $0.029$ & $0.026$ & $0.026$ & $0.023$ & $0.022$ \\
    \textsc{MatMul-Opt} \citep{ang2026matmul}
                      & $0.139$ & $0.033$ & $0.022$ & $0.019$ & $0.017$ & $0.014$ & $0.012$ \\
    \midrule
    \textsc{CPTC-Gauss}     & $\mathbf{0.093}$ & $0.024$ & $0.019$ & $0.017$ & $0.018$ & $0.018$ & $0.018$ \\
    \textsc{CPTC-MMaware}   & $\mathbf{0.093}$ & $\mathbf{0.022}$ & $\mathbf{0.016}$ & $\mathbf{0.014}$ & $\mathbf{0.013}$ & $\mathbf{0.011}$ & $\mathbf{0.010}$ \\
    \bottomrule
  \end{tabular}
  
\end{table}

\paragraph{Calibration objective.} Per-entry MSE is the wrong
calibration objective for the matmul-aware codebook, which
deliberately spreads codepoints into the source tails; it selects
too tight a scale. By the symmetry of the data-generating model the
optimal scales satisfy $c_A^{\star} = c_B^{\star}$, so we calibrate
one shared scale per $\rho$ by minimizing matmul Frobenius error
over five held-out matrix pairs on a $19$-point grid over
$[0.4, 1.3] \cdot \hat\sigma$.

\paragraph{Comparison with \citet{ang2026matmul}.}
\citet{ang2026matmul} present a similar plot (their Figure 2). Our
\textsc{MatMul-Opt}
curve reproduces their result at $K{=}4$: it tracks NF4 at
$\rho{=}0$ and pulls ahead as $|\rho|$ grows, with
$1.86\times$ lower error than NF4 by $\rho{=}0.9$;
for \textsc{CPTC-MMaware} the trellis gain persists on top
($1.5\times$ at $\rho{=}0$, $1.2\times$ at $\rho{=}0.9$).

\subsection{LLM weight-only quantization details}
\label{app:llm}

Every LLM result in this section uses
the YAQA Sketch-B pipeline of \citet{tseng2025yaqa} without
fine-tuning: all quantized rows share its RHT preprocessing,
two-sided Kronecker-factored Hessians, per-block scales, and the
pipeline's published loading factor (\texttt{scale\_override}
$0.9$, from the authors' released recipe), with only the
trellis codebook swapped, so the comparison isolates the format
from the rounding algorithm. The Llama 3.1 Instruct results use
the authors' released Hessians; the Qwen3.8-27B results at the
end of this section use Hessians we collected ourselves, since
none are published for that model. All of them use the authors'
evaluation protocol (GPTQ Wikitext-2 and C4 splits, sequence
length $8{,}192$). $D_{\mathrm{KL}}$ is the divergence to the
unquantized model's next-token distribution, the metric YAQA
optimizes and the most sensitive of the three.

\paragraph{Harness and protocol validation.} Our reproduction of
the YAQA pipeline matches its authors' published numbers wherever the same
artifacts can be evaluated, at both scales.
Our 8B BF16 baselines match theirs to displayed precision
($6.502/8.022$ vs.\ $6.50/8.02$ at 8B), and running the QTIP quantizer
ourselves under their recipe (seed $0$ of the QTIP HYB rows of
\cref{tab:llm_main}) reproduces the corresponding row of YAQA's Table~1 to all displayed
digits at $K{=}4$ ($0.0130$, $6.608/8.118$ vs.\ $0.013$,
$6.61/8.12$), with $K{=}3/2$ regenerations at $0.0452/0.2403$
vs.\ their $0.044/0.241$ and the 70B regenerations at
$0.0298/0.0967/0.3281$ vs.\ their $0.030/0.094/0.335$ at
$K{=}4/3/2$. Two controls localize everything to the codebook: re-collecting the
full $65{,}536$-sequence Sketch-B Hessians from scratch reproduces our
results to three decimals, and six independent draws of HYB's
$k$-means table span $0.006$ effective bits. The loading factor
matters: at unit scale the QTIP quantizer degrades to $0.0172$ at
$K{=}4$, so we adopt the authors' released $0.9$; it was tuned by
them for HYB, and CPTC is nearly insensitive to it ($0.00967$ at
unit loading vs.\ $0.00955$ at $0.9$, seed $0$, \cref{tab:llm_rtn}), so the
shared setting does not favor CPTC. The factor was not varied for 1MAD. Every KL and perplexity cell of \cref{tab:llm_main} but 70B HYB is the mean of three runs that differ only in the Random Hadamard
Transform's sign vectors: the pipeline is deterministic given the
Hessians and this seed (re-collecting the Hessians moved results
by under $0.001$ KL), so we exposed the per-layer RHT seed and
reran seeds $1$ and $2$ (seed $0$ is the original run;
\cref{tab:llm_seeds}). The spread across seeds is $0.3$--$4.1\%$
of the value, comparable across codebooks, and both CPTC arms have a
lower KL than 1MAD on every individual seed at every rate and
scale (the closest call is CPTC$_{Q5}$ at 70B $K{=}3$, seed $1$: $0.0871$ vs.\
$0.0872$), so the ordering of \cref{tab:llm_main} does not rest
on a seed. HYB has three seeds at 8B and one at 70B because its runs,
on QTIP's own encoder, take about six times longer.

\begin{table}[t]
  \centering
  \small
  \caption{KL to BF16 ($\downarrow$) on Wikitext-2 for each RHT seed
    behind the means of \cref{tab:llm_main}: seed $0$ is the
    original run, seeds $1$ and $2$ change only the RHT sign
    vectors; range is $(\max - \min)/\text{mean}$. At every rate
    and scale both CPTC arms are below 1MAD on each seed; HYB was
    rerun at 8B only.}
  \label{tab:llm_seeds}
  \begin{tabular}{llcccccc}
    \toprule
    Model & Codebook & $K$ & seed $0$ & seed $1$ & seed $2$ & mean & range \\
    \midrule
    8B & QTIP HYB & 4 & $0.01301$ & $0.01294$ & $0.01324$ & $0.01306$ & $2.3\%$ \\
    8B & QTIP 1MAD & 4 & $0.01019$ & $0.00978$ & $0.01004$ & $0.01000$ & $4.1\%$ \\
    8B & CPTC$_{Q5}$ & 4 & $0.00975$ & $0.00958$ & $0.00951$ & $0.00961$ & $2.5\%$ \\
    8B & CPTC & 4 & $0.00955$ & $0.00947$ & $0.00950$ & $0.00951$ & $0.8\%$ \\
    8B & QTIP HYB & 3 & $0.04520$ & $0.04509$ & $0.04486$ & $0.04505$ & $0.8\%$ \\
    8B & QTIP 1MAD & 3 & $0.04047$ & $0.04092$ & $0.04010$ & $0.04050$ & $2.0\%$ \\
    8B & CPTC$_{Q5}$ & 3 & $0.03953$ & $0.03996$ & $0.03920$ & $0.03956$ & $1.9\%$ \\
    8B & CPTC & 3 & $0.03822$ & $0.03912$ & $0.03863$ & $0.03866$ & $2.3\%$ \\
    8B & QTIP HYB & 2 & $0.24026$ & $0.24487$ & $0.24209$ & $0.24241$ & $1.9\%$ \\
    8B & QTIP 1MAD & 2 & $0.23204$ & $0.22568$ & $0.22348$ & $0.22707$ & $3.8\%$ \\
    8B & CPTC$_{Q5}$ & 2 & $0.22073$ & $0.22044$ & $0.22252$ & $0.22123$ & $0.9\%$ \\
    8B & CPTC & 2 & $0.21820$ & $0.22043$ & $0.22191$ & $0.22018$ & $1.7\%$ \\
    \midrule
    70B & QTIP 1MAD & 4 & $0.02352$ & $0.02404$ & $0.02387$ & $0.02381$ & $2.2\%$ \\
    70B & CPTC$_{Q5}$ & 4 & $0.02264$ & $0.02255$ & $0.02224$ & $0.02248$ & $1.8\%$ \\
    70B & CPTC & 4 & $0.02265$ & $0.02272$ & $0.02265$ & $0.02267$ & $0.3\%$ \\
    70B & QTIP 1MAD & 3 & $0.08818$ & $0.08719$ & $0.08744$ & $0.08760$ & $1.1\%$ \\
    70B & CPTC$_{Q5}$ & 3 & $0.08449$ & $0.08706$ & $0.08518$ & $0.08558$ & $3.0\%$ \\
    70B & CPTC & 3 & $0.08381$ & $0.08496$ & $0.08481$ & $0.08453$ & $1.4\%$ \\
    70B & QTIP 1MAD & 2 & $0.32468$ & $0.32101$ & $0.32053$ & $0.32207$ & $1.3\%$ \\
    70B & CPTC$_{Q5}$ & 2 & $0.31241$ & $0.31374$ & $0.31727$ & $0.31447$ & $1.5\%$ \\
    70B & CPTC & 2 & $0.31936$ & $0.31884$ & $0.31453$ & $0.31758$ & $1.5\%$ \\
    \bottomrule
  \end{tabular}
\end{table}

A full 8B quantization run takes about six minutes on
one 8-GPU node (fused Viterbi encoder; a banded reformulation of
the rounding solver with identical results), which is what makes
recipe-uniform grids and these controls affordable. We did not run
LC or RPTC in the LLM pipeline;
\cref{tab:qtip_k} remains the locus of that comparison.

\paragraph{Results (\cref{tab:llm_main}).} Beyond the margins of
\cref{sec:exp_llm}: against the published HYB numbers CPTC's
$D_{\mathrm{KL}}$ margin at 8B is $8.6/12.1/26.9\%$ at $K{=}2/3/4$, and
against 1MAD it is $3/5/5\%$ at 8B and
$1/4/5\%$ at 70B (full-table arm; $3/2/4\%$ and $2/2/6\%$ for
CPTC$_{Q5}$). At $K{=}5$ (single runs, seed $0$) quantization
is nearly free and the ordering holds: at 8B CPTC $0.0031$ (perplexity
$6.516$ vs.\ $6.502$), CPTC$_{Q5}$ $0.0032$, 1MAD $0.0032$ and HYB
$0.0075$, whose table saturates at $3.94$ effective bits at $5$ bits
(\cref{thm:ceiling}); at 70B CPTC $0.0068$, CPTC$_{Q5}$ $0.0069$
and 1MAD $0.0070$.
The $Q{=}5$ arm, whose $63$ FP16
codepoints make the GPU kernel's table gather bank-conflict-free
(\cref{app:gpu_gemm}), costs at most $0.001$ $D_{\mathrm{KL}}$ at
8B ($0.0001$ at $K{=}4$), costs $0.001$ at 70B $K{=}3$, and
\emph{beats} the full table at 70B $K \in \{2,4\}$
($0.3145$ vs.\ $0.3176$, $0.0225$ vs.\ $0.0227$). Zero-shot accuracy over ARC-\{c,e\}, BoolQ, HellaSwag,
PIQA sits within $1.1$ points of BF16 ($71.38$ at 8B, $77.09$ at
70B) for every codebook at $K \ge 3$ and drops together at $K{=}2$
($4.0$--$4.7$ points at 8B, $2.4$--$2.8$ at 70B), so it
separates rates but not formats and we do not rank by it. Absolute zero-shot numbers do not transfer across harness versions (our harness
scores the identical BF16 model $1.56$ points higher on average
than their published row), so every zero-shot number we report is
self-measured.

\paragraph{Codebook or rounding? (\cref{tab:llm_rtn}).}
Replacing YAQA's adaptive rounding with round-to-nearest (RTN; both
codebooks at unit loading) isolates the two effects. CPTC's
advantage survives without adaptive rounding ($37\%$ lower
$D_{\mathrm{KL}}$ than HYB under RTN, $44\%$ under YAQA at the
same loading), so it is a property of the codebook, not an
interaction with the rounding algorithm; conversely YAQA
rounding is worth $3.1$--$3.5\times$ for either codebook.

\begin{table}[htbp]
  \centering
  \small
  \caption{The codebook effect with and without adaptive rounding: KL
    to BF16 ($\downarrow$) at $K{=}4$ on 8B, same pipeline at unit
    loading, single runs, with YAQA's adaptive rounding and with
    plain round-to-nearest. The codebook advantage survives
    removing adaptive rounding. Unit loading is HYB's unfavorable
    setting ($0.0130$ at its released $0.9$, seed $0$).}
  \label{tab:llm_rtn}
  \begin{tabular}{lccc}
    \toprule
    Rounding & CPTC & HYB & CPTC advantage \\
    \midrule
    RTN     & $0.03363$ & $0.05334$ & $-37\%$ \\
    YAQA-B  & $\mathbf{0.00967}$ & $0.01720$ & $-44\%$ \\
    \midrule
    YAQA gain & $3.5\times$ & $3.1\times$ & \\
    \bottomrule
  \end{tabular}
\end{table}

\paragraph{KL tracks effective bits (\cref{fig:kl_vs_keff}).}
\label{app:kl_vs_keff}
I.i.d.\ effective bits under the protocol of \cref{tab:qtip_k} predict the
quantized model's quality. For CPTC, QTIP 1MAD and QTIP HYB at
$K \in \{2,3,4\}$, nine cells per model (the CPTC$_{Q5}$ arm lies
within $0.007$ KL of CPTC on seed $0$ and is omitted for
legibility), a quadratic least-squares fit of $\log_2 D_{\mathrm{KL}}$
against $K_{\text{eff}}$ has $r^2 = 0.999$ at 8B and $0.997$ at
70B. Because the codebook points
cluster near the three rates, each point is weighted
by the inverse count of its $0.1$-bit $K_{\text{eff}}$ bin, so the
fit is uniform over the covered range rather than a curve through
three clusters, and the 8B fit also includes the depth-sweep
points below, $44$ distinct cells in all. A slope of $-2$ is what a KL
proportional to weight MSE predicts, since one effective bit is a
$4\times$ factor in distortion (\cref{eq:keff}); across rates 70B follows it
(fitted slope $-2.0$ at $2$ bits to $-1.9$ at $4$),
while at 8B the slope runs from $-3.0$ at $2$ bits through $-2.3$
at $3$ to $-1.7$ at $4$: the KL gain per effective bit shrinks
with rate. Within a single rate the three codebooks give slopes of
$-1.1$/$-2.0$/$-1.3$ at 70B and $-3.8$/$-2.4$/$-1.4$ at 8B for
$K = 2$/$3$/$4$, from three points each. Rank order among them is
exact at both scales: in
every same-rate pair the codebook with more effective bits has
the lower KL ($9/9$ pairs each). The relation does not order the two CPTC arms:
CPTC$_{Q5}$ has the lower KL at 70B for $K \in \{2, 4\}$
(\cref{tab:llm_main}). Nor do the LLM data resolve gaps whose KL effect is within the seed range ($0.3$--$4.1\%$): the LC margins of \cref{tab:qtip_k} for $K \in [2, 4]$ and INT6's cost (at most $0.01$ bits) move KL by about $1\%$. $K_{\text{eff}}$
values are the $L{=}16$ Gaussian measurements; KL values are the
seed-0 runs of \cref{tab:llm_main}.

\paragraph{Trellis-depth sweep.} Varying $L$ at fixed $K$ moves a
single codebook along the same law. We quantized Llama 3.1 8B with
CPTC at every $L$ from $K$ to $16$ for $K \in \{2,3,4\}$ (per-$(L, K)$ coefficients, RHT seed $0$, same recipe), and measured
each configuration's i.i.d.\ effective bits under the protocol of
\cref{tab:qtip_k}. Within each rate, KL falls with depth up to seed noise: at $K{=}4$ it falls from
$0.047$ ($L{=}4$, $2.80$ effective bits) to $0.00955$ ($L{=}16$,
$3.93$); at $K{=}3$ from $0.169$ ($L{=}3$, $2.16$) to $0.038$; at
$K{=}2$ from $0.78$ ($L{=}2$, $1.48$) to $0.218$. The $L{=}K$
cells use the conditional-mean fill of \cref{eq:table_rule} rather than
a Lloyd--Max table, which is why $K{=}4$, $L{=}4$ sits below INT4;
the exact scalar reproduction of \cref{sec:constants} needs the
scalar codebook loaded.
\Cref{fig:kl_vs_keff} plots the $38$ cells with
$K_{\text{eff}} \ge 1.75$ (all but $L \le 5$ at $K{=}2$):
$\log_2 D_{\mathrm{KL}}$ against $K_{\text{eff}}$ has Spearman
$\rho = -0.999$ and a uniformly weighted slope of $-2.4$
($r^2 = 0.99$); per rate the slope over $L$ is $-2.0$ at $K{=}4$,
$-2.7$ at $K{=}3$ and $-3.9$ at $K{=}2$, the steepening described
above. The depth points land on the codebook comparison
of the previous paragraph without any change of fit, which is the
transfer claim in its strongest form: what \cref{tab:qtip_k}
measures on Gaussian samples is what the model sees.

\paragraph{A preliminary result on a different architecture:
  Qwen3.8-27B (\cref{tab:llm_qwen}).} Everything above is
dense-attention Llama. Qwen3.8-27B interleaves $48$ gated-DeltaNet
linear-attention layers with $16$ gated-attention layers in a
$3{:}1$ pattern and carries a vision tower, so it tests whether the
codebook advantage is architecture-specific. No Sketch-B Hessians
are published for it, so we collected the full $65{,}536$-sequence
set ourselves (about $6.5$ days on one 8-GPU node, text only with
the vision tower detached). We quantize $400$ linears (six per
DeltaNet layer, seven per attention layer); the small decay-gate
projections, the short convolutions and all norms stay in BF16, and
the $17{,}408$-wide MLP dimension, which admits no power-of-two
Hadamard, uses an order-$68$ Paley construction. This is a single
run on our own Hessians, and we report it as preliminary. CPTC
lowers $D_{\mathrm{KL}}$ at every rate ($0.0563/0.1060/0.2630$
vs.\ $0.0719/0.1175/0.3063$ at $K{=}4/3/2$) and C4 perplexity at
$K{=}3$ and $K{=}2$, tying at $K{=}4$ ($6.8171$ vs.\ $6.8165$); at
$K{=}5$ (CPTC only; HYB's $512$-pair table caps at $3.94$
effective bits there, \cref{tab:qtip_k}) C4
perplexity is within $0.005$ of BF16 and $D_{\mathrm{KL}}$ halves
again to $0.0295$. As on Llama, zero-shot accuracy does not separate the formats:
every quantized model sits within a point of BF16 ($74.1$), and even
$K{=}2$ costs only $0.7$--$1.0$ points for both codebooks.

\paragraph{Why we do not rank this model by Wikitext-2
  perplexity.} On this split every quantized model scores
\emph{below} BF16, contradicting $D_{\mathrm{KL}}$ and C4. Raw
Wikipedia text is far from this model's post-training data, and the
BF16 model is overconfident on it: softening its logits by a
temperature $\tau$ takes its Wikitext-2 perplexity from $7.685$ at $\tau{=}1$
to a minimum of $7.440$ at $\tau{=}1.15$, which reproduces CPTC's
$K{=}4$ score of $7.447$ with no quantization, while the best
temperature for C4 ($\tau{=}1.05$) lowers its perplexity by only
$0.016$. Four of the quantized scores
are lower still, so temperature does not account for all of the
effect. The same
harness shows no inversion on Llama (\cref{tab:llm_main}), so we
report the column for completeness and rank
by $D_{\mathrm{KL}}$, which is also the argument of \citet{tseng2025yaqa}
for the metric.

\begin{table}[htbp]
  \centering
  \small
  \caption{Qwen3.8-27B (hybrid gated-DeltaNet/attention) in YAQA's
    pipeline with our own $65{,}536$-sequence Hessians, no
    fine-tuning, single run: KL and perplexity ($\downarrow$), Acc
    ($\uparrow$, mean 0-shot accuracy as in \cref{tab:llm_main}).
    CPTC has the lower KL at every rate and the lower or equal C4
    perplexity. $^\dagger$Wikitext-2 perplexity is not informative
    on this model (softening the BF16 model's logits by a
    temperature alone lowers it to CPTC's $K{=}4$ score, see text), so that column is
    unbolded. W2: Wikitext-2.}
  \label{tab:llm_qwen}
  \begin{tabular}{clcccc}
    \toprule
    \multirow{2}{*}{$K$} & \multirow{2}{*}{Codebook} &
    $D_{\mathrm{KL}}$ $\downarrow$ & \multicolumn{2}{c}{Perplexity $\downarrow$} &
    Acc $\uparrow$ \\
    \cmidrule(lr){3-3} \cmidrule(lr){4-5} \cmidrule(lr){6-6}
    & & W2 & W2$^\dagger$ & C4 & 0-shot \\
    \midrule
    16 & BF16 & $0$ & $7.685$ & $6.800$ & $74.1$ \\
    \midrule
    5 & CPTC & $0.0295$ & $7.563$ & $6.805$ & $73.9$ \\
    \midrule
    \multirow{2}{*}{4} & HYB & $0.0719$ & $7.263$ & $\mathbf{6.817}$ & $73.7$ \\
    & CPTC & $\mathbf{0.0563}$ & $7.447$ & $\mathbf{6.817}$ & $74.0$ \\
    \midrule
    \multirow{2}{*}{3} & HYB & $0.1175$ & $7.270$ & $6.879$ & $74.1$ \\
    & CPTC & $\mathbf{0.1060}$ & $7.475$ & $\mathbf{6.870}$ & $73.8$ \\
    \midrule
    \multirow{2}{*}{2} & HYB & $0.3063$ & $6.802$ & $7.222$ & $73.1$ \\
    & CPTC & $\mathbf{0.2630}$ & $7.088$ & $\mathbf{7.192}$ & $73.4$ \\
    \bottomrule
  \end{tabular}
\end{table}

\section{Systems details}
\label{app:part_systems}

\subsection{Fused decode + tensor-core GEMM on stock GPUs}
\label{app:gpu_gemm}

This appendix details the decode-side kernel of
\cref{sec:stock_decode}. It is a separate claim from the accelerator
result of \cref{sec:hardware}: that one is a post-layout datapath; this one is what a GPU with no CPTC support can do today with
general-purpose integer instructions. The two share only the
decode primitive of \cref{alg:decode}.

\paragraph{Structure.} The kernel targets weights times
low-batch activations ($\le 8$ columns), the regime weight-only
quantization is for. Each warp decodes a $16 \times 64$ tile of the weight matrix ($16$ output rows by $64$ positions along the contraction);
each lane reads a $128$-bit window of the code stream and
decodes $32$ values, and \emph{the decoded values land directly
in the \texttt{mma.sync.m16n8k16} operand registers}. This costs
nothing at run time because the map from stream position to
fragment slot is a fixed permutation, applied to the weights
once, before quantization, so no lane shuffles or shared-memory
transposes are needed to reach tensor-core layout. Only one word
per lane crosses a lane boundary (the window straddling the next
lane's data), supplied by a single \texttt{shfl}. The $Q{=}5$ table's $63$ FP16
entries occupy $32$ four-byte words, one per shared-memory bank,
so the per-lane table gather is conflict-free without replication. Accumulation is FP32 throughout, including the
cross-warp reduction and any CTA-level split-K atomics;
only the final store narrows.

\paragraph{The INT4 baseline.} Speedup over cuBLAS FP16 conflates
two effects: moving $4\times$ fewer weight bytes, and paying to
decode them. To separate these we run an INT4 baseline that is the
same kernel with the same layout, staging, MMA and epilogue, whose
decode is a shift and a mask instead of \cref{alg:decode}. It is
verified to be the $L{=}4$ CPTC decoder with an identity table, so
it is bit-exact-comparable, and it is an \emph{upper bound} on any
$4$-bit weight-only kernel using this data path. The gap between
CPTC and this baseline is the cost of the trellis and nothing else.

\paragraph{Results (\cref{tab:gpu_gemm}).} The rows are two decode
schemes over one GEMM pipeline. \emph{Two multiplies} is the full
$L{=}16$ decoder of \cref{alg:decode} with the $Q{=}5$ table.
\emph{State table} indexes a $2^L$-entry
table by the $L{=}10$ state itself; it is
the small-$L$ design point of \cref{sec:encode} ($3.84$ effective
bits at $K{=}4$), not the $L{=}16$ quantizer of \cref{tab:qtip_k}.
\Cref{tab:gpu_gemm} isolates decode cost per weight, the only thing the
schemes differ in. Decode time does not depend on the table
contents. Two GPUs give opposite answers. On the L40S
the two-multiply form and the baseline both sit at $0.99$--$1.00\times$
a measured pure-streaming ceiling at the saturated size: the memory
system binds and the decode hides inside it, the two-multiply form
at $0.995\times$ the baseline's speed ($0.98\times$ across the Llama shapes). On
the H100, with about $4\times$ the bandwidth, the decode ALU becomes
the limiter and the same form falls to $0.48\times$ at the saturated size ($0.62\times$ across the Llama shapes); the state-table
scheme, which spends Hopper's larger shared memory instead of
arithmetic (a $4$\,KiB state table of $2^{10}$ $32$-bit words, replicated $32$-fold into $128$\,KiB so the
warp-wide gather is bank-conflict-free), reduces decode to a
shift, a mask-or and one load and reaches $0.73\times$ ($0.79\times$ across the Llama shapes). That table
exceeds Ada's $99$\,KiB per-CTA cap, hence no L40S state-table row in \cref{tab:gpu_gemm}.
All measurements are cold-L2 (the code stream is rotated over
enough copies to exceed L2), CUDA-event timed, autotuned over CTA
width, warps per tile, split-K and pipeline depth, and checked
against an FP32 reference matmul on the decoded weights (maximum absolute error about $10^{-5}$).

\begin{table}[t]
  \centering
  \small
  \caption{Decode fused into the tensor-core GEMM at batch $1$ ($K{=}4$, $T{=}256$, the $Q{=}5$ table except in the state-table rows; FP16
    in/out, FP32 accumulation). ``stack'' ($\downarrow$): time per
    token for the $32$-layer Llama 3.1 8B linear stack; ``vs.\
    cuBLAS'' ($\uparrow$): its speedup over cuBLAS FP16;
    ``saturated'' ($\downarrow$): a $229{,}376 \times 4{,}096$ matrix
    large enough that cuBLAS itself is bandwidth-bound; ``$\times$ ceil''
    ($\uparrow$): throughput relative to the measured pure-streaming read
    ceiling at that size (above $1$ is possible because a GEMM's
    small activation operand comes from L2). The INT4 baseline
    shares the whole pipeline without any trellis decode, so its
    gap to the CPTC rows is the decode cost. The state-table
    scheme needs $128$\,KiB of shared memory per CTA and does not
    run on Ada. $^\dagger$QTIP's released HYB kernel at $K{=}4$ on
    the seven unfused per-projection shapes, timed as its own
    harness does, kernel time only. $^\ddagger$CPTC on those same
    shapes, kernel-only / with the output-zeroing launch; speedup
    from the kernel-only time. Speedups of the $^\dagger$ and
    $^\ddagger$ rows are against the fused-stack cuBLAS time.}
  \label{tab:gpu_gemm}
  \begin{tabular}{llcccc}
    \toprule
    GPU & Decode & stack (ms) & vs.\ cuBLAS & saturated ($\mu$s) & $\times$ ceil \\
    \midrule
    L40S & cuBLAS FP16      & $19.08$ & ---            & $2{,}595$ & $0.959$ \\
    (Ada)& two multiplies   & $5.24$  & $3.64\times$   & $626$  & $0.993$ \\
         & INT4 baseline    & $5.12$  & $3.73\times$   & $623$  & $0.998$ \\
         & QTIP HYB kernel$^\dagger$ & $5.8$ & $3.3\times$ & --- & --- \\
         & two multiplies, unfused$^\ddagger$ & $5.35$ / $5.84$ & $3.57\times$ & --- & --- \\
    \midrule
    H100 & cuBLAS FP16      & $5.51$  & ---            & $657$  & $1.011$ \\
 (Hopper)& two multiplies   & $3.40$  & $1.62\times$   & $330$  & $0.504$ \\
         & state table      & $2.67$  & $2.06\times$   & $219$  & $0.759$ \\
         & INT4 baseline    & $2.11$  & $2.61\times$   & $159$  & $1.048$ \\
         & QTIP HYB kernel$^\dagger$ & $3.2$ & $1.7\times$ & --- & --- \\
         & two multiplies, unfused$^\ddagger$ & $3.54$ / $4.31$ & $1.56\times$ & --- & --- \\
         & state table, unfused$^\ddagger$ & $2.85$ / $3.65$ & $1.93\times$ & --- & --- \\
    \bottomrule
  \end{tabular}
\end{table}

\paragraph{What binds each scheme.} Profiling attributes the two
regimes to different pipes. The arithmetic schemes are ALU-issue-bound
on Hopper ($78$--$83\%$ of peak ALU; variants spanning
$4.5$--$5.5$ ALU operations per value all measured within a factor
of $1.005$ of each other), and the state-table scheme saturates the
shared-memory path at $85.5\%$ of peak L1TEX with one load per
decoded value; an oracle that halves its lookups for free
bounds the available speedup at $1.26\times$. Neither limit can be removed without changing the decode method (fewer lookups per value for the state table, less ALU work for the arithmetic schemes),
which is the concrete sense in which \cref{sec:hardware}'s decode
unit is the remaining lever on bandwidth-rich parts.

\paragraph{Against QTIP's released kernels.} QTIP ships CUDA decode
kernels for its HYB codebook at $K \in \{2,3,4\}$, fixed to Llama's
per-projection shapes at batch $1$. We built them unmodified, timed
them the way their own test harness does (inside a CUDA graph
with per-launch events, cold-L2), and ran our kernel on the same seven
shapes. At $K{=}4$ their kernel takes
$5.8$\,ms per token for the $32$-layer Llama 3.1 8B stack on the
L40S and $3.2$\,ms on the H100; our two-multiply form takes $5.35$
and $3.54$\,ms kernel-only ($5.84$ and $4.31$\,ms including the
output-zeroing launch its split-K atomics currently require), and
the Hopper state-table scheme $2.85$\,ms ($3.65$\,ms with that launch). The fused-stack rows of
\cref{tab:gpu_gemm} run fewer, larger shapes, hence their
$5.24$ and $3.40$\,ms. On the
bandwidth-bound L40S both are near the memory wall (HYB $5.8$\,ms against CPTC's $5.35$\,ms kernel-only); on the
H100, kernel-only, the portable form runs at $0.90\times$ HYB's speed and the state
table at $1.12\times$ ($0.74\times$ and $0.88\times$ with the zeroing launch). HYB's is the cheaper GPU decode:
one $32$-bit shared load per \emph{two} values with no
arithmetic, which is possible only because its output alphabet is a $512$-pair table ($1{,}024$ pairs with the sign flip), the restriction behind its quality gap in
\cref{tab:qtip_k,tab:llm_main}. CPTC keeps the full alphabet and
pays with the multiplier, which the accelerator of
\cref{sec:hardware} provides natively.

\subsection{CPU decode kernels}
\label{app:cpu_decode}

The decode primitive also vectorizes on general-purpose CPUs. We
implemented it in C++17 for AVX-512 (with the VBMI byte permutes),
for AVX2 and for AArch64 NEON, with a portable scalar fallback, and
verified every path bit-exact against the reference decoder for
$K \in [1,6]$, for the E2M5 and the INT8 table types and for both
table configurations ($Q{=}6$ with the mirror fold, and the $Q{=}5$,
$63$-entry table stored in full), on x86 (Zen~3, Zen~4) and on Arm
(Graviton5, with Arm Neoverse~V3 cores). The AVX-512 kernel decodes $64$ values
per step in about $15$ instructions (a byte permute and a multishift to cut out the
$16$-bit windows, two $16$-bit lane multiplies and shifts per $32$
windows, and one table permute); the $Q{=}5$
form needs no compare, subtract or sign flip for the fold, and
the table permute costs the same for the $31$-, $63$- and folded $127$-codepoint tables, so
a smaller table would not speed the CPU up. \cref{tab:cpu_decode}
gives single-core throughput; the kernel is issue-bound (removing
one multiply in a timing-only experiment gains only $1.15\times$). The
NEON kernel spends $16$ instructions per $16$ values on $128$-bit
vectors; on the Neoverse~V3 its multiplies, dot product and table
lookup share two of the four vector pipes, which caps it at
about $8.7$ billion values/s per core in the GEMM path ($7.6$ measured at $K{=}3$; the standalone rates
of \cref{tab:cpu_decode} also pay a natural-order permute and a
store per value) and puts $K{=}2$ and $K{=}3$ out of reach of the
memory roof on that machine.

\begin{table}[htbp]
  \centering
  \small
  \caption{Single-core decode throughput on CPUs ($\uparrow$, billions of values per second)
    from an L2-resident stream of $2^{17}$ values at $K{=}4$, $T{=}256$, best of $100$--$200$ runs (other $K$ at $0.97$--$1.04\times$
    these rates on x86 and $0.87$--$1.00\times$ on NEON; the AVX-512
    row writes its output in the GEMM's tile order). FP16 output uses the E2M5 table,
    INT8 output an INT8 table. Zen~4: AMD EPYC 9R14 (AVX-512);
    Zen~3: AMD EPYC 7R13 (AVX2); Graviton5: AWS Graviton5, Arm Neoverse~V3 cores (NEON).
    The last rows unpack INT4 nibbles through a $16$-entry table,
    the cheapest $4$-bit decode.}
  \label{tab:cpu_decode}
  \begin{tabular}{llcccc}
    \toprule
    & & \multicolumn{2}{c}{FP16 out} & \multicolumn{2}{c}{INT8 out} \\
    \cmidrule(lr){3-4} \cmidrule(lr){5-6}
    Kernel & CPU & $Q{=}6$ fold & $Q{=}5$ & $Q{=}6$ fold & $Q{=}5$ \\
    \midrule
    AVX-512 & Zen 4 & $15.0$ & $24.4$ & $20.1$ & $25.3$ \\
    AVX2    & Zen 4 & $2.8$  & $3.1$  & $4.1$  & $4.7$ \\
    AVX2    & Zen 3 & $2.4$  & $2.7$  & $3.7$  & $4.1$ \\
    NEON    & Graviton5 & $3.9$ & $5.0$ & $5.1$ & $6.8$ \\
    \midrule
    INT4 unpack (FP16 out), AVX-512 / AVX2 & Zen 4 & \multicolumn{4}{c}{$47$ / $29$} \\
    INT4 unpack (FP16 out), NEON & Graviton5 & \multicolumn{4}{c}{$23$} \\
    \bottomrule
  \end{tabular}
\end{table}

\paragraph{End to end.} \cref{tab:cpu_stack} times the weight
stream of a Llama 3.1 8B forward pass (the seven linear layers of each of
$8$ transformer layers, scaled to the model's $32$; weights streamed from DRAM, activations
resident) on the $64$-core Zen~4 server in one parallel region,
against FP16 weights, INT8 weights with VNNI dot products, and the
INT4 nibble unpack feeding the same VNNI path. CPTC decodes to INT8
codes and uses the same VNNI dot product ($Q{=}5$ table in INT8). The
machine sustains about $270$\,GB/s, which FP16, INT8 and INT4 all
reach: they are bandwidth-bound. The comparison of interest is at
equal bits: CPTC at $K{=}4$ reaches the same bandwidth as INT4 and
takes the same time at batch $1$, so the trellis decode is hidden
under the memory stream. At $K{=}3$ and $K{=}2$ the stream is $0.75\times$
and $0.50\times$ as long and the time falls to $0.83\times$ and
$0.60\times$ the $K{=}4$ time, against $0.75\times$ and
$0.50\times$ for a purely bandwidth-bound format: the decode is
partly exposed at $K{=}3$ ($233$ of the $270$\,GB/s roof) and more
so at $K{=}2$ ($217$\,GB/s). The
INT4 baseline is a pure nibble lookup without scale handling, the
strongest form of it; run-to-run spread is about $5\%$. The lower
block repeats the measurement on one $96$-core NUMA node of a
Graviton5 (Neoverse~V3, NEON with SDOT dot products), whose node
roof is about $405$\,GB/s. At $K{=}4$ the picture is the same:
CPTC reaches $0.97\times$ the INT4 bandwidth and takes the same time
at batch $1$ ($8.9$ vs.\ $8.6$\,ms). At $K{=}3$ and $K{=}2$ the
$128$-bit NEON decoder, capped at about $8.7$ billion values/s per core, cannot
feed the shorter streams at the roof ($11.3$ and $16.9$ billion values/s per core would be needed), so those rates are decode-bound at $275$
and $208$\,GB/s: $K{=}2$ finishes in the INT4 time and $K{=}3$
runs at $0.91\times$ its speed, rather than faster. With the
mirror-folded $Q{=}6$ table, the accelerator's format, the $K{=}4$
stack takes $14.1$\,ms on Zen~4 and $11.1$\,ms on Graviton5,
$0.94\times$ and $0.78\times$ the INT4 speed.

\begin{table}[htbp]
  \centering
  \small
  \caption{Time to stream the weights of a $32$-layer Llama 3.1 8B forward pass (ms, $\downarrow$) at batch $1$ and the bandwidth
    reached, on a $64$-core Zen~4
    server (top) and one $96$-core Graviton5 NUMA node (bottom).
    CPTC decodes an INT8 table ($Q{=}5$) into the INT8 dot
    product (VNNI on x86, SDOT on Arm); INT4 unpacks nibbles into
    the same dot product. Best of $7$ passes over $8$ transformer layers, scaled to $32$.}
  \label{tab:cpu_stack}
  \begin{tabular}{lccc}
    \toprule
    Weights & bits & ms & GB/s \\
    \midrule
    \multicolumn{4}{l}{\emph{Zen~4, $64$ cores (AVX-512, VNNI)}} \\
    FP16 & $16$ & $51.9$ & $269$ \\
    INT8, VNNI & $8$ & $25.8$ & $271$ \\
    INT4 unpack, VNNI & $4$ & $13.3$ & $262$ \\
    \midrule
    CPTC $K{=}6$ & $6$ & $19.9$ & $263$ \\
    CPTC $K{=}4$ & $4$ & $13.6$ & $257$ \\
    CPTC $K{=}3$ & $3$ & $11.3$ & $233$ \\
    CPTC $K{=}2$ & $2$ & $8.1$ & $217$ \\
    \midrule
    \multicolumn{4}{l}{\emph{Graviton5, one $96$-core node (NEON, SDOT)}} \\
    FP16 & $16$ & $36.0$ & $388$ \\
    INT8, SDOT & $8$ & $17.5$ & $399$ \\
    INT4 unpack, SDOT & $4$ & $8.6$ & $405$ \\
    \midrule
    CPTC $K{=}6$ & $6$ & $13.8$ & $379$ \\
    CPTC $K{=}4$ & $4$ & $8.9$ & $391$ \\
    CPTC $K{=}3$ & $3$ & $9.5$ & $275$ \\
    CPTC $K{=}2$ & $2$ & $8.4$ & $208$ \\
    \bottomrule
  \end{tabular}
\end{table}

\subsection{Viterbi encoder: reference implementation and fused kernels}
\label{app:viterbi}
\label{app:viterbi_gpu}

\cref{lst:viterbi} is the single-iteration Viterbi update in
PyTorch notation; the full $T$-step encoder applies this sequentially over the
block's values, and tail-biting wraps that in an outer
pass that fixes the starting state. The rest of this subsection
describes the fused GPU kernel that accelerates this reference by $33$--$134\times$ (\cref{tab:enc_matrix}) and its Trainium~4 counterpart. The decoder-side counterpart, a per-value golden
reference decoder, is the oracle against which the dominance rule of \cref{prop:dominance}
and the hardware kernels are validated bit-exactly.

\begin{lstlisting}[language=Python,caption={Viterbi iteration for
CPTC encoding.},label=lst:viterbi,basicstyle=\small\ttfamily]
# a:              running cost per (L-K)-bit prefix, shape (2**(L-K),)
# decoded_states: codebook, shape (2**L,); W_t: next value; K: bits/value
# The reference walks the block in reverse, so each new code lands in the
# low K bits of its window in the packed stream of Alg. 2.
def viterbi_iteration(a, decoded_states, W_t, K):
    # squared error of every full state s = (prefix << K) | u
    current_cost = (decoded_states - W_t).square()
    # state s extends prefix s >> K, whose running cost a holds
    total = a.unsqueeze(-1).expand(-1, 2**K).reshape(-1) + current_cost
    # next prefix = low L-K bits of s; minimize over the high K bits;
    # a_new and b_t have shape (2**(L-K),)
    a_new, b_t = total.reshape(2**K, -1).min(dim=0)
    return a_new, b_t
\end{lstlisting}

\paragraph{What the kernel computes.} For each block the encoder
solves the tail-biting minimum-distortion path problem of
\cref{sec:trellis}: over all state sequences that follow the
bitshift transition and close into a cycle, it minimizes
$\sum_t (w_t - \phi(q_t))^2$ with $\phi$ the decode map of
\cref{alg:decode}, and emits the per-value codes $c_t$. Launched as
framework-level tensor operations (\cref{lst:viterbi} under
\texttt{torch.compile}) the recursion is latency-bound: each of the
$T$ trellis steps and $T$ backtrace steps is a separate kernel
launch. We fuse the whole
pipeline (forward pass, $\arg\min$, half backtrace to the
tail-biting midpoint state, second forward pass, full backtrace)
into one persistent Triton launch that keeps the running costs
in registers across all $T$ steps. Four optimizations
matter: the per-step minimum packs the FP16 path cost and the
predecessor index into one $32$-bit word, so one unsigned
minimum returns both; for $K \le 4$
the $K$-bit predecessor indices pack $\lfloor 8/K \rfloor$ per byte; the
first $T/2$ steps of the first forward pass skip the backpointer store,
which the half backtrace never reads; and
at small $2^{L-K}$ the kernel packs several independent blocks per
program.

\cref{tab:enc_matrix} reports the full $(L, K)$ throughput matrix
on an NVIDIA L40S at $T{=}128$: $33$--$134\times$ faster than the
\texttt{torch.compile} reference at matching MSE. The kernel accumulates path costs in FP16 where the reference uses FP32,
which flips $2$--$4\%$ of codes on near-tie paths (measured at $K{=}4$ and at $L{=}16$) and moves the
MSE by at most $0.02\%$ ($L \in [10, 16]$, $K \in [1, 6]$, both encoding the
same FP16 input). At the main design point ($L{=}16$, $K{=}4$) it
sustains ${\sim}60$ million values/s, putting $7.0$ billion values, the size of
Llama 3.1 8B's linear layers, under two GPU-minutes; smaller trellises encode
proportionally faster ($1.1$ billion values/s at $L{=}12$). The
kernel is compute-limited: an Nsight Compute profile at $L{=}10$, $K{=}4$
shows $73\%$ SM throughput against under $1\%$ DRAM utilization, with
the integer ALU pipe the busiest at $70\%$.

\paragraph{Against QTIP's shipped encoder.} QTIP's released
repository encodes with a batched PyTorch Viterbi
(\texttt{bitshift\_codebook.quantize}, commit \texttt{e90c668}),
the same framework-level form as our reference. We timed it
unmodified on the same L40S at its shipped settings ($L{=}16$,
$T{=}256$, i.e.\ its $16 \times 16$ tiles), $2^{22}$ standard-normal
values, median of $5$ runs: $0.76$ million values/s for 1MAD, $0.77$
for 3INST and $1.5$ for HYB, whose $V{=}2$ pairs halve the trellis
steps. Our fused encoder at $T{=}256$ runs $59$ million values/s at
$L{=}16$, $1{,}819$ at $L{=}11$ and $3{,}110$ at $L{=}10$. By
\cref{tab:scaling_L}, CPTC at $L{=}11$ scores $3.865$
effective bits against 1MAD's $3.864$ at $L{=}16$ (CPTC $L{=}16$:
$3.934$). Encoding at equal quality is therefore about $2{,}400\times$ faster than QTIP's shipped encoder: $32\times$ less work from running $L{=}11$ instead of $L{=}16$, which is method-level, and the remaining $75\times$ from the fused kernel ($78\times$ when both encoders run at $L{=}16$), which a fused Viterbi for QTIP's codebooks would recover.

\begin{table}[htbp]
  \centering
  \small
  \caption{Fused Viterbi encoder throughput ($\uparrow$, millions of
    values per second) on one L40S at every trellis depth and bit
    width ($T{=}128$, $2^{22}$ values, median of $5$; reference: median of $3$), with the
    speedup over the \texttt{torch.compile} reference of
    \cref{lst:viterbi} in parentheses. Work grows as $2^L$; the
    $K$ dependence comes from backpointer traffic and occupancy.}
  \label{tab:enc_matrix}
  \begin{tabular}{lcccccc}
    \toprule
    $L \backslash K$ & $1$ & $2$ & $3$ & $4$ & $5$ & $6$ \\
    \midrule
    $10$ & $2{,}960$ ($114\times$) & $3{,}539$ ($60\times$) & $3{,}381$ ($35\times$) & $3{,}253$ ($33\times$) & $3{,}526$ ($37\times$) & $3{,}666$ ($37\times$) \\
    $11$ & $1{,}642$ ($127\times$) & $1{,}864$ ($62\times$) & $2{,}002$ ($46\times$) & $1{,}935$ ($43\times$) & $1{,}985$ ($44\times$) & $2{,}080$ ($46\times$) \\
    $12$ & $864$ ($133\times$) & $992$ ($66\times$) & $1{,}020$ ($45\times$) & $1{,}100$ ($49\times$) & $1{,}149$ ($50\times$) & $1{,}098$ ($47\times$) \\
    $13$ & $436$ ($134\times$) & $505$ ($67\times$) & $528$ ($45\times$) & $562$ ($49\times$) & $579$ ($48\times$) & $570$ ($48\times$) \\
    $14$ & $218$ ($134\times$) & $244$ ($65\times$) & $255$ ($44\times$) & $276$ ($48\times$) & $277$ ($47\times$) & $263$ ($44\times$) \\
    $15$ & $84$ ($103\times$) & $105$ ($56\times$) & $116$ ($38\times$) & $127$ ($44\times$) & $133$ ($45\times$) & $131$ ($43\times$) \\
    $16$ & $32$ ($79\times$) & $48$ ($51\times$) & $53$ ($35\times$) & $60$ ($41\times$) & $63$ ($43\times$) & $61$ ($41\times$) \\
    \bottomrule
  \end{tabular}
  
\end{table}

\paragraph{On Trainium~4.} \cref{tab:trn4_enc_matrix} is the
counterpart of \cref{tab:enc_matrix} for the Trainium~4 encoder of
\cref{sec:encode}: the per-chip throughput of the best launch at
every $(L, K)$ in a cycle-level simulation of the pre-silicon design.
These simulated per-chip rates ($T{=}256$) are a reference point,
not a comparison with the measured per-GPU rates of
\cref{tab:enc_matrix} ($T{=}128$).

\begin{table}[htbp]
  \centering
  \small
  \caption{Trainium~4 Viterbi encoder throughput per chip ($\uparrow$,
    millions of values per second) at every trellis depth and bit
    width in a cycle-level simulation ($T{=}256$, best launch per
    cell). ``---'': not
    supported, as the backtrace needs $L - K \ge 3$;
    $^{\dagger}$rows split across partitions, the only layout that
    fits.}
  \label{tab:trn4_enc_matrix}
  \begin{tabular}{lcccccc}
    \toprule
    $L \backslash K$ & $1$ & $2$ & $3$ & $4$ & $5$ & $6$ \\
    \midrule
    $8$ & $14{,}620$ & $15{,}518$ & $16{,}464$ & $16{,}586$ & $16{,}328$ & --- \\
    $9$ & $7{,}814$ & $8{,}333$ & $8{,}774$ & $8{,}924$ & $8{,}846$ & $8{,}624$ \\
    $10$ & $3{,}895$ & $4{,}163$ & $4{,}476$ & $4{,}558$ & $4{,}583$ & $4{,}546$ \\
    $11$ & $1{,}921$ & $2{,}063$ & $2{,}194$ & $2{,}284$ & $2{,}321$ & $2{,}321$ \\
    $12$ & $989$ & $1{,}058$ & $1{,}063$ & $1{,}139$ & $1{,}159$ & $1{,}170$ \\
    $13$ & $498$ & $529$ & $546$ & $564$ & $575$ & $584$ \\
    $14$ & $252$ & $263$ & $274$ & $281$ & $283$ & $290$ \\
    $15$ & $120$ & $129$ & $138$ & $140$ & $141$ & $141$ \\
    $16$ & $42$$^{\dagger}$ & $63$ & $67$ & $69$ & $70$ & $70$ \\
    \bottomrule
  \end{tabular}
\end{table}

\end{document}